\def\WITHAPPENDIX{}

\ifdefined\WITHAPPENDIX
\documentclass[acmsmall, nonacm]{acmart}
\else
\documentclass[acmsmall]{acmart}
\fi
\AtBeginDocument{%
  }

\usepackage{etoolbox}
\usepackage{stmaryrd}
\usepackage{amstext}
\usepackage{xspace}
\usepackage[svgnames]{xcolor}
\usepackage{mathtools}
\usepackage{hyperref}
\usepackage{cleveref}
\usepackage{mathpartir}

\usepackage{amsfonts}
\usepackage{booktabs}
\usepackage{calc}
\usepackage{listings}
\usepackage{subcaption}
\usepackage{tikz}
\usetikzlibrary{arrows,automata}
\usepackage{subcaption}
\usepackage{twoopt}
\usepackage{upgreek}
\usepackage{upquote}
\usepackage{xspace}
\usepackage{thmtools} 

\ifdefined\NOCOLOR%
  \newcommand{\withcolor}[2]{#2} 
\else
  \newcommand{\withcolor}[2]{{\color{#1}{#2}}}
\fi

\definecolor{amethyst}{rgb}{0.6, 0.4, 0.8}

\newcommand{\colorexp}{DarkRed!50} 
\newcommand{\colorpart}{DarkMagenta} 
\newcommand{\colorlbl}{DarkMagenta!50} 
\newcommand{\colorproc}{DarkRed} 
\newcommand{\colortpexp}{DarkBlue!50} 
\newcommand{\colorgt}{DarkBlue} 

\newcommand{\code}[1]{\texttt{#1}}

\newcommand{\oft}[0]{\ensuremath{\mathrel{:}}}
\renewcommand{\emptyset}{\varnothing}

\newcommand{\globalt}[1]{\withcolor{\colorgt}{#1}} 

\newcommand{\expt}[1]{\withcolor{\colortpexp}{#1}} 

\newcommand{\dlbl}[1]{\withcolor{\colorlbl}{#1}} 

\newcommand{\dpart}[1]{\withcolor{\colorpart}{\sf #1}} 

\newcommand{\expr}[1]{\withcolor{\colorexp}{#1}} 
\newcommand{\proc}[1]{\withcolor{\colorproc}{#1}} 

\newcommand{\Rp}{\dpart{p}}
\newcommand{\Rq}{\dpart{q}}
\newcommand{\Rr}{\dpart{r}}
\newcommand{\Rs}{\dpart{s}}

\newcommand{\lbl}{\ensuremath\dlbl{\ell}}

\newcommand{\vargu}{\ensuremath{\globalt{\code{g}}}}

\newcommand{\gt}[1]{\ensuremath{\globalt{#1}}}

\newcommand{\gstep}[3]{\ensuremath{\gt{\globalt{#1} \globalt{\xrightarrow{\globalt{#2}}} {\globalt{#3}}}}}
\renewcommand{\gstep}[3]{\ensuremath{\sttransxxx{#1}{#2}{#3}}}

\newcommand{\poft}[6]{\ensuremath{\denvx{#1}; \senvx{#2} \vdash \dpart{#6} \triangleleft \proc{#4} \oft \gt{#5}}}
\newcommand{\poftemptyenvs}[4]{\ensuremath{{\vdash \dpart{#4} \triangleleft \proc{#2} \oft \gt{#3}}}}

\newcommand{\indep}{\ensuremath{\mathrel{\Diamond}}}

\newcommand{\todo}[1]{}

\usepackage{ifthen}
\usepackage{subcaption}
\usepackage{tikz}
\usepackage{twoopt}
\usepackage{centernot}

\usetikzlibrary{calc}

\newcommand{\lnamex}[1]{\withcolor{DarkMagenta!50}{#1}}
\newcommand{\rnamex}[2][\sf]{\withcolor{DarkMagenta}{#1#2}}

\newcommand{\rolexdefault}{\small}
\newcommandtwoopt{\rolex}[3][\rolexdefault][\color{DarkMagenta}]{{\pmb{\text{\upshape#1#2\texttt{#3}}}}}

\newcommand{\datax}[2][\dataxdefault]{\smash{\texttt{{\renewcommand{\ }{\phantom{x}}\upshape#1\fontdimen2\font=.375em#2}}}}

\newcommand{\pre}{\mspace{1mu}{\textsf{.}}\mspace{1mu}}
\newcommand{\prewide}{\mskip\medmuskip{\textsf{.}}\mskip\medmuskip}
\newcommand{\commawide}{\mskip\medmuskip{{,}}\mskip\medmuskip}

\newcommand{\mtypexx}[2]{\withcolor{DarkBlue}{\lnamex{#1}\ifthenelse{\equal{#2}{}}{}{(\dtypex{#2})}}}
\newcommand{\dtypex}[1]{\withcolor{DarkBlue!50}{#1}}
\newcommand{\stypex}[1]{\withcolor{DarkBlue}{#1}}

\newcommand{\one}{\stypex{\mathbf{end}}}

\newcommand{\isa}[1][\stypex{\mspace{1mu}}]{#1{:}#1}
\newcommand{\comxx}[3][\isa]{\stypex{\rnamex{#2} \unbuf \rnamex{#3} #1}}
\newcommand{\comxxxx}[4]{\stypex{\rnamex{#1} \unbuf \rnamex{#2} \isa \mtypexx{#3}{#4}}}
\newcommand{\comxxxxx}[5]{\stypex{\rnamex{#1} \unbuf \rnamex{#2} \isa \mtypexx{#3}{#4} \pre #5}}
\newcommand{\comxxxxxx}[6]{\stypex{\rnamex{#1} \unbuf \rnamex{#2} \isa \famxx{\mtypexx{#3}{#4} \pre #5}{#6}}}

\newcommand{\mux}[1]{\stypex{\upmu#1}}
\newcommand{\muxx}[2]{\stypex{\mux{#1} \pre #2}}
\newcommand{\parxx}[2]{\stypex{#1 \mspace\medmuskip{\parallel}\mspace\medmuskip #2}}

\newcommand{\sel}[1][\stypex{\mspace{1mu}}]{#1{\oplus}#1}
\newcommand{\selx}[2][\sel]{\stypex{\rnamex{#2} #1}}
\newcommand{\selxxx}[3]{\stypex{\rnamex{#1} \sel \mtypexx{#2}{#3}}}
\newcommand{\selxxxxx}[5]{\stypex{\rnamex{#1} \sel \famxx{\mtypexx{#2}{#3} \pre #4}{#5}}}

\newcommand{\brn}[1][\stypex{\mspace{1mu}}]{#1{\&}#1}
\newcommand{\brnx}[2][\brn]{\stypex{\rnamex{#2} #1}}
\newcommand{\brnxxx}[3]{\stypex{\rnamex{#1} \brn \mtypexx{#2}{#3}}}
\newcommand{\brnxxxxx}[5]{\stypex{\rnamex{#1} \brn \famxx{\mtypexx{#2}{#3} \pre #4}{#5}}}

\newcommandtwoopt{\sttransx}[3][][]{{\stypex{#3} \mathrel{{#1{\rightarrow}}\ifthenelse{\equal{#2}{}}{}{{}^{#2}}}}}
\newcommandtwoopt{\sttransxx}[4][][]{{\stypex{#3} \mathrel{{#1{\xrightarrow{\stypex{#4}}}}\ifthenelse{\equal{#2}{}}{}{{}^{#2}}}}}
\newcommandtwoopt{\sttransxxx}[5][][]{\stypex{#3} \mathrel{{#1{\xrightarrow{\stypex{#4}}}}\ifthenelse{\equal{#2}{}}{}{{}^{#2}}} \stypex{#5}}

\newcommandtwoopt{\sttransinclxx}[4][][]{{\stypex{#3} \mathrel{{#1{\xrightarrow{\rnamex[]{#4}}}}\ifthenelse{\equal{#2}{}}{}{{}^{#2}}}}}
\newcommandtwoopt{\sttransinclxxx}[5][][]{\stypex{#3} \mathrel{{#1{\xrightarrow{\rnamex[]{#4}}}}\ifthenelse{\equal{#2}{}}{}{{}^{#2}}} \stypex{#5}}

\newcommandtwoopt{\sttransexclxx}[4][][]{{\stypex{#3} \mathrel{{#1{\xrightarrow{\overline{\rnamex[]{#4}}}}}\ifthenelse{\equal{#2}{}}{}{{}^{#2}}}}}
\newcommandtwoopt{\sttransexclxxx}[5][][]{\stypex{#3} \mathrel{{#1{\xrightarrow{\overline{\rnamex[]{#4}}}}}\ifthenelse{\equal{#2}{}}{}{{}^{#2}}} \stypex{#5}}

\newcommandtwoopt{\stTransexclxx}[4][][]{{\stypex{#3} \mathrel{{#1{\xRightarrow{\overline{\rnamex[]{#4}}}}}\ifthenelse{\equal{#2}{}}{}{{}^{#2}}}}}
\newcommandtwoopt{\stTransexclxxx}[5][][]{\stypex{#3} \mathrel{{#1{\xRightarrow{\overline{\rnamex[]{#4}}}}}\ifthenelse{\equal{#2}{}}{}{{}^{#2}}} \stypex{#5}}

\newcommand{\mexprxx}[2]{\lnamex{#1}\ifthenelse{\equal{#2}{}}{}{(\dexprx{#2})}}
\newcommand{\dexprx}[1]{\withcolor{DarkRed!50}{#1}}
\newcommand{\sexprx}[1]{\withcolor{DarkRed}{#1}}

\newcommand{\nil}{\sexprx{\mathbf{end}}}

\newcommand{\snd}[1][\sexprx{\mspace{1mu}}]{#1{!}#1}

\newcommand{\sndxxx}[3]{\sexprx{\rnamex{#1} \snd \mexprxx{#2}{#3}}}
\newcommand{\sndxxxx}[4]{\sexprx{\rnamex{#1} \snd \mexprxx{#2}{#3} \pre #4}}

\newcommand{\rcv}[1][\sexprx{\mspace{1mu}}]{#1{?}#1}
\newcommand{\rcvx}[2][\rcv]{\sexprx{\rnamex{#2} #1}}
\newcommand{\rcvxx}[2]{\sexprx{\rnamex{#1} \rcv \setx{#2}}}
\newcommand{\rcvxxx}[3]{\sexprx{\rnamex{#1} \rcv \mexprxx{#2}{#3}}}
\newcommand{\rcvxxxxx}[5]{\sexprx{\rnamex{#1} \rcv \famxx{\mexprxx{#2}{#3} \pre #4}{#5}}}

\newcommand{\ifxxx}[3]{\sexprx{\mathbf{if}\mskip\medmuskip\dexprx{#1}\mskip\medmuskip\mathbf{then}\mskip\medmuskip#2\mskip\medmuskip\mathbf{else}\mskip\medmuskip#3}}

\newcommand{\letxx}[2]{\sexprx{\mathbf{let}\mskip\medmuskip\dexprx{#1}{=}\dexprx{#2}\mskip\medmuskip\proc{\mathbf{in}}\mskip\medmuskip}}
\newcommand{\letxxx}[3]{\sexprx{\mathbf{let}\mskip\medmuskip\dexprx{#1}{=}\dexprx{#2}\mskip\medmuskip\proc{\mathbf{in}}\mskip\medmuskip#3}}

\newcommand{\recx}[1]{\sexprx{\mathbf{rec}\mskip\medmuskip#1}}
\newcommand{\recxx}[2]{\sexprx{\recx{#1} \pre #2}}

\newcommand{\atxx}[2]{\sexprx{\rnamex{#1} \triangleleft #2}}
\newcommand{\compxx}[2]{\sexprx{#1\mskip\medmuskip{\mid}\mskip\medmuskip#2}}
\newcommand{\compxxx}[3]{\sexprx{#1\mskip\medmuskip{\mid}\mskip\medmuskip#2\mskip\medmuskip{\mid}\mskip\medmuskip#3}}

\newcommand{\evalxx}[2]{\dexprx{#1} \Downarrow \dexprx{#2}}

\newcommandtwoopt{\setransx}[3][][]{{\sexprx{#3} \mathrel{{#1{\rightarrow}}\ifthenelse{\equal{#2}{}}{}{{}^{#2}}}}}
\newcommandtwoopt{\setransxxx}[5][][]{\sexprx{#3} \mathrel{{#1{\xrightarrow{\stypex{#4}}}}\ifthenelse{\equal{#2}{}}{}{{}^{#2}}} \sexprx{#5}}

\newcommand{\denvx}[1]{\withcolor{DarkMagenta!50}{#1}}
\newcommand{\denvxxx}[3]{\denvx{#1, \dexprx{#2} : \dtypex{#3}}}
\newcommand{\senvx}[1]{\withcolor{DarkMagenta}{#1}}
\newcommand{\senvxxx}[3]{\senvx{#1, \sexprx{#2} : \stypex{#3}}}

\newcommand{\dtypedxx}[2]{{\vdash \dexprx{#1} : \dtypex{#2}}}
\newcommand{\dtypedxxx}[3]{\denvx{#1} \vdash \dexprx{#2} : \dtypex{#3}}
\newcommand{\stypedxx}[2]{{\vdash \sexprx{#1} : \stypex{#2}}}
\newcommand{\stypedxxx}[3]{\denvx{#1} \vdash \sexprx{#2} : \stypex{#3}}
\newcommand{\stypedxxxx}[4]{\denvx{#1}; \senvx{#2} \vdash \sexprx{#3} : \stypex{#4}}

\newcommand{\denotx}[1]{\llbracket#1\rrbracket}
\newcommand{\substxx}[2]{\withcolor{black}{[}#1 \mathbin{\withcolor{black}{:=}} #2\withcolor{black}{]}}

\newcommand{\unbuf}{\stypex{\mspace{1mu}{\unbufsym}\mspace{1mu}}}
\newcommand{\unbufsym}{\stypex{{\rightarrowtriangle}}}

\newcommand{\typed}{\mathrel{\typedsym}}
\newcommand{\typedsym}{{:}}

\newcommand{\famxx}[2]{\setx{#1}_{\smash{#2}}}
\newcommand{\setxx}[2]{\{#1\mid#2\}}
\newcommand{\setx}[1]{\{#1\}}

\newcommand{\rulexxx}[3]{\dfrac{#2}{#3}\ifthenelse{\equal{#1}{}}{}{\, \text{\upshape\scriptsize[\hypertarget{\detokenize{#1}}{\textsc{#1}}]}}}

\newcommand{\GRAMMAR}{\enspace{::=}\enspace}
\newcommand{\PIPE}{\enspace{\big|}\enspace}

\newenvironment{tree}{%
	\begin{tikzpicture}[remember picture, anchor=base west, baseline={(0,0)}, inner sep=0pt, line width=0pt]%
		\newcommand{\holex}[1]{{\tikz{\coordinate (##1)}}}%
		\newcommand{\branchx}[2][0,0]{\node at (##1) {\ensuremath{##2}};}%
		\newcommandtwoopt{\branchxx}[4][0,0][]{\node[##2] at (##1) {\ensuremath{\branchherexx{##3}{##4}}};}%
		\newcommandtwoopt{\branchxxx}[5][0,0][]{\node[##2] at (##1) {\ensuremath{\branchherexxx{##3}{##4}{##5}}};}%
		\newcommand{\branchherexx}[2]{{\left\{\begin{aligned}{}&##1\\&##2\end{aligned}\!\right.}}%
		\newcommand{\branchherexxx}[3]{{\left\{\begin{aligned}{}&##1\\&##2\\&##3\end{aligned}\!\right.}}%
}{%
	\end{tikzpicture}%
}

\newcommand{\autorefrule}[1]{\ensuremath{\text{\hyperlink{\detokenize{#1}}{[\textsc{#1}]}}}}

\setcopyright{acmlicensed}
\copyrightyear{2018}
\acmYear{2018}
\acmDOI{XXXXXXX.XXXXXXX}

\ifdefined\WITHAPPENDIX
\else
\acmJournal{JACM}
\acmVolume{37}
\acmNumber{4}
\acmArticle{111}
\acmMonth{8}
\fi

\begin{document}

\newcommand{\agdalogo}{\raisebox{-.15\baselineskip}{\includegraphics[height=.85\baselineskip]{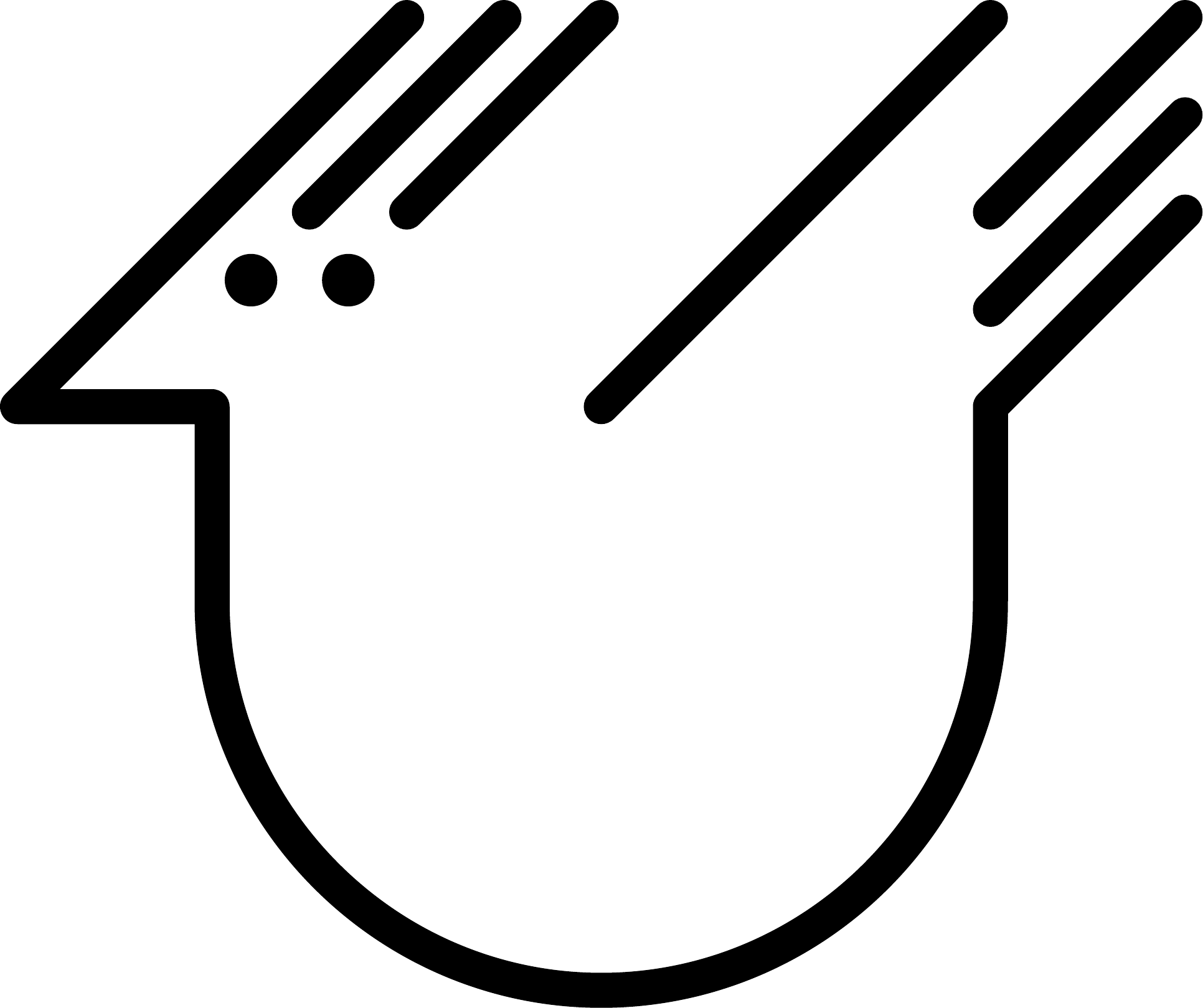}}}

\newcounter{agdaenvs}
\newcommand{\agdaenvname}{}
\newcommand{\agdaenvlabel}{}
\newenvironment{agdaenv}[3]{
  \renewcommand{\agdaenvname}{agdaenv\arabic{agdaenvs}}
  \ifthenelse{\equal{#1}{theorem}}{\renewcommand{\agdaenvlabel}{Theorem}}{}
  \ifthenelse{\equal{#1}{lemma}}{\renewcommand{\agdaenvlabel}{Lemma}}{}
  \ifthenelse{\equal{#1}{corollary}}{\renewcommand{\agdaenvlabel}{Corollary}}{}
  \newtheorem{\agdaenvname}[#1]{\smash{\llap{\href{#3}{\agdalogo}\enspace}}\agdaenvlabel}
  \ifthenelse{\equal{#2}{}}{
    \begin{\agdaenvname}
  }{
    \begin{\agdaenvname}[#2]
  }
}{
  \end{\agdaenvname}
  \stepcounter{agdaenvs}
}

\definecolor{RED}{rgb}{1, 0, 0}
\newcommand{\rev}[1]{{\color{red}#1}}
\newcommand{\startrev}{\begingroup\color{red}}
\newcommand{\finishrev}{\endgroup}

\title{A Synthetic Reconstruction of Multiparty Session Types\ifdefined\WITHAPPENDIX{ (with Appendix)}\fi}

\author{David Castro-Perez}
\orcid{0000-0002-6939-4189}
\affiliation{%
  \institution{University of Kent}
  \city{Canterbury}
  \country{UK}}
\email{d.castro-perez@kent.ac.uk}

\author{Francisco Ferreira}
\orcid{0000-0001-8494-7696}
\affiliation{%
  \institution{Royal Holloway, University of London}
  \city{Egham}
  \country{UK}}
\email{francisco.ferreiraruiz@rhul.ac.uk}

\author{Sung-Shik Jongmans}
\orcid{0000-0002-4394-8745}
\affiliation{%
  \institution{University of Groningen}
  \city{Groningen}
  \country{The Netherlands}}
\email{s.s.t.q.jongmans@rug.nl}

\begin{abstract}
  Multiparty session types (MPST) provide a rigorous foundation for verifying
  the safety and liveness of concurrent systems. However, existing approaches
  often force a difficult trade-off: classical, projection-based techniques are
  compositional but limited in expressiveness, while more recent techniques
  achieve higher expressiveness by relying on non-compositional, whole-system
  model checking, which scales poorly.

  This paper introduces a new approach to MPST that delivers both expressiveness
  and compositionality, called the synthetic approach. Our key innovation is a
  type system that verifies each process directly against a global protocol
  specification, represented as a labelled transition system (LTS) in general,
  with global types as a special case. This approach uniquely avoids the need for
  intermediate local types and projection.

  We demonstrate that our approach, while conceptually simpler, supports a
  benchmark of challenging protocols that were previously beyond the reach of
  compositional techniques in the MPST literature. We generalise our type
  system, showing that it can validate processes against any specification that
  constitutes a ``well-behaved'' LTS, supporting protocols not expressible with
  the standard global type syntax. The entire framework, including all theorems
  and many examples, has been formalised and mechanised in Agda, and we have
  developed a prototype implementation as an extension to VS Code.
\end{abstract}




\maketitle

\begin{acks}
  This work is partially supported by the
  \grantsponsor{epsrc}{EPSRC}{https://www.ukri.org/councils/epsrc/} grants
  \grantnum{epy00339x1}{EP/Y00339X/1} and \grantnum{ept0145121}{EP/T014512/1}.
\end{acks}

\renewcommand{\rev}[1]{#1}
\renewcommand{\startrev}{}
\renewcommand{\finishrev}{}

\raggedbottom


\section{Introduction}\label{sect:intr}

Programming of concurrent systems is hard. One of the challenges is to
prove---broadly construed---that implementations of \textit{protocols} among
message-passing processes are \textit{safe} and \textit{live} relative to
specifications. Safety means that ``bad'' communications never happen: {if} a
communication happens in the implementation, {then} it is allowed to happen by
the specification. Liveness means that ``good'' communications eventually
happen. \textit{Multiparty session typing}
(MPST)~\citep{DBLP:conf/popl/HondaYC08} is a method to automatically prove the
safety and liveness of protocol implementations relative to specifications. The
idea is to write specifications as \emph{behavioural
types}~\citep{DBLP:journals/ftpl/AnconaBB0CDGGGH16,DBLP:journals/csur/HuttelLVCCDMPRT16} against which implementations are type-checked. Well-typedness, then, implies safety and liveness.

In this paper, we present \textbf{a new approach to MPST, called the
\textit{synthetic approach}}. Inspired by the recent concept of
\textit{synthetic behavioural typing} \citep{DBLP:conf/ecoop/JongmansF23}, the
synthetic approach to MPST has \textbf{a unique way of combining high
expressiveness and compositional verification}, significantly beyond the state
of the art in the MPST literature. Moreover, we show that the synthetic approach
can be generalised to verify protocol implementations relative to specifications
expressed as \textit{labelled transition systems} (semantic objects) instead of
as behavioural types (syntactic objects). This makes the synthetic approach very
broadly applicable.

In the rest of this section, we explain in more detail the MPST method
(\Cref{sect:intr:mpst}), the state-of-the-art (\Cref{sect:intr:sota}), and our
contributions (\Cref{sect:intr:contrib}).

\subsection{Multiparty Session Typing (MPST) -- \textit{Classical Approach}}\label{sect:intr:mpst}


To explain the MPST method, \Cref{fig:mpst:popl08} visualises the idea \rev{(while \Cref{fig:mpst:popl19} and \Cref{fig:mpst:this} are discussed in \Cref{sect:intr:sota} and \Cref{sect:overv})}:

\begin{figure}[t]
	\begin{minipage}[b]{\linewidth*1/3}\centering
		\begin{tikzpicture}[x=.5cm, y=-1cm, font=\footnotesize]
			\tikzstyle{object} = [inner sep=0mm, anchor=base, outer sep=1mm]
			\node [object] (G) at (0,0) {$\stypex{G}$};
			\node [object] (L1) at (-1.5,1) {$\stypex{L_1}$};
			\node [object] (L2) at (-.5,1) {$\stypex{L_2}$};
			\node [object] (Ldots) at (.5,1) {$\cdots$};
			\node [object] (Ln) at (1.5,1) {$\stypex{L_n}$};
			\node [object] (P1) at (-1.5,2) {$\sexprx{P_1}$};
			\node [object] (P2) at (-.5,2) {$\sexprx{P_2}$};
			\node [object] (Pdots) at (.5,2) {$\cdots$};
			\node [object] (Pn) at (1.5,2) {$\sexprx{P_n}$};

			\tikzstyle{arrow} = [-stealth, rounded corners]
			\tikzstyle{arrow up} = [stealth-, rounded corners]
			\draw [arrow] (G.south) to (L1.north);
			\draw [arrow] (G.south) to (L2.north);
			\draw [arrow] (G.south) to (Ln.north);
			\draw [arrow up] (L1.south) to (P1.north);
			\draw [arrow up] (L2.south) to (P2.north);
			\draw [arrow up] (Ln.south) to (Pn.north);

			\tikzstyle{label} = [inner sep=0mm, anchor=base east, minimum width=1.75cm]
			\node [label] at (-2,0) {global type};
			\node [label] at (-2,.5) {project};
			\node [label] at (-2,1) {local types};
			\node [label] at (-2,1.5) {type-check};
			\node [label] at (-2,2) {processes};
		\end{tikzpicture}
		\subcaption{Classical \citep{DBLP:conf/popl/HondaYC08}}
		\label{fig:mpst:popl08}
	\end{minipage}%
	\begin{minipage}[b]{\linewidth*1/3}\centering
		\begin{tikzpicture}[x=.5cm, y=-1cm, font=\footnotesize]
			\tikzstyle{object} = [inner sep=0mm, anchor=base, outer sep=1mm]
			\node [object] (Phi) at (0,0) {$\stypex{\varphi}$};
			\node [object] (L1) at (-1.5,1) {$\stypex{L_1}$};
			\node [object] (L2) at (-.5,1) {$\stypex{L_2}$};
			\node [object] (Ldots) at (.5,1) {$\cdots$};
			\node [object] (Ln) at (1.5,1) {$\stypex{L_n}$};
			\node [object] (P1) at (-1.5,2) {$\sexprx{P_1}$};
			\node [object] (P2) at (-.5,2) {$\sexprx{P_2}$};
			\node [object] (Pdots) at (.5,2) {$\cdots$};
			\node [object] (Pn) at (1.5,2) {$\sexprx{P_n}$};

			\pgfinterruptboundingbox
			\draw [thick, lightgray, dashed, rounded corners] ([yshift=-.25cm]L1.south west) to (L1.north west) to node (X) {} (Ln.north east) to ([yshift=-.25cm]Ln.south east) to cycle;
			\endpgfinterruptboundingbox

			\tikzstyle{arrow} = [-stealth, rounded corners]
			\tikzstyle{arrow up} = [stealth-, rounded corners]
			\draw [arrow up] (Phi.south) to (X.north);
			\draw [arrow up] (L1.south) to (P1.north);
			\draw [arrow up] (L2.south) to (P2.north);
			\draw [arrow up] (Ln.south) to (Pn.north);

			\tikzstyle{label} = [inner sep=0mm, anchor=base east, minimum width=1.75cm]
			\node [label] at (-2,0) {consistency};
			\node [label] at (-2,.5) {model-check};
			\node [label] at (-2,1) {local types};
			\node [label] at (-2,1.5) {type-check};
			\node [label] at (-2,2) {processes};
		\end{tikzpicture}
		\subcaption{``Less Is More'' \citep{DBLP:journals/pacmpl/ScalasY19}}
		\label{fig:mpst:popl19}
	\end{minipage}%
	\begin{minipage}[b]{\linewidth*1/3}
		\begin{tikzpicture}[x=.5cm, y=-1cm, font=\footnotesize]
			\tikzstyle{object} = [inner sep=0mm, anchor=base, outer sep=1mm]
			\node [object] (G) at (0,0) {$\stypex{G}$};
			\node [object] (P1) at (-1.5,2) {$\sexprx{P_1}$};
			\node [object] (P2) at (-.5,2) {$\sexprx{P_2}$};
			\node [object] (Pdots) at (.5,2) {$\cdots$};
			\node [object] (Pn) at (1.5,2) {$\sexprx{P_n}$};

			\tikzstyle{arrow} = [-stealth, rounded corners]
			\tikzstyle{arrow up} = [stealth-, rounded corners]
			\draw [arrow up] (G) to (P1.north);
			\draw [arrow up] (G) to (P2.north);
			\draw [arrow up] (G) to (Pn.north);

			\tikzstyle{label} = [inner sep=0mm, anchor=base east, minimum width=1.75cm]
			\node [label] at (-2,0) {global type};
			\node [label] at (-2,1.5) {type-check};
			\node [label] at (-2,2) {processes};
		\end{tikzpicture}
		\subcaption{Synthetic [this paper]}
		\label{fig:mpst:this}
	\end{minipage}%

	\caption{MPST approaches}
	\Description{}
	\label{fig:mpst}
\end{figure}
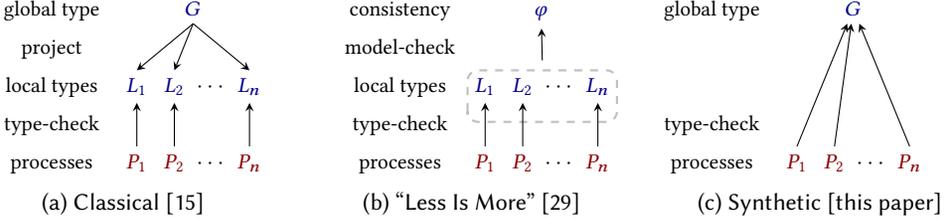

\begin{figure}[t]
	\tikzstyle{process} = [inner sep=.75mm, draw, anchor=south, rounded corners=.5mm, minimum height=3.75mm, minimum width=3.75mm, font=\scriptsize]
	\tikzstyle{lifeline} = [-|, dashed]
	\tikzstyle{com} = [-latex]
	\tikzstyle{value} = [inner sep=0pt, anchor=base, yshift=.5mm, font=\scriptsize]
	\begin{minipage}{\linewidth*1/2}\centering
		\begin{tikzpicture}[x=2.25cm, y=-1cm*1/3, font=\footnotesize]
			\node [process] (a) at (0,-.75) {$\rolex[]{a}$};
			\node [process] (b) at (1,-.75) {$\rolex[]{b}$};
			\node [process] (c) at (2,-.75) {$\rolex[]{c}$};
			\draw [lifeline] (0,-.5) to (0,2.75);
			\draw [lifeline] (1,-.5) to (1,2.75);
			\draw [lifeline] (2,-.5) to (2,2.75);
			\draw [lifeline, draw=none] (0,-.5) to (0,3.75);

			\draw [com] ([xshift=.5mm]0,0) to node [value] {$\mexprxx{\datax[]{AppThenGet}}{\datax[]{5}}$} ([xshift=-.5mm]1,0);
			\draw [com] ([xshift=.5mm]1,1) to node [value] {$\mexprxx{\datax[]{AppThenGet}}{\datax[]{6}}$} ([xshift=-.5mm]2,1);
			\draw [com] ([xshift=-.5mm]2,2) to node [value, pos=.75] {$\mexprxx{\datax[]{Val}}{\datax[]{12}}$} ([xshift=.5mm]0,2);
		\end{tikzpicture}
		\subcaption{Final number pushed by Carol}
		\label{fig:ring:push}
	\end{minipage}%
	\begin{minipage}{\linewidth*1/2}\centering
		\begin{tikzpicture}[x=2.25cm, y=-1cm*1/3, font=\footnotesize]
			\node [process] (a) at (0,-.75) {$\rolex[]{a}$};
			\node [process] (b) at (1,-.75) {$\rolex[]{b}$};
			\node [process] (c) at (2,-.75) {$\rolex[]{c}$};
			\draw [lifeline] (0,-.5) to (0,3.75);
			\draw [lifeline] (1,-.5) to (1,3.75);
			\draw [lifeline] (2,-.5) to (2,3.75);

			\draw [com] ([xshift=.5mm]0,0) to node [value] {$\mexprxx{\datax[]{App}}{\datax[]{5}}$} ([xshift=-.5mm]1,0);
			\draw [com] ([xshift=.5mm]1,1) to node [value] {$\mexprxx{\datax[]{App}}{\datax[]{6}}$} ([xshift=-.5mm]2,1);
			\draw [com] ([xshift=.5mm]0,2) to node [value, pos=.25] {$\mexprxx{\datax[]{Get}}{}$} ([xshift=-.5mm]2,2);
			\draw [com] ([xshift=-.5mm]2,3) to node [value, pos=.75] {$\mexprxx{\datax[]{Val}}{\datax[]{12}}$} ([xshift=.5mm]0,3);
		\end{tikzpicture}
		\subcaption{Final number pulled by Alice}
		\Description{}
		\label{fig:ring:pull}
	\end{minipage}

	\caption{Example runs of the Ring protocol}
	\label{fig:ring}
\end{figure}
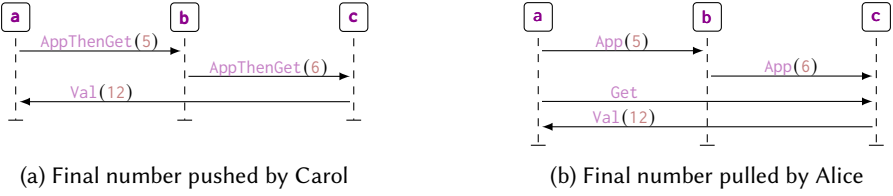

\begin{enumerate}
	\item First, a protocol among roles $\rnamex{r_1}, \ldots, \rnamex{r_n}$ is
	implemented as a family of \textbf{processes} $\sexprx{P_1}, \ldots,
	\sexprx{P_n}$, while it is specified as a \textbf{global type} $\stypex{G}$.
	The global type models the behaviour of all processes together (e.g., ``a
	number from Alice to Bob, followed by a boolean from Bob to Carol'').

	\item Next, $\stypex{G}$ is decomposed into a family of \textbf{local types}
	$\stypex{L_1}, \ldots, \stypex{L_n}$ by \textbf{projecting} $\stypex{G}$ onto
	each role. Each local type models the behaviour of one process alone (e.g., for
	Bob, ``a number from Alice, followed by a boolean to Carol'').

	\item Last, the family of processes is verified by \textbf{type-checking}
	$\sexprx{P_i}$ against $\stypex{L_i}$ for each role. The main result is that
	well-typedness implies safety and liveness: if each process is statically
	well-typed at compile-time, then the parallel composition of the family of
	processes is dynamically safe and live at run-time.
\end{enumerate}
The following example demonstrates the MPST method.

\begin{example}\label{exmp:ring}
	The \textit{Ring} protocol consists of roles \textit{Alice}, \textit{Bob}, and
	\textit{Carol}:
	\begin{itemize}
		\item Alice sends initial number $n$ to Bob.

		\item Bob receives $n$, applies function $f$ (e.g., increment), and sends $f(n)$ to Carol.

		\item Carol receives $f(n)$, applies function $g$ (e.g., double), and sends
		$g(f(n))$ to Alice.

		\item Alice receives final number $g(f(n))$.
	\end{itemize}
	There are two ``modes'' in which the protocol can run: either Carol
	\textit{pushes} the final number to Alice immediately after applying $g$, or
	Alice \textit{pulls} the final number from Carol sometime later. The choice
	between the modes is Alice's and communicated along the ring.
	\Cref{fig:ring} visualises example runs.

	The following \textbf{global type} specifies the protocol:
	\begin{gather*}
		\stypex{G^\text{Ring}} = \stypex{\begin{tree}
			\branchx{
				\comxx{\rolex{a}}{\rolex{b}} \holex{h1}
			}
			\branchxx[h1]{
				\mtypexx{\datax{AppThenGet}}{\datax{Nat}}
				\prewide
				\comxxxx{\rolex{b}}{\rolex{c}}{\datax{AppThenGet}}{\datax{Nat}}
				\prewide
				\comxxxx{\rolex{c}}{\rolex{a}}{\datax{Val}}{\datax{Nat}}
				\prewide
				\one
			}{
				\mtypexx{\datax{App}}{\datax{Nat}}
				\prewide
				\comxxxx{\rolex{b}}{\rolex{c}}{\datax{App}}{\datax{Nat}}
				\prewide
				\comxxxx{\rolex{a}}{\rolex{c}}{\datax{Get}}{}
				\prewide
				\comxxxx{\rolex{c}}{\rolex{a}}{\datax{Val}}{\datax{Nat}}
				\prewide
				\one
			}
		\end{tree}}
	\qquad
		\begin{gathered}
			\text{(\textit{push})}
		\\
			\text{(\textit{pull})}
		\end{gathered}
	\end{gather*}
	Global type $\comxxxxxx{p}{q}{\ell_i}{t_i}{G_i}{i \in I}$ specifies the
	communication of a message labelled $\lnamex{\ell_j}$, with a payload of type
	$\dtypex{t_j}$, from role $\rnamex{p}$ to role $\rnamex{q}$, followed by
	$\stypex{G_j}$, for some $j \in I$.%
	\footnote{%
		\rev{We adopt the same notation to represent a collection of
		branches---including the usage of \textit{index set} $I$---as in the
		original MPST paper \cite{DBLP:conf/popl/HondaYC08}. Similar notation
		for branching dates back at least as far as Milner's work on CCS (e.g.,
		\cite{DBLP:conf/podc/Milner82}) and remains standard in recent work
		(e.g., \cite{DBLP:journals/tocl/Glabbeek24}). We stipulate that there is
		a one-to-one correspondence between $I$ and $\setxx{\ell_i}{i \in I}$.}%
	} %
	We omit braces when $I$ is a singleton, and
	we write ``$\mtypexx{\ell}{}$'' instead of ``$\mtypexx{\ell}{\datax{Unit}}$''.
	The following \textbf{local types, projected from the global type,} specify
	Alice and Bob. Let $\lnamex{\ell_1} = \lnamex{\datax{AppThenGet}}$ and
	$\lnamex{\ell_2} = \lnamex{\datax{App}}$:
	\begin{gather*}
		\stypex{L_{\smash{\rolex[][]{a}}}} = \stypex{\begin{tree}
			\branchx{
				\selx{\rolex{b}} \holex{h1}
			}
			\branchxx[h1]{
				\mtypexx{\ell_1}{\datax{Nat}}
				\prewide
				\brnxxx{\rolex{c}}{\datax{Val}}{\datax{Nat}}
				\prewide
				\one
			}{
				\mtypexx{\ell_2}{\datax{Nat}}
				\prewide
				\selxxx{\rolex{c}}{\datax{Get}}{}
				\prewide
				\brnxxx{\rolex{c}}{\datax{Val}}{\datax{Nat}}
				\prewide
				\one
			}
		\end{tree}}
	\qquad
		\stypex{L_{\smash{\rolex[][]{b}}}} = \stypex{\begin{tree}
			\branchx{
				\brnx{\rolex{a}} \holex{h1}
			}
			\branchxx[h1]{
				\mtypexx{\ell_1}{\datax{Nat}}
				\prewide
				\selxxx{\rolex{c}}{\ell_1}{\datax{Nat}}
				\prewide
				\one
			}{
				\mtypexx{\ell_2}{\datax{Nat}}
				\prewide
				\selxxx{\rolex{c}}{\ell_2}{\datax{Nat}}
				\prewide
				\one
			}
		\end{tree}}
	\end{gather*}
	Local types $\selxxxxx{q}{\ell_i}{t_i}{L_i}{i \in I}$ and
	$\brnxxxxx{p}{\ell_i}{t_i}{L_i}{i \in I}$ specify the send and receive of a
	message labelled $\lnamex{\ell_j}$, with a payload of type $\dtypex{t_j}$, from
	role $\rnamex{p}$ to role $\rnamex{q}$, followed by $\stypex{L_j}$, for some $j
	\in I$. The following \textbf{processes, well-typed by the local types,}
	implement Alice and Bob in \Cref{fig:ring:push}:
	\begin{gather*}
		\proc{P_{\smash{\rolex[][]{a}}}^\text{Ring}} = \sexprx{
			\sndxxx{\rolex{b}}{\ell_1}{\datax{5}}
			\prewide
			\rcvxxx{\rolex{c}}{\datax{Val}}{\datax{z}}
			\prewide
			\nil
		}
	\qquad
		\proc{P_{\smash{\rolex[][]{b}}}^\text{Ring}} = \sexprx{\begin{tree}
			\branchx{
				\rcvx{\rolex{a}} \holex{h1}
			}
			\branchxx[h1]{
				\mexprxx{\ell_1}{\datax{x}}
				\prewide
				\sndxxx{\rolex{c}}{\ell_1}{\datax{x+1}}
				\prewide
				\nil
			}{
				\mexprxx{\ell_2}{\datax{x}}
				\prewide
				\sndxxx{\rolex{c}}{\ell_2}{\datax{x+1}}
				\prewide
				\nil
			}
		\end{tree}}
	\end{gather*}
	Process $\sndxxxx{q}{\ell}{e}{P}$ implements the send of a message labelled
	$\lnamex{\ell}$, with (the value of) expression $\dexprx{e}$ as the payload, to
	role $\rnamex{q}$, followed by $\sexprx{P}$. Process
	$\rcvxxxxx{p}{\ell_i}{x_i}{P_i}{i \in I}$ implements the receive of the payload
	of a message labelled $\lnamex{\ell_j}$, from role $\rnamex{p}$, into variable
	$\dexprx{x_j}$, followed by $\sexprx{P_j}$, for some $j \in I$.
	\rev{Communication is \textit{synchronous} in this paper: a send blocks the sender until the receiver is ready to perform a corresponding receive. For instance, if Alice is ready to send a $\mexprxx{\datax[]{Get}}{}$ message before Carol has finished her computation, then Alice needs to wait until Carol is ready to receive.}
	\qed
\end{example}


\subsection{State of the Art -- \textit{``Less Is More'' Approach}\texorpdfstring{ \citep{DBLP:journals/pacmpl/ScalasY19}}{}}\label{sect:intr:sota}

For well-typedness to imply safety and liveness, \textbf{a family of local types
needs to be \textit{consistent}}. Intuitively, consistency means that if the
local type of Alice specifies a send to Bob, then the local type of Bob should
specify a corresponding receive from Alice. That is, consistency is the
multiparty generalisation of binary \textit{duality}
\citep{DBLP:conf/concur/Honda93,DBLP:conf/esop/HondaVK98}.

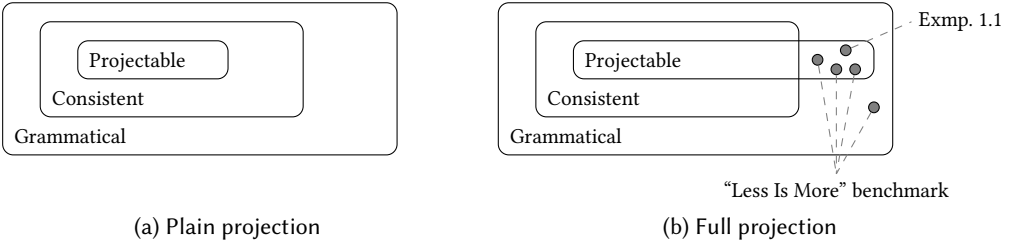
\begin{figure}[t]
	\begin{minipage}[b]{.5\linewidth}\centering
		\begin{tikzpicture}[x=.5cm, y=.5cm, font=\footnotesize]
			\tikzstyle{label} = [pos=0, anchor=base west, xshift=1.5mm, yshift=1.5mm, inner sep=0pt]
			\draw [rounded corners] (0,0) rectangle node [label] {Grammatical}  (10.5,4);
			\draw [rounded corners] (1,1) rectangle node [label] {Consistent}  (8,3.5);
			\draw [rounded corners] (2,2) rectangle node [label] {Projectable} (6,3);
			\node [inner sep=0mm, outer sep=1mm, anchor=north] (Ylabel) at (9,-.5) {\phantom{``Less Is More'' benchmark}};
		\end{tikzpicture}
		\subcaption{Plain projection}
		\label{fig:proj:plain}
	\end{minipage}%
	\begin{minipage}[b]{.5\linewidth}\centering
		\begin{tikzpicture}[x=.5cm, y=.5cm, font=\footnotesize]
			\tikzstyle{label} = [pos=0, anchor=base west, xshift=1.5mm, yshift=1.5mm, inner sep=0pt]
			\draw [rounded corners] (0,0) rectangle node [label] {Grammatical}  (10.5,4);
			\draw [rounded corners] (1,1) rectangle node [label] {Consistent}  (8,3.5);
			\draw [rounded corners] (2,2) rectangle node [label] {Projectable} (10,3);
			\node [inner sep=.5mm, circle, draw=black, fill=black!50] (X) at (9.25,2.75) {};
			\node [inner sep=.5mm, circle, draw=black, fill=black!50] (Y1) at (8.5,2.5) {};
			\node [inner sep=.5mm, circle, draw=black, fill=black!50] (Y2) at (9.0,2.25) {};
			\node [inner sep=.5mm, circle, draw=black, fill=black!50] (Y3) at (9.5,2.25) {};
			\node [inner sep=.5mm, circle, draw=black, fill=black!50] (Y4) at (10,1.25) {};
			\node [inner sep=0mm, outer sep=1mm, anchor=north] (Ylabel) at (9,-.5) {``Less Is More'' benchmark};
			\node [inner sep=0mm, outer sep=1mm, anchor=west] (Xlabel) at (11,3.5) {Exmp.~\ref{exmp:ring}};
			\draw [dashed, black!50] (X) to (Xlabel.west);
			\draw [dashed, black!50] (Y1) to (Ylabel.north);
			\draw [dashed, black!50] (Y2) to (Ylabel.north);
			\draw [dashed, black!50] (Y3) to (Ylabel.north);
			\draw [dashed, black!50] (Y4) to (Ylabel.north);
		\end{tikzpicture}
		\subcaption{Full projection}
		\label{fig:proj:full}
	\end{minipage}
	\caption{(Sub)sets of global types in the classical approach: ``Grammatical'' indicates the set of all global types; ``Consistent'' indicates the subset of global types for which consistent families of local types exist; ``Projectable'' indicates the subset of global types for which families of local types can be constructed through projection.}
	\label{fig:proj}
	\Description{}
\end{figure}

In the original paper in the MPST literature \citep{DBLP:conf/popl/HondaYC08},
\textbf{projection implies consistency}: if a family of local types is projected
from a global type, then that family is consistent. Thus, well-typedness implies
safety and liveness. The trouble with the original paper, though, is that only
few global types can be projected. Formally, projection is a function
from global type--role pairs, but its domain in the original paper is small.
\Cref{fig:proj:plain} visualises the issue. The following example demonstrates
that it affects the Ring protocol in \Cref{exmp:ring}.

\begin{example}\label{exmp:plain} The projections onto Alice and Bob of
	$\smash{\stypex{G^\text{Ring}}}$ are defined as
	$\smash{\stypex{L_{\smash{\rolex[][]{a}}}^\text{Ring}}}$ and
	$\smash{\stypex{L_{\smash{\rolex[][]{b}}}^\text{Ring}}}$ in
	\Cref{exmp:ring}, but the pro\-jection onto Carol is undefined.
	\rev{Intuitively, this is because the basic ``plain projection'' of the
	original paper demands that Carol has exactly the same behaviour in each of
	the branches (i.e., even though Carol can actually learn which branch is
	taken based on the label of the message she receives, the plain projection
	does not leverage this additional information).} As a result, in the absence
	of a local type for Carol, the implementation cannot be fully type-checked,
	so safety and liveness cannot be proved. Thus, the Ring protocol is not
	actually supported.
	\qed
\end{example}

To address this issue, instead of using the basic ``plain projection'' of the
original paper, a more advanced ``full projection'' is used in many later papers
in the MPST literature.\footnote{Plain projection is based on \textit{plain
merge}. Full projection is based on \textit{full merge}. The details do not matter in this
paper.} The key benefit of using full projection instead of plain projection is
that many more global types become projectable. Formally, the domain of the
function is larger. Against conventional wisdom at the time, though,
\textbf{projection \textit{no longer} implies consistency}. Thus, well-typedness
no longer implies safety and liveness: whether or not it does, depends on
whether or not the family of local types happens to be consistent, which needs
to be proved separately. This surprising discovery was made by Scalas and Yoshida
\rev{in an influential paper,} colloquially called the \textit{``Less
Is More'' paper} \citep{DBLP:journals/pacmpl/ScalasY19}. \Cref{fig:proj:full} visualises the issue. The following
example demonstrates that it, too, affects the Ring protocol of
\Cref{exmp:ring}.

\begin{example}\label{exmp:ring-carol}
	The following local types, projected from $\smash{\stypex{G^\text{Ring}}}$ in
	\Cref{exmp:ring} using full projection instead of plain projection, specify
	Alice, Bob, and Carol in the Ring protocol:
	\begin{gather*}
		\begin{gathered}
			\stypex{L_{\smash{\rolex[][]{a}}}^\text{Ring}} = \text{... (\Cref{exmp:ring})}
		\\
			\smash{\stypex{L_{\smash{\rolex[][]{b}}}^\text{Ring}}} = \text{... (\Cref{exmp:ring})}
		\end{gathered}
	\qquad
		\stypex{L_{\smash{\rolex[][]{c}}}^\text{Ring}} = \stypex{\begin{tree}
			\branchx{
				\brnx{\rolex{b}} \holex{h1}
			}
			\branchxx[h1]{
				\mtypexx{\datax{AppThenGet}}{\datax{Nat}}
				\prewide
				\selxxx{\rolex{a}}{\datax{Val}}{\datax{Nat}}
				\prewide
				\one
			}{
				\mtypexx{\datax{App}}{\datax{Nat}}
				\prewide
				\brnxxx{\rolex{a}}{\datax{Get}}{}
				\prewide
				\selxxx{\rolex{a}}{\datax{Val}}{\datax{Nat}}
				\prewide
				\one
			}
		\end{tree}}
	\end{gather*}
	The following processes, well-typed by the local types, implement Alice, Bob,
	and Carol in \Cref{fig:ring:push}:
	\begin{gather*}
		\begin{gathered}
			\sexprx{P_{\smash{\rolex[][]{a}}}^\text{Ring}} = \text{... (\Cref{exmp:ring})}
		\\
			\smash{\sexprx{P_{\smash{\rolex[][]{b}}}^\text{Ring}}} = \text{... (\Cref{exmp:ring})}
		\end{gathered}
	\qquad
		\sexprx{P_{\smash{\rolex[][]{c}}}^\text{Ring}} = \begin{tree}
			\branchx{
				\rcvx{\rolex{b}} \holex{h1}
			}
			\branchxx[h1]{
				\mexprxx{\datax{AppThenGet}}{\datax{y}}
				\prewide
				\sndxxx{\rolex{a}}{\datax{Val}}{\datax{y*2}}
				\prewide
				\nil
			}{
				\mexprxx{\datax{App}}{\datax{y}}
				\prewide
				\letxx{\datax{z}}{\datax{y*2}}
				\rcvxxx{\rolex{a}}{\datax{Get}}{\datax{\_}}
				\prewide
				\sndxxx{\rolex{a}}{\datax{Val}}{\datax{z}}
				\prewide
				\nil
			}
		\end{tree}
	\end{gather*}

	Now, not only the projections onto Alice and Bob are defined (cf. \Cref{exmp:plain}), but also the
	projection onto Carol. As a result, in the presence of a local type for each of
	Alice, Bob, and Carol, the implementation can be fully type-checked:
	$\smash{\sexprx{P_{\smash{\rolex[][]{a}}}^\text{Ring}}},
	\smash{\sexprx{P_{\smash{\rolex[][]{b}}}^\text{Ring}}},
	\smash{\sexprx{P_{\smash{\rolex[][]{c}}}^\text{Ring}}}$ are, in fact,
	well-typed by
	$\smash{\stypex{L_{\smash{\rolex[][]{a}}}^\text{Ring}}},\allowbreak
	\smash{\stypex{L_{\smash{\rolex[][]{b}}}^\text{Ring}}},\allowbreak
	\smash{\stypex{L_{\smash{\rolex[][]{c}}}^\text{Ring}}}$. However, projection
	no longer implies consistency,%
	\footnote{%
		\rev{Technically, the reason why the family of local types is
		inconsistent, is that an auxiliary partial function on local types
		(roughly: a second-order projection that takes the projection of a local
		type), which is used to compute consistency, is undefined for
		$\smash{\stypex{L_{\smash{\rolex[][]{a}}}^\text{Ring}}}$ and
		$\smash{\stypex{L_{\smash{\rolex[][]{c}}}^\text{Ring}}}$; this function,
		its effect on (in)consistency, and more examples, appear elsewhere
		\cite{DBLP:journals/pacmpl/ScalasY19}.}%
	} %
	so well-typedness no longer implies safety and liveness, so safety and
	liveness cannot be proved. (The parallel composition of the family
	of processes \textit{is} safe and live, though.)
	Thus, the Ring protocol is still not actually supported. \qed
\end{example}

Essentially, well-typedness is meaningless until consistency has been proved separately. Scalas and Yoshida propose a new
approach to MPST based on this observation in the ``Less Is More'' paper,
independent of global types and projection
\citep{DBLP:journals/pacmpl/ScalasY19}. The idea is to model consistency as a
temporal logic formula $\stypex{\varphi}$ such that the family of local types is
consistent if, and only if, its \textit{operational semantics} in the form of a
\textit{labelled transition system} (LTS) satisfies $\stypex{\varphi}$.
\Cref{fig:mpst:popl19} visualises the idea:
\begin{enumerate}
	\item First, a protocol among roles $\rnamex{r_1}, \ldots, \rnamex{r_n}$ is
	implemented as a family of \textbf{processes} $\sexprx{P_1}, \ldots,
	\sexprx{P_n}$ (like the classical approach), while it is specified as a family
	of \textbf{local types} $\stypex{L_1}, \ldots, \stypex{L_n}$, but without a
	global type and projection (unlike the classical approach).

	\item Next, the family of local types is verified by \textbf{model checking}
	the operational semantics of $\stypex{L_1}, \ldots, \stypex{L_n}$ for
	satisfaction of a \textbf{consistency property} $\stypex{\varphi}$.

	\item Last, the family of processes is verified by \textbf{type-checking}
	$\sexprx{P_i}$ against $\stypex{L_i}$ for each role. The main result is that
	consistency and well-typedness imply safety and liveness.
\end{enumerate}

To demonstrate the effectiveness of the ``Less Is More'' approach, Scalas and
Yoshida introduce a set of four challenging example protocols: whereas the classical
approach cannot be used to prove the safety and liveness of implementations of
these protocols, the ``Less Is More'' approach can. We call this set
the \textit{``Less Is More'' benchmark}. \Cref{fig:proj:full} visualises that
three protocols in the benchmark are projectable (using full projection) but not
consistent, while one is not even projectable.

\subsection{This Paper: ``Less Is More'', Compositionally -- \textit{Synthetic Approach}}\label{sect:intr:contrib}

The main strength of the ``Less Is More'' approach is that it supports many more
protocols than the classical approach does: to date, it remains the only
approach in the MPST literature that passes the ``Less Is More'' benchmark. The
main weakness, though, is that \textbf{the ``Less Is More'' approach is
\textit{non-compositional}}: as part of the model checking step, multiple
``small'' LTSs $\denotx{\stypex{L_1}}, \ldots,\allowbreak\denotx{\stypex{L_n}}$
(operational semantics of local types $\stypex{L_1}, \ldots, \stypex{L_n}$) need
to be composed into a single ``large'' LTS $\denotx{\stypex{L_1}} \times
{\cdots} \times \denotx{\stypex{L_n}}$ (operational semantics of the family).
This is computationally expensive
\citep{DBLP:journals/pacmpl/UdomsrirungruangY25}: in the worst case, the size of
the large LTS is exponential in the sizes of the small LTSs. Moreover, it goes
against the compositional nature of concurrent systems programming in general.

For several years now, it has been an open problem to develop an approach that
passes the ``Less Is More'' benchmark compositionally. \textbf{This paper
presents the first solution to this open problem}: the \textit{synthetic
approach}. It leverages a recent style of behavioural type systems, called
\textit{synthetic behavioural typing}, in which the operational semantics of
behavioural types is used not only to prove type soundness (as usual), but also
to define the typing rules \citep{DBLP:conf/ecoop/JongmansF23}. Concretely, we
make the following novel contributions:
\begin{itemize}
	\item \textbf{The special case of the synthetic approach:} We present a new
	MPST-specific type system to type-check processes against \textit{global types,
	without local types and projection}. In this way, a key innovation of the synthetic approach
	is that it is, essentially, the opposite of the ``Less Is More'' approach (in
	which processes are type-checked against \textit{local types, without global
	types and projection}).

	The main theoretical result is that well-typedness implies safety and liveness,
	\textit{without the need to prove consistency separately}. The main practical
	result is that the synthetic approach is the first one to pass the ``Less Is
	More'' benchmark compositionally.

	\item \textbf{The general case:} We present a new generic type system---beyond
	MPST---to type-check processes against arbitrary \textit{well-behaved} LTSs
	instead of only global types.

	The main theoretical results are that: (a) well-behavedness and well-typedness
	imply safety and liveness; (b) the LTSs of all global types are well-behaved.
	The key advantage of the general case is that it is strictly more expressive
	than the special case. For instance, beyond ``Less Is More'',
	protocols are supported in which a sender chooses between different receivers.
\end{itemize}
Furthermore:
\begin{itemize}
	\item We formalised all definitions/theorems/examples, and mechanised all
	proofs, in Agda.

	\item We developed a prototype language and tooling as an extension to VS Code.
\end{itemize}

In \Cref{sect:overv}, we present a detailed overview of our contributions.
In \Cref{sect:prelim}, we recall some preliminaries from the MPST literature to set the stage for our theoretical development in later sections.
Then, in \Cref{sec:global}, we discuss the details of
typing with global types first (the special case), while in \Cref{sec:behaviour}, we
generalise the session classifiers from global types to LTSs (the general case). In
\Cref{sec:formalisation} and \Cref{sec:proto}, we discuss the
mechanisation and implementation of our theory, respectively. We finish
with a discussion of related work and a conclusion in \Cref{sect:relw,sect:concl}.

Throughout the paper, we continue to use colours to emphasise syntactic categories of
objects: shades of $\sexprx{\text{red}}$ for data/process expressions, shades of $\stypex{\text{blue}}$ for types, and shades of $\rnamex{\text{magenta}}$ for objects common to both expressions and types (e.g., role names, message labels). The colours are just syntax highlighting: they do not have additional meaning.
\rev{Furthermore, all lemmas and theorems that have been formalised in Agda are
explicitly marked with the Agda logo: \agdalogo.}


\section{Overview of the Contributions}\label{sect:overv}

\Cref{fig:mpst:this} visualises the idea behind the synthetic approach of this
paper.
\begin{enumerate}
	\item First, a protocol among roles $\rnamex{r_1}, \ldots, \rnamex{r_n}$ is
	implemented as a family of \textbf{processes} $\sexprx{P_1}, \ldots,
	\sexprx{P_n}$, while it is specified as a \textbf{global type} $\stypex{G}$.

	\item Next, the family of processes is verified by \textbf{type-checking}
	$\sexprx{P_i}$ against $\stypex{G}$ for each role. The main result is that
	well-typedness implies safety and liveness.
\end{enumerate}
Thus, the synthetic approach to MPST works \textbf{without local types and
projection} (cf. the ``Less Is More'' approach) and \textbf{without the need to
prove consistency separately} (cf. the classical approach). In fact, the
synthetic approach can be further generalised to work \textbf{without global
types}: processes can be type-checked directly against well-behaved
\textit{labelled transition systems} (LTS), regardless of any particular syntax
to express such LTSs. Global types are just one instantiation (i.e.,
type-checking against global types is a special case of type-checking against
well-behaved LTSs). We clarify the special case and the general case in the
remaining subsections.

\subsection{The Special Case: Type Checking against Global Types}

Our technique to type-check processes against global types consists of two parts:
\begin{itemize}
	\item First, we associate each global type $\stypex{G}$ with
	\textit{operational semantics} in the form of an LTS. Each state models a
	continuation of the protocol specified by $\stypex{G}$, while each transition
	models a possible communication.

	\item Second, we use LTSs to define the typing rules. For instance:
	\begin{gather*}
		\dfrac{
			\dtypedxxx{\varGamma}{e}{t}
		\qquad
			\stypedxxx{\varGamma}{\atxx{p}{P}}{G'}
		\qquad
			\sttransxxx{G}{\comxxxx{p}{q}{\ell}{t}}{G'}
		}{
			\stypedxxx{\varGamma}{\atxx{p}{\sndxxxx{q}{\ell}{e}{P}}}{G}
		}
	\end{gather*}
	This simplified typing rule---we present the actual one later in this
	paper---states that, as an implementation of role $\rnamex{p}$, process
	$\sndxxxx{q}{\ell}{e}{P}$ is well-typed by global type $\stypex{G}$ in
	environment $\denvx{\varGamma}$ when: (1) expression $\dexprx{e}$ is well-typed
	by payload type $\dtypex{t}$; (2) process $\sexprx{P}$ is well-typed by global
	type $\stypex{G'}$; \textbf{(3) $\stypex{G}$ has a transition to
	$\stypex{G'}$}. That is, $\stypex{G}$ and $\stypex{G'}$ are treated as states
	of an LTS.

	We note that $\stypex{G'}$ occurs in the premise of the rule, but not in the
	conclusion. Thus, from a bottom-up perspective, $\stypex{G'}$ is a free
	meta-variable that needs to be \textit{synthesised} to apply the rule. In
	philosophical logic, rules with free meta-variables in the premises are
	called ``synthetic'' \citep{Hintikka1973-HINLLA,VanBenthem1974-VANHOA-3};
	this is where the name ``synthetic approach'' comes
	from.\footnote{\rev{Unrelated to ``synthetic approaches'' in dependent type
	theory}}
\end{itemize}

\begin{table}[t]
	\caption{Principles of the typing rules}
	\label{tab:principles}

	\begin{tabular}{@{}l@{\quad}l@{\quad}r@{}}
		\toprule
		& \textbf{Principle} & \llap{\textbf{Exmp.}}

	\\	\midrule
		\textbf{Sending}
		& \rev{P1:} Each send implemented needs to be specified & \ref{exmp:ring-typing-a}
	\\	& \rev{P2:} Not each send specified needs to be implemented (at least one, though) & \ref{exmp:ring-typing-a}
	\\	\midrule
		\textbf{Receiving}
		& \rev{P3:} Not each receive implemented needs to be specified & \ref{exmp:ring-typing-c}
	\\	& \rev{P4:} Each receive specified needs to be implemented & \ref{exmp:ring-typing-b}
	\\	\midrule
		\textbf{Skipping}
		& \rev{P5:} Each communication specified needs to be skipped by each ``third party'' & \ref{exmp:ring-typing-c}
	\\	\bottomrule
	\end{tabular}
\end{table}

To further clarify our technique, we present a series of examples that revisit
the Ring protocol of \Cref{exmp:ring}. The first example in the series
demonstrates an LTS; the remaining examples demonstrate the typing rules. In
each of the latter examples, we construct a typing derivation for one of the
processes in the Ring protocol. Different examples highlight different principles enforced by the
typing rules. \Cref{tab:principles} summarises the principles. The asymmetry
between the principles for sending and receiving is further explained in the examples.

\begin{figure}[t]\centering
	\begin{tikzpicture}[x=1.5cm, y=.67cm]
		\tikzstyle{state} = []
		\tikzstyle{trans} = [-stealth, rounded corners]
		\tikzstyle{label} = [inner sep=0pt, anchor=base, yshift=1.5mm, font=\footnotesize]

		\node [state] (G1) at (0,0) {$\stypex{G_{\smash{1}}^\text{Ring}}$};
		\node [state] (G4) at (8,0) {$\stypex{G_{\smash{4}}^\text{Ring}}$};
		\node [state] (G2) at (3,+1) {$\stypex{G_{\smash{2}}^\text{Ring}}$};
		\node [state] (G3) at (6,0) {$\stypex{G_{\smash{3}}^\text{Ring}}$};
		\node [state] (G5) at (2,-1) {$\stypex{G_{\smash{5}}^\text{Ring}}$};
		\node [state] (G6) at (4,-1) {$\stypex{G_{\smash{6}}^\text{Ring}}$};

		\draw [trans] ([xshift=-.5cm]G1.west) to (G1);
		\draw [trans] (G1) to (G1|-G2) to node [label] {$\begin{array}[b]{@{}c@{}}\comxx{\rolex[]{a}}{\rolex[]{b}} \mtypexx{\datax[]{AppThenGet}}{\datax[]{Nat}}\end{array}$} (G2);
		\draw [trans] (G2) to node [label] {$\begin{array}[b]{@{}c@{}}\comxx{\rolex[]{b}}{\rolex[]{c}} \mtypexx{\datax[]{AppThenGet}}{\datax[]{Nat}}\end{array}$} (G3|-G2) to (G3);
		\draw [trans] (G3) to node [label] {$\begin{array}[b]{@{}c@{}}\comxx{\rolex[]{c}}{\rolex[]{a}} \mtypexx{\datax[]{Val}}{\datax[]{Nat}}\end{array}$} (G4);
		\draw [trans] (G1) to (G1|-G5) to node [label] {$\begin{array}[b]{@{}c@{}}\comxx{\rolex[]{a}}{\rolex[]{b}}\\\mtypexx{\datax[]{App}}{\datax[]{Nat}}\end{array}$} (G5);
		\draw [trans] (G5) to node [label] {$\begin{array}[b]{@{}c@{}}\comxx{\rolex[]{b}}{\rolex[]{c}}\\\mtypexx{\datax[]{App}}{\datax[]{Nat}}\end{array}$} (G6);
		\draw [trans] (G6) to node [label] {$\begin{array}[b]{@{}c@{}}\comxx{\rolex[]{a}}{\rolex[]{c}}\\\mtypexx{\datax[]{Get}}{}\end{array}$} (G3|-G6) to (G3);
	\end{tikzpicture}

	\caption{Operational semantics of $\smash{\protect\stypex{G^\text{Ring}}}$ in \Cref{exmp:ring}. Let $\smash{\protect\stypex{G_{\smash{1}}^\text{Ring}} = \protect\stypex{G^\text{Ring}}}$.}
	\label{fig:ring-lts}
	\Description{}
\end{figure}

\begin{example}\label{exmp:ring-lts}
	\Cref{fig:ring-lts} visualises the operational semantics of
	$\smash{\stypex{G^\text{Ring}}}$ in \Cref{exmp:ring} as an LTS with six states.
	Global action $\comxxxx{p}{q}{\ell}{t}$ specifies the communication of a
	message labelled $\lnamex{\ell}$, with a payload of type $\dtypex{t}$, from
	role $\rnamex{p}$ to role $\rnamex{q}$. \qed
\end{example}

\begin{example}\label{exmp:ring-typing-a}
	We prove that $\smash{\sexprx{P_{\smash{\rolex[][]{a}}}^\text{Ring}}}$ in
	\Cref{exmp:ring} is well-typed by the LTS of $\smash{\stypex{G^\text{Ring}}}$
	in \Cref{fig:ring-lts}.

	First, the following derivation states that, as an implementation of Alice, the
	empty process is well-typed by ``state''
	$\smash{\stypex{G_{\smash{4}}^\text{Ring}}}$ in type environments where
	variable $\dexprx{\datax{z}}$ is a number:
	\begin{gather}\label{eqn:01}
		\dfrac{
			\sttransx[\centernot]{G_{\smash{4}}^\text{Ring}}
		}{
			\stypedxxxx{\denvxxx{\varnothing}{\datax{z}}{\datax{Nat}}}{\varnothing}{\atxx{\rolex{a}}{\nil}}{G_{\smash{4}}^\text{Ring}}
		}
	\end{gather}
	The intuition is that, as $\smash{\stypex{G_{\smash{4}}^\text{Ring}}}$ does not
	have any transitions, \textbf{it allows Alice to terminate}. Generally, a
	global type allows a role to terminate when none of the reachable successor
	``states'' of that global type---after zero-or-more transitions---has a
	transition with that role. For instance,
	$\smash{\stypex{G_{\smash{6}}^\text{Ring}}}$ allows Bob to terminate (because
	neither $\smash{\stypex{G_{\smash{3}}^\text{Ring}}}$ nor
	$\smash{\stypex{G_{\smash{4}}^\text{Ring}}}$ has a transition with Bob), but
	$\smash{\stypex{G_{\smash{2}}^\text{Ring}}}$ does not allow Alice to terminate
	(because $\smash{\stypex{G_{\smash{3}}^\text{Ring}}}$ has a transition with
	Alice).

	Next, the following derivation states that, as an implementation of Alice,
	process $\sexprx{\rcvxxx{\rolex{c}}{\datax{Val}}{\datax{z}} \pre \nil}$ is
	well-typed by ``state'' $\smash{\stypex{G_{\smash{3}}^\text{Ring}}}$ in empty
	type environments:
	\begin{gather}\label{eqn:02}
		\dfrac{
		  \rev{%
		  \big\{\;
			{\sttransxxx{G_{\smash{3}}^\text{Ring}}{\comxxxx{\rolex[]{c}}{\rolex[]{a}}{\datax[]{Val}}{\datax[]{Nat}}}{G_{\smash{4}}^\text{Ring}}}
			\quad {\mapsto} \quad
			{\stypedxxxx{\denvxxx{\varnothing}{\datax{z}}{\datax{Nat}}}{\varnothing}{\atxx{\rolex{a}}{\nil}}{G_{\smash{4}}^\text{Ring}} \text{ (\ref{eqn:01})}}
		  \;\big\}
		  }
		\qquad
		}{
			\stypedxxxx{\varnothing}{\varnothing}{\atxx{\rolex{a}}{\rcvxxx{\rolex{c}}{\datax{Val}}{\datax{z}} \pre \nil}}{G_{\smash{3}}^\text{Ring}}
		}
	\end{gather}
	The intuition is that, as $\smash{\stypex{G_{\smash{3}}^\text{Ring}}}$ has a
	transition that models a communication from Carol to Alice of a message
	labelled $\lnamex{\datax{Val}}$, with a payload of type $\dtypex{\datax{Nat}}$,
	\textbf{it allows Alice to perform such a receive}. As the payload is received
	into variable $\dexprx{\datax{z}}$, the successor process must be well-typed in
	type environments that map $\dexprx{\datax{z}}$ to $\dtypex{\datax{Nat}}$; this
	was proved by \Cref{eqn:01}.

	Next, the following derivation states that, as an implementation of Alice,
	process $\sexprx{\rcvxxx{\rolex{c}}{\datax{Val}}{\datax{z}} \pre \nil}$---the
	same as in the previous derivation---is well-typed by ``state''
	$\smash{\stypex{G_{\smash{2}}^\text{Ring}}}$ in empty type environments:
	\begin{gather}\label{eqn:03}
		\dfrac{
			\sttransinclxx[\centernot]{G_{\smash{2}}^\text{Ring}}{\setx{\rolex[]{a}}}
		\qquad
		\rev{%
			\big\{\;
			{\sttransxxx{G_{\smash{2}}^\text{Ring}}{\comxxxx{\rolex[]{b}}{\rolex[]{c}}{\datax[]{AppThenGet}}{\datax[]{Nat}}}{G_{\smash{3}}^\text{Ring}}}
			\quad {\mapsto} \quad
			{\stypedxxxx{\varnothing}{\varnothing}{\atxx{\rolex{a}}{\rcvxxx{\rolex{c}}{\datax{Val}}{\datax{z}} \pre \nil}}{G_{\smash{3}}^\text{Ring}} \text{ (\ref{eqn:02})}}
			\;\big\}
			}
		}{
			\stypedxxxx{\varnothing}{\varnothing}{\atxx{\rolex{a}}{\rcvxxx{\rolex{c}}{\datax{Val}}{\datax{z}} \pre \nil}}{G_{\smash{2}}^\text{Ring}}
		}
	\end{gather}
	The intuition is that, as $\smash{\stypex{G_{\smash{2}}^\text{Ring}}}$ has one
	transition, but none with Alice, \textbf{it allows Alice to skip the
	communication modelled by that transition}. Skipping communications in this way
	subsumes the concept of \textit{merging}---a key ingredient of projection---in
	the classical approach to MPST \citep{DBLP:journals/pacmpl/ScalasY19}.

	Last, the following derivation states that, as an implementation of Alice,
	process $\sexprx{\sndxxx{\rolex{b}}{\datax{AppThenGet}}{\datax{5}} \pre}
	\allowbreak \sexprx{\rcvxxx{\rolex{c}}{\datax{Val}}{\datax{z}} \pre \nil}$ is
	well-typed by ``state'' $\smash{\stypex{G_{\smash{1}}^\text{Ring}}}$ in empty
	type environments \rev{(henceforth omitted)}:
	\begin{gather}\label{eqn:04}
		\dfrac{
			\dtypedxx{\datax{5}}{\datax{Nat}}
		\qquad
			\stypedxx{\atxx{\rolex{a}}{\rcvxxx{\rolex{c}}{\datax{Val}}{\datax{z}} \pre \nil}}{G_{\smash{2}}^\text{Ring}} \text{ (\ref{eqn:03})}
		\qquad
			\sttransxxx{G_{\smash{1}}^\text{Ring}}{\comxxxx{\rolex[]{a}}{\rolex[]{b}}{\datax[]{AppThenGet}}{\datax[]{Nat}}}{G_{\smash{2}}^\text{Ring}}
		}{
			\stypedxx{\atxx{\rolex{a}}{\sndxxx{\rolex{b}}{\datax{AppThenGet}}{\datax{5}} \pre \rcvxxx{\rolex{c}}{\datax{Val}}{\datax{z}} \pre \nil}}{G_{\smash{1}}^\text{Ring}}
		}
	\end{gather}
	The intuition is that, as $\smash{\stypex{G_{\smash{1}}^\text{Ring}}}$ has a
	transition that models a communication from Alice to Bob of a message labelled
	$\lnamex{\datax{AppThenGet}}$, with a payload of type $\dtypex{\datax{Nat}}$,
	\textbf{it allows Alice to perform such a send}.

	We note that $\smash{\stypex{G_{\smash{1}}^\text{Ring}}}$ has two transitions,
	but the well-typed process has only one corresponding output alternative. This
	demonstrates \rev{principles P1/P2} that \textbf{each send implemented needs to be
	specified}, but \textbf{\textit{not} each send specified needs to be
	implemented} (at least one, though).

	Given the definitions of
	$\smash{\sexprx{P_{\smash{\rolex[][]{a}}}^\text{Ring}}}$ in \Cref{exmp:ring}
	and $\smash{\stypex{G^\text{Ring}}}$ in \Cref{fig:ring-lts}, we conclude from
	\Cref{eqn:04}:
	\begin{gather}
		\stypedxx{P_{\smash{\rolex[][]{a}}}^\text{Ring}}{G^\text{Ring}}
		\tag*{\qed}
	\end{gather}
\end{example}

\begin{example}\label{exmp:ring-typing-b}
	We prove that $\smash{\sexprx{P_{\smash{\rolex[][]{b}}}^\text{Ring}}}$ in
	\Cref{exmp:ring} is well-typed by the LTS of $\smash{\stypex{G^\text{Ring}}}$
	in \Cref{fig:ring-lts}.

	First, using similar derivations as in \Cref{exmp:ring-typing-a}, we can
	prove:
	\begin{align}
	&\label{eqn:05}
		\stypedxxxx{\denvxxx{\varnothing}{\datax{x}}{\datax{Nat}}}{\varnothing}{\atxx{\rolex{b}}{\sndxxx{\rolex{c}}{\datax{AppThenGet}}{\datax{x+1}} \pre \nil}}{G_{\smash{2}}^\text{Ring}}
	\\&\label{eqn:06}
		\smash{\stypedxxxx{\denvxxx{\varnothing}{\datax{x}}{\datax{Nat}}}{\varnothing}{\atxx{\rolex{b}}{\sndxxx{\rolex{c}}{\datax{App}}{\datax{x+1}} \pre \nil}}{G_{\smash{5}}^\text{Ring}}}
	\end{align}
	Next, the following derivation states that, as an implementation of Bob, his
	input process $\smash{\sexprx{P_{\smash{\rolex[][]{b}}}^\text{Ring}}}$ in
	\Cref{exmp:ring} is well-typed by ``state''
	$\smash{\stypex{G_{\smash{1}}^\text{Ring}}}$. Let $\lnamex{\ell_1} =
	\lnamex{\datax{AppThenGet}}$ and $\lnamex{\ell_2} = \lnamex{\datax{App}}$:
	\begin{gather}\label{eqn:07}
		\dfrac{
		  \rev{\left\{\begin{array}{@{}c@{}}\strut\\[1ex]\strut\end{array}\right.
		  \begin{array}{@{}c@{\quad}c@{\quad}l@{}}
			{\sttransxxx{G_{\smash{1}}^\text{Ring}}{\comxxxx{\rolex[]{a}}{\rolex[]{b}}{\datax[]{AppThenGet}}{\datax[]{Nat}}}{G_{\smash{2}}^\text{Ring}}}
			& {\mapsto} &
			{\stypedxxxx{\denvxxx{\varnothing}{\datax{x}}{\datax{Nat}}}{\varnothing}{\atxx{\rolex{b}}{\sndxxx{\rolex{c}}{\datax{AppThenGet}}{\datax{x+1}} \pre \nil}}{G_{\smash{2}}^\text{Ring}} \text{ (\ref{eqn:05})}}
			\\
			{\sttransxxx{G_{\smash{1}}^\text{Ring}}{\comxxxx{\rolex[]{a}}{\rolex[]{b}}{\datax[]{App}}{\datax[]{Nat}}}{G_{\smash{5}}^\text{Ring}}}
			& {\mapsto} &
			{\stypedxxxx{\denvxxx{\varnothing}{\datax{x}}{\datax{Nat}}}{\varnothing}{\atxx{\rolex{b}}{\sndxxx{\rolex{c}}{\datax{App}}{\datax{x+1}} \pre \nil}}{G_{\smash{5}}^\text{Ring}} \text{ (\ref{eqn:06})}}
		  \end{array}
		  \left.\begin{array}{@{}c@{}}\strut\\[1ex]\strut\end{array}\right\}}
		}{
			\stypedxxxx{\varnothing}{\varnothing}{\atxx{\rolex{b}}{\rcvxx{\rolex{a}}{\mexprxx{\ell_1}{\datax{x}} \pre \sndxxx{\rolex{c}}{\ell_1}{\datax{x+1}} \pre \nil, \mexprxx{\ell_2}{\datax{x}} \pre \sndxxx{\rolex{c}}{\ell_2}{\datax{x+1}} \pre \nil}}}{G_{\smash{1}}^\text{Ring}}
		}
	\end{gather}
	The intuition is that, as $\smash{\stypex{G_{\smash{1}}^\text{Ring}}}$ has
	transitions that model communications from Alice to Bob of a message labelled
	$\lnamex{\datax{AppThenGet}}$ or $\lnamex{\datax{App}}$, with a payload of type
	$\dtypex{\datax{Nat}}$, \textbf{it allows Bob to perform such receives}. As
	the payload is received into variable $\dexprx{\datax{x}}$, the successor
	process must be well-typed in type environments that map $\dexprx{\datax{x}}$
	to $\dtypex{\datax{Nat}}$; this was proved by \Cref{eqn:05,eqn:06}.

	We note that $\smash{\stypex{G_{\smash{1}}^\text{Ring}}}$ has two transitions,
	and the well-typed process has two corresponding input alternatives. This
	demonstrates \rev{principle P4} that \textbf{each receive specified needs to be
	implemented}. Thus, there is asymmetry between typing input processes (e.g.,
	Bob in this example) and typing output processes (e.g., Alice in
	\Cref{exmp:ring-typing-a}): a sender must be able to offer at least one message
	label specified in the LTS, while the receiver must be able to accept all of
	them.

	Given the definitions of
	$\smash{\sexprx{P_{\smash{\rolex[][]{b}}}^\text{Ring}}}$ in \Cref{exmp:ring}
	and $\smash{\stypex{G^\text{Ring}}}$ in \Cref{fig:ring-lts}, we conclude from
	\Cref{eqn:07}:
	\begin{gather}
		\stypedxx{P_{\smash{\rolex[][]{b}}}^\text{Ring}}{G^\text{Ring}}
		\tag*{\qed}
	\end{gather}
\end{example}

\begin{example}\label{exmp:ring-typing-c}
	We prove that $\smash{\sexprx{P_{\smash{\rolex[][]{c}}}^\text{Ring}}}$ in
	\Cref{exmp:ring-carol} is well-typed by the LTS of
	$\smash{\stypex{G^\text{Ring}}}$ in \Cref{fig:ring-lts}.

	First, using a similar derivation as in \Cref{exmp:ring-typing-a}, we can
	prove:
	\begin{align}
	&\label{eqn:08}
		\stypedxxxx{\denvxxx{\varnothing}{\datax{y}}{\datax{Nat}}}{\varnothing}{\atxx{\rolex{c}}{\sndxxx{\rolex{a}}{\datax{Val}}{\datax{y*2}} \pre \nil}}{G_{\smash{3}}^\text{Ring}}
	\\&\label{eqn:09}
		\smash{\stypedxxxx{\denvxxx{\varnothing}{\datax{y}}{\datax{Nat}}}{\varnothing}{\atxx{\rolex{c}}{\letxx{\datax{z}}{\datax{y*2}} \rcvxxx{\rolex{a}}{\datax{Get}}{\datax{\_}} \pre \sndxxx{\rolex{a}}{\datax{Val}}{\datax{z}} \pre \nil}}{G_{\smash{6}}^\text{Ring}}}
	\end{align}

	Next, the following derivations state that, as an implementation of Carol, her
	input process $\smash{\sexprx{P_{\smash{\rolex[][]{c}}}^\text{Ring}}}$ in
	\Cref{exmp:ring-carol} is well-typed by both ``state''
	$\smash{\stypex{G_{\smash{2}}^\text{Ring}}}$ and ``state''
	$\smash{\stypex{G_{\smash{5}}^\text{Ring}}}$. Let $\lnamex{\ell_1} =
	\lnamex{\datax{AppThenGet}}$ and $\lnamex{\ell_2} = \lnamex{\datax{App}}$.
	Also, let $\sexprx{P_1} = \sexprx{\sndxxx{\rolex{a}}{\datax{Val}}{\datax{y*2}}
	\pre \nil}$ and $\sexprx{P_2} = \sexprx{\letxx{\datax{z}}{\datax{y*2}}
	\rcvxxx{\rolex{a}}{\datax{Get}}{\datax{\_}} \pre
	\sndxxx{\rolex{a}}{\datax{Val}}{\datax{z}} \pre \nil}$:
	\begin{gather}\label{eqn:10}
		\dfrac{
		  \rev{%
		  \big\{\;
		  {\sttransxxx{G_{\smash{2}}^\text{Ring}}{\comxxxx{\rolex[]{b}}{\rolex[]{c}}{\ell_1}{\datax[]{Nat}}}{G_{\smash{3}}^\text{Ring}}}
		  \quad {\mapsto} \quad
		  {\stypedxxxx{\denvxxx{\varnothing}{\datax{y}}{\datax{Nat}}}{\varnothing}{\atxx{\rolex{c}}{P_1}}{G_{\smash{3}}^\text{Ring}} \text{ (\ref{eqn:08})}}
		  \;\big\}
		  }
		}{
			\stypedxx{\atxx{\rolex{c}}{\rcvxx{\rolex{b}}{\mexprxx{\ell_1}{\datax{y}} \pre P_1, \mexprxx{\ell_2}{\datax{y}} \pre P_2}}}{G_{\smash{2}}^\text{Ring}}
		}
	\\\label{eqn:11}
		\dfrac{
		  \rev{%
			\big\{\;
			{\sttransxxx{G_{\smash{5}}^\text{Ring}}{\comxxxx{\rolex[]{b}}{\rolex[]{c}}{\ell_2}{\datax[]{Nat}}}{G_{\smash{6}}^\text{Ring}}}
			\quad {\mapsto} \quad
			{\stypedxxxx{\denvxxx{\varnothing}{\datax{y}}{\datax{Nat}}}{\varnothing}{\atxx{\rolex{c}}{P_2}}{G_{\smash{6}}^\text{Ring}} \text{ (\ref{eqn:09})}}
			\;\big\}
			}
		}{
			\stypedxx{\atxx{\rolex{c}}{\rcvxx{\rolex{b}}{\mexprxx{\ell_1}{\datax{y}} \pre P_1, \mexprxx{\ell_2}{\datax{y}} \pre P_2}}}{G_{\smash{5}}^\text{Ring}}
		}
	\end{gather}
	The intuition is that, as $\smash{\stypex{G_{\smash{2}}^\text{Ring}}}$ (resp.
	$\smash{\stypex{G_{\smash{5}}^\text{Ring}}}$) has a transition that models a
	communication from Bob to Carol of a message labelled
	$\lnamex{\datax{AppThenGet}}$ (resp. $\lnamex{\datax{App}}$), \textbf{it allows
	Carol to perform such a receive}.

	We note that the well-typed process has two input alternatives, but
	$\smash{\stypex{G_{\smash{2}}^\text{Ring}}}$ (resp.
	$\smash{\stypex{G_{\smash{5}}^\text{Ring}}}$) has only one corresponding
	transition. This demonstrates \rev{principle P3} that \textbf{\textit{not} each
	receive implemented needs to be specified}: a receiver may be able to accept
	more message labels than just those specified in the LTS. Reminiscent of
	\textit{subtyping} in the MPST literature
	\citep{DBLP:journals/jlp/GhilezanJPSY19}, this is fine because the
	sender---assuming it is well-typed---is guaranteed to offer only message labels
	specified in the LTS.

	Last, the following derivation states that, as an implementation of Carol, her
	input process $\smash{\sexprx{P_{\smash{\rolex[][]{c}}}^\text{Ring}}}$ in
	\Cref{exmp:ring-carol} is well-typed by ``state''
	$\smash{\stypex{G_{\smash{1}}^\text{Ring}}}$:
	\begin{gather}\label{eqn:12}
		\dfrac{
			\sttransinclxx[\centernot]{G_{\smash{1}}^\text{Ring}}{\setx{\rolex[]{c}}}
		\qquad
		\rev{%
		\left\{\begin{array}{@{}c@{}}\strut\\[1ex]\strut\end{array}\right.
		\begin{array}{@{}r@{\quad}c@{\quad}l@{}}
				{\sttransxxx{G_{\smash{1}}^\text{Ring}}{\comxxxx{\rolex[]{a}}{\rolex[]{b}}{\ell_1}{\datax[]{Nat}}}{G_{\smash{2}}^\text{Ring}}}
				& {\mapsto} &
				{\stypedxx{\atxx{\rolex{c}}{\rcvxx{\rolex{b}}{\mexprxx{\ell_1}{\datax{y}} \pre P_1, \mexprxx{\ell_2}{\datax{y}} \pre P_2}}}{G_{\smash{2}}^\text{Ring}} \text{ (\ref{eqn:10})}}
				\\
				{\sttransxxx{G_{\smash{1}}^\text{Ring}}{\comxxxx{\rolex[]{a}}{\rolex[]{b}}{\ell_2}{\datax[]{Nat}}}{G_{\smash{5}}^\text{Ring}}}
				& {\mapsto} &
				{\stypedxx{\atxx{\rolex{c}}{\rcvxx{\rolex{b}}{\mexprxx{\ell_1}{\datax{y}} \pre P_1, \mexprxx{\ell_2}{\datax{y}} \pre P_2}}}{G_{\smash{5}}^\text{Ring}}} \text{ (\ref{eqn:11})}
		\end{array}
		\left.\begin{array}{@{}c@{}}\strut\\[1ex]\strut\end{array}\right\}
		}
		}{
			\stypedxx{\atxx{\rolex{c}}{\rcvxx{\rolex{b}}{\mexprxx{\ell_1}{\datax{y}} \pre P_1, \mexprxx{\ell_2}{\datax{y}} \pre P_2}}}{G_{\smash{1}}^\text{Ring}}
		}
	\end{gather}
	The intuition is that, as $\smash{\stypex{G_{\smash{1}}^\text{Ring}}}$ has two
	transitions, but none with Carol, \textbf{it allows Carol to skip the
	communications modelled by those transitions}.

	We note that $\smash{\stypex{G_{\smash{1}}^\text{Ring}}}$ has two successor
	``states'', and the process is well-typed by each of them. This demonstrates
	\rev{principle P5} that \textbf{each communication specified needs to be skipped by each ``third
	party''} that does not participate in that communication: regardless of which
	communications happen between whichever senders and receivers, third parties
	that do not participate in those communications must be able to behave as
	specified in any continuation.

	Given the definitions of
	$\smash{\sexprx{P_{\smash{\rolex[][]{c}}}^\text{Ring}}}$ in
	\Cref{exmp:ring-carol} and $\smash{\stypex{G^\text{Ring}}}$ in
	\Cref{fig:ring-lts}, we conclude from \Cref{eqn:12}:
	\begin{gather}
		\stypedxx{P_{\smash{\rolex[][]{c}}}^\text{Ring}}{G^\text{Ring}}
		\tag*{\qed}
	\end{gather}
\end{example}

The main theoretical result for the special case of our synthetic approach to
MPST is that \textbf{well-typedness implies safety and liveness}.

\begin{example}
	\Cref{exmp:ring-typing-a,exmp:ring-typing-b,exmp:ring-typing-c} demonstrated that
	$\smash{\sexprx{P_{\smash{\rolex[][]{a}}}^\text{Ring}}},
	\smash{\sexprx{P_{\smash{\rolex[][]{b}}}^\text{Ring}}},
	\smash{\sexprx{P_{\smash{\rolex[][]{c}}}^\text{Ring}}}$ are well-typed by
	$\smash{\stypex{G^\text{Ring}}}$. Thus, by \Cref{thm:preservation-glob,thm:progress-glob}, we conclude that the
	parallel composition of this family of processes---the \textit{session}---is
	safe and live. Safety means that each communication that happens in the session
	is allowed by the global type. Liveness means that after any number of
	communications, either the session has successfully terminated, or another
	communication can happen. It is the first time in the MPST literature that this
	is proved compositionally for \Cref{exmp:ring}. \qed
\end{example}

The main practical result is that the synthetic approach is \textbf{the first
one to pass the ``Less Is More'' benchmark compositionally}, as we demonstrate fully in \Cref{sect:glob:benchm}.

\subsection{The General Case: Type Checking against LTSs}\label{sect:overv:general}

A key observation of the previous examples is that \textbf{the syntax of global
types does \textit{not} matter} at all in the typing rules; only \textbf{the
operational semantics does}. That is, the syntactic structure of global types is
never inspected in the typing rules. Instead, global types are treated as opaque
states of an LTS, whose transitions are the only objects of
significance. This observation is pushed forward in the general case of our
synthetic approach to MPST.

The idea of the general case is to define a predicate to judge whether or not an
LTS is \textit{well-behaved}. Processes can subsequently be type-checked against
well-behaved LTSs, regardless of how those LTSs are generated, and independent
of the syntactic structure of states---if any. Intuitively, an LTS is
well-behaved when it fulfils the following requirements:
\begin{itemize}
	\item \textbf{Sender determinacy:} If a state has multiple transitions, then
	these transitions model communications either with different senders and
	different receivers, or with the same sender but different receivers, or
	with the same sender and the same receiver---but never with different
	senders and the same receiver.

	\item \textbf{Determinism:} If a source state has multiple transitions that
	model the same communication, then they have the same target state.

	\item \textbf{Conditional commutativity and confluence (diamond):} Subject to
	additional conditions (see \Cref{def:mb} for details), if a state has multiple
	transitions that model independent communications, then those communications
	commute (i.e., the transitions form a diamond).
\end{itemize}
The following example demonstrates these requirements.

\begin{example}
	We argue that the LTS in \Cref{fig:ring-lts} is well-behaved. State
	$\smash{\stypex{G_{\smash{1}}^\text{Ring}}}$ is the only state that has
	multiple transitions. These transitions have the same sender and the same
	receiver, so sender determinacy is fulfilled. Moreover, these transitions do
	not model the same communication (i.e., the message labels are different), nor
	are they independent, so determinism, conditional commutativity, and confluence
	are fulfilled, too. The remaining states have only a single transition, so they
	trivially fulfil the well-behavedness requirements. \qed
\end{example}

The main theoretical results for the general case of our synthetic approach are
that: \textbf{(a) well-behavedness and well-typedness imply safety and liveness;
(b) the LTSs of all global types are well-behaved.} Thus, when processes are
type-checked against LTSs of global types, well-behavedness of those LTSs does
not need to be proved separately; it is already implied. This makes type
checking against global types really a special case of type checking against
LTSs.

%

The key advantage is that \textbf{the general case is strictly more
expressive than the special case}: more families of processes can be
successfully type-checked by well-behaved LTSs than by global types. In
particular, there exist well-behaved LTSs that cannot be expressed as a global
type, but they are \textit{inhabited} in our type system. An LTS is inhabited
when there exists a family of processes each of which is well-typed by that LTS.
The following example demonstrates a protocol that is not supported in the
``Less Is More'' paper, nor is it supported by the special case in this paper,
but it is supported by the general case.

\begin{example}
	The following well-behaved LTS specifies a protocol in which a
	$\lnamex{\datax{Foo}}$ message is communicated from Alice to Bob, and a
	$\lnamex{\datax{Bar}}$ message from Alice to Carol, in any order:
	\begin{center}
		\begin{tikzpicture}[x=1cm, y=.67cm]
			\tikzstyle{state} = []
			\tikzstyle{trans} = [-stealth, rounded corners]
			\tikzstyle{label} = [inner sep=0pt, anchor=base, yshift=1.5mm, font=\footnotesize]

			\node [state] (G1) at (0,0) {$\stypex{S_{\smash{1}}^\text{Diam}}$};
			\node [state] (G2) at (3,+1) {$\stypex{S_{\smash{2}}^\text{Diam}}$};
			\node [state] (G3) at (6,0) {$\stypex{S_{\smash{3}}^\text{Diam}}$};
			\node [state] (G4) at (3,-1) {$\stypex{S_{\smash{4}}^\text{Diam}}$};

			\draw [trans] ([xshift=-.5cm]G1.west) to (G1);
			\draw [trans] (G1) to (G1|-G2) to node [label] {$\comxxxx{\rolex[]{a}}{\rolex[]{b}}{\datax[]{Foo}}{}$} (G2);
			\draw [trans] (G2) to node [label] {$\comxxxx{\rolex[]{a}}{\rolex[]{c}}{\datax[]{Bar}}{}$} (G3|-G2) to (G3);
			\draw [trans] (G1) to (G1|-G4) to node [label] {$\comxxxx{\rolex[]{a}}{\rolex[]{c}}{\datax[]{Bar}}{}$} (G4);
			\draw [trans] (G4) to node [label] {$\comxxxx{\rolex[]{a}}{\rolex[]{b}}{\datax[]{Foo}}{}$} (G3|-G4) to (G3);
		\end{tikzpicture}
	\end{center}
	The following processes, well-typed by the LTS, implement Alice, Bob, and Carol:
	\begin{gather*}
		\sexprx{P_{\smash{\rolex[][]{a}}}^\text{Diam}} = \sexprx{\sndxxx{\rolex{b}}{\datax{Foo}}{} \prewide \sndxxx{\rolex{c}}{\datax{Bar}}{} \prewide \nil}
	\qquad
		\sexprx{P_{\smash{\rolex[][]{b}}}^\text{Diam}} = \sexprx{\rcvxxx{\rolex{a}}{\datax{Foo}}{\datax{\_}} \prewide \nil}
	\qquad
		\sexprx{P_{\smash{\rolex[][]{c}}}^\text{Diam}} = \sexprx{\rcvxxx{\rolex{a}}{\datax{Bar}}{\datax{\_}} \prewide \nil}
		\tag*{\qed}
	\end{gather*}
\end{example}

\subsection{Formalisation and Mechanisation in Agda}

We formalised all the theorems presented in this paper, as well as many examples, using the
Agda proof assistant. The formalisation follows the definitions given here
directly, without requiring any modifications or simplifications to facilitate
the proofs. This reinforces the claim that the synthetic approach to multiparty
session types, as introduced above, is particularly amenable to formalisation
and mechanisation -- a point we substantiate further in
\Cref{sec:formalisation}.

\subsection{Prototype Language and Tooling in VS Code}\label{sect:overv:vscode}

We developed a prototype language and tooling as an extension of \textit{VS
Code}, including a dedicated
\textit{LSP
server}.
The language consists of textual versions of global types and processes, while
the tooling consists of a parser, a syntax highlighter, and a type checker---all
running in the LSP server---that leverage the synthetic approach to MPST of this
paper. The prototype shows that there exists an algorithm to apply our typing
rules in practice, opening up the door towards integration of our type system in
mainstream languages. The following example demonstrates the prototype.

\begin{figure}[t]
	\begin{minipage}{.5\linewidth-.25cm}
		\includegraphics[width=\linewidth]{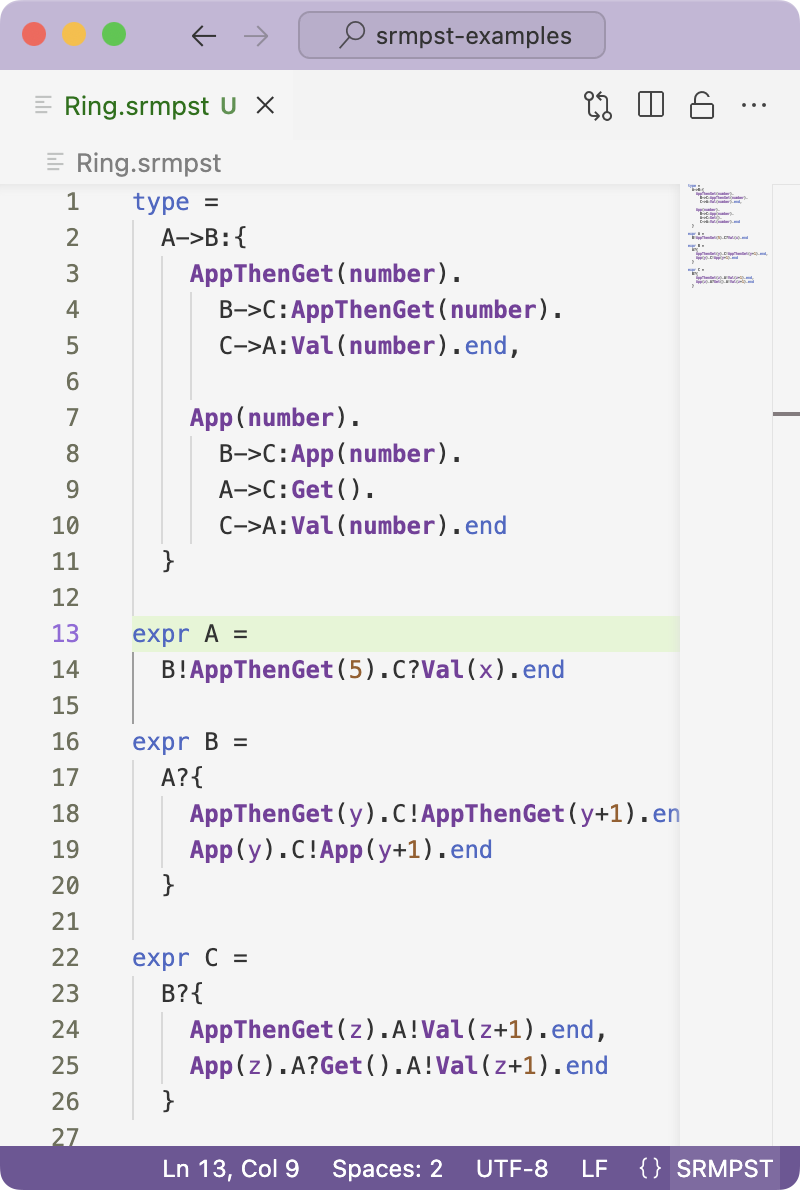} 
		\subcaption{Well-typed}
		\label{fig:vscode:well}
	\end{minipage}%
	\hfill%
	\begin{minipage}{.5\linewidth-.25cm}
		\includegraphics[width=\linewidth]{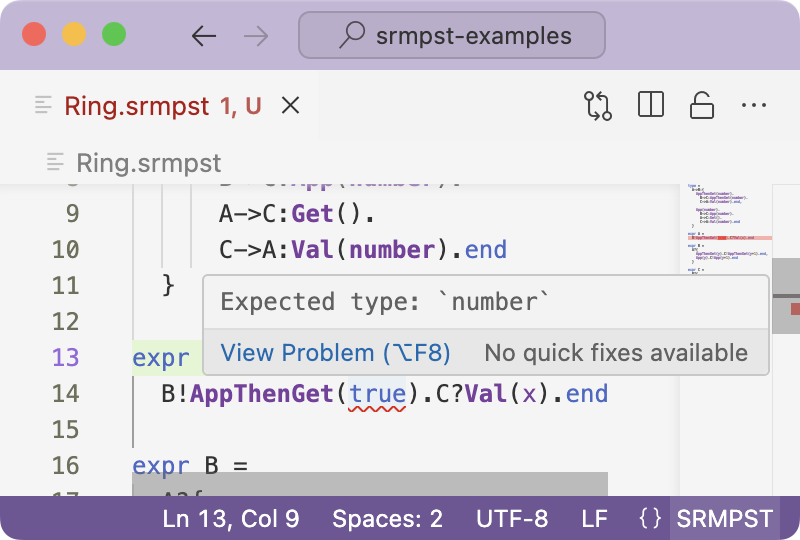}
		\subcaption{Ill-typed: Wrong payload type} 
		\label{fig:vscode:ill1}
		\bigbreak
		\includegraphics[width=\linewidth]{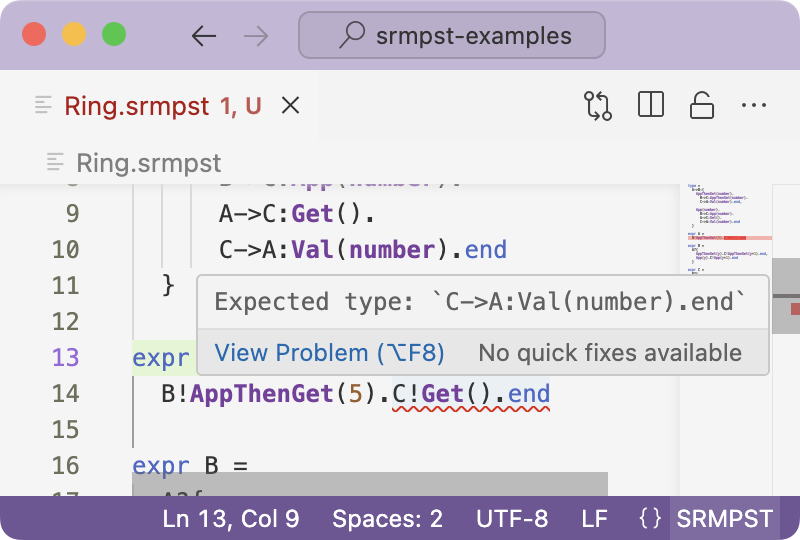} 
		\subcaption{Ill-typed: Wrong action}
		\label{fig:vscode:ill2}
	\end{minipage}
	\caption{Screenshots of the prototype language and tooling in VS Code}
	\label{fig:vscode}
	\Description{}
\end{figure}

\begin{example}
	\Cref{fig:vscode} shows three VS Code screenshots of the Ring specification and
	implementation, as a global type (lines 1-11) and as processes (lines 13-26),
	in our prototype language; they correspond with
	$\smash{\stypex{G^\text{Ring}}}$ and
	$\smash{\sexprx{P_{\smash{\rolex[][]{a}}}^\text{Ring}}},
	\smash{\sexprx{P_{\smash{\rolex[][]{b}}}^\text{Ring}}},
	\smash{\sexprx{P_{\smash{\rolex[][]{c}}}^\text{Ring}}}$ in \Cref{exmp:ring,exmp:ring-carol}. The processes in \Cref{fig:vscode:well} are well-typed,
	while the process for Alice in \Cref{fig:vscode:ill1,fig:vscode:ill2} is ill-typed. The error messages
	give the programmer actionable feedback about how/why the protocol
	is violated. \qed
\end{example}


\section{Preliminaries}\label{sect:prelim}

We first recall
the existing syntax and operational semantics of global types and processes. The
definitions in this section are standard in the MPST literature (e.g.,
\citep{DBLP:journals/jlp/GhilezanJPSY19,DBLP:journals/pacmpl/ScalasY19}). For
instance, as commonly done, we stipulate that each protocol implementation consists of a
fixed set of processes (one for every role) and channels (two between every pair
of roles; one in every direction).

\subsection{Global Types}\label{sec:global-types}

\subsubsection*{Syntax}

Regarding the syntax of global types:
\begin{itemize}
	\item Let $\rnamex{\mathbb{R}} = \setx{\rolex{a}, \rolex{b}, \rolex{c},
	\ldots}$ denote the set of \textit{roles}, ranged over by $\rnamex{p},
	\rnamex{q}, \rnamex{r}, \rnamex{s}$.

	\item Let $\lnamex{\mathcal{L}} = \setx{\lnamex{\datax{App}},
	\lnamex{\datax{Get}}, \lnamex{\datax{Val}}, \ldots}$ denote the set of
	\textit{message labels}, ranged over by $\lnamex{\ell}$.

	\item Let $\dtypex{\mathbb{T}} = \setx{\dtypex{\datax{Unit}},
	\dtypex{\datax{Bool}}, \ldots}$ denote the set of \textit{payload types},
	ranged over by $\dtypex{t}$.

	\item Let $\setx{\stypex{\datax{X}}, \stypex{\datax{Y}}, \stypex{\datax{Z}},
	\ldots}$ denote the set of \textit{recursion variables}, ranged over by
	$\stypex{X}, \stypex{Y}, \stypex{Z}$.

	\item Let $\stypex{\mathbb{G}}$ denote the set of \textit{global types}, ranged
	over by $\stypex{G}$. It is induced by the following~grammar:
	\begin{gather*}
		\stypex{G} \GRAMMAR \comxxxxxx{p}{q}{\ell_i}{t_i}{G_i}{i \in I} \PIPE \muxx{X}{G} \PIPE \stypex{X} \PIPE \one \PIPE \parxx{G_1}{G_2}
	\end{gather*}
	Global type $\comxxxxxx{p}{q}{\ell_i}{t_i}{G_i}{i \in I}$ specifies the
	communication of a message labelled $\lnamex{\ell_j}$, with a payload of type
	$\dtypex{t_j}$, from role $\rnamex{p}$ to role $\rnamex{q}$, followed by
	$\stypex{G_j}$, for some $j \in I$. Each $G_i$ is called ``a branch''. Global
	types $\muxx{X}{G}$ and $\stypex{X}$ specify recursion. As usual, recursion
	must be guarded in the sense of \citet{Yoshida:2020}. Global type $\one$
	specifies the empty protocol. Global type $\parxx{G_1}{G_2}$ specifies the
	interleaving of $\stypex{G_1}$ and $\stypex{G_2}$.
	\rev{As in the original MPST paper \cite{DBLP:conf/popl/HondaYC08}, we stipulate that the roles in $\stypex{G_1}$ and $\stypex{G_2}$ are disjoint (straightforward to syntactically check).}

	\item Let $\stypex{\mathbb{A}} = \setxx{\comxxxx{p}{q}{\ell}{t}}{\rnamex{p},
	\rnamex{q} \in \rnamex{\mathbb{R}} \text{ and } \lnamex{\ell} \in
	\lnamex{\mathcal{L}} \text{ and } \dtypex{t} \in \dtypex{\mathbb{T}}}$ denote
	the set of \textit{global actions}, ranged over by $\stypex{\alpha}$. Global
	action $\comxxxx{p}{q}{\ell}{t}$ specifies the communication of a message \rev{labelled} $\lnamex{\ell}$, with a payload of type $\dtypex{t}$, from role
	$\rnamex{p}$ to role $\rnamex{q}$.
\end{itemize}

\subsubsection*{Operational semantics}

\begin{figure}[t]
	\centering$\begin{gathered}
		\rulexxx{$\rightarrow$G-Com1}{
			j \in I
		}{
			\sttransxxx{\comxxxxxx{p}{q}{\ell_i}{t_i}{G_i}{i \in I}}{\comxxxx{p}{q}{\ell_j}{t_j}}{G_j}
		}
	\\\\
		\rulexxx{$\rightarrow$G-Com2}{
			\setx{\rnamex{p}, \rnamex{q}} \cap \setx{\rnamex{r}, \rnamex{s}} = \emptyset
		\qquad
			\sttransxxx{G_i}{\comxxxx{r}{s}{\ell}{t}}{G_i'} \text{ for each } i \in I
		}{
			\sttransxxx{\comxxxxxx{p}{q}{\ell_i}{t_i}{G_i}{i \in I}}{\comxxxx{r}{s}{\ell}{t}}{\comxxxxxx{p}{q}{\ell_i}{t_i}{G_i'}{i \in I}}
		}
	\\\\
		\rulexxx{$\rightarrow$G-Rec}{
			\sttransxxx{G \substxx{X}{\muxx{X}{G}}}{\alpha}{G'}
		}{
			\sttransxxx{\muxx{X}{G}}{\alpha}{G'}
		}
	\qquad
		\rulexxx{$\rightarrow$G-Par1}{
			\sttransxxx{G_1}{\alpha}{G_1'}
		}{
			\sttransxxx{\parxx{G_1}{G_2}}{\alpha}{\parxx{\rev{G_1'}}{G_2}}
		}
	\qquad
		\rulexxx{$\rightarrow$G-Par2}{
			\sttransxxx{G_2}{\alpha}{G_2'}
		}{
			\sttransxxx{\parxx{G_1}{G_2}}{\alpha}{\parxx{G_1}{\rev{G_2'}}}
		}
	\end{gathered}$

	\caption{Transition rules for global types}
	\label{fig:trans-glob}
	\Description{}
\end{figure}

Regarding the operational semantics of global types, let
$\sttransxxx{G}{\alpha}{G'}$ denote the transition from global type $\stypex{G}$
to global type $\stypex{G'}$ through global action $\stypex{\alpha}$. It is the
smallest relation induced by the rules in \Cref{fig:trans-glob}:
\begin{itemize}
	\item Rule \autorefrule{$\rightarrow$G-Com1} states that a communication global
	type can make a transition to any one of its branches through the corresponding
	global action.

	\item Rule \autorefrule{$\rightarrow$G-Com2} states that a communication global
	type can also make a transition when: each of its branches can make a
	transition through the same global action (second premise); this ``lexically
	next'' global action is \textit{independent} of the ``lexically first'' global
	action (first premise). Thus, independent global actions may happen
	out-of-order.
	\rev{This feature \cite{DBLP:conf/popl/HondaYC08} is needed
	to ensure that global types are not unnecessarily restrictive. Without
	allowing out-of-order execution of independent global actions, for instance,
	there would exist no global type for the following processes (trailing
	$\nil$ omitted to save space):
	\begin{gather*}
		\sexprx{P_{\smash{\rolex[][]{a}}}^\text{Com2}} = \sexprx{\sndxxx{\rolex{b1}}{\datax{Foo}}{} \prewide \sndxxx{\rolex{b2}}{\datax{Foo}}{}}
		\qquad
		\sexprx{P_{\smash{\rolex[][]{b1},\rolex[][]{b2}}}^\text{Com2}} = \sexprx{\rcvxxx{\rolex{a}}{\datax{Foo}}{\datax{\_}} \prewide \sndxxx{\rolex{c}}{\datax{Bar}}{}}
		\qquad
		\sexprx{P_{\smash{\rolex[][]{c}}}^\text{Com2}} = \sexprx{\rcvxxx{\rolex{b1}}{\datax{Bar}}{\datax{\_}} \prewide \rcvxxx{\rolex{b2}}{\datax{Bar}}{\datax{\_}}}
	\end{gather*}
	Using rule \autorefrule{$\rightarrow$G-Com2}, though, the following global type
	precisely specifies the protocol:
	\begin{gather*}
		\stypex{G^\text{Com2}} = \stypex{\comxxxx{\rolex{a}}{\rolex{b1}}{\datax{Foo}}{} \prewide \comxxxx{\rolex{a}}{\rolex{b2}}{\datax{Foo}}{} \prewide \comxxxx{\rolex{b1}}{\rolex{c}}{\datax{Bar}}{} \prewide \comxxxx{\rolex{b2}}{\rolex{c}}{\datax{Bar}}{} \prewide \one}
	\end{gather*}
	Crucially, the two middle communications can happen out-of-order.}

	\item Rule \autorefrule{$\rightarrow$G-Rec} states that a recursive global type
	can make a transition when its unfolding can. In this rule, $\stypex{G
	\substxx{X}{\muxx{X}{G}}}$ denotes the substitution of $\stypex{X}$ by
	$\muxx{X}{G}$ in $\stypex{G}$.

	\item Rules \autorefrule{$\rightarrow$G-Par1} and
	\autorefrule{$\rightarrow$G-Par2} state that an interleaving global type can
	make a transition when one of its operands can.
\end{itemize}

\begin{figure}[t]
	\centering$\begin{gathered}
		\rulexxx{}{
			\rnamex[]{R} \subseteq \setx{\rnamex{p}, \rnamex{q}}
		\qquad
			\sttransxxx{G}{\comxxxx{p}{q}{\ell}{t}}{G'}
		}{
			\sttransinclxxx{G}{R}{G'}
		}
	\qquad
		\rulexxx{}{
			\rnamex[]{R} \cap \setx{\rnamex{p}, \rnamex{q}} = \emptyset
		\qquad
			\sttransxxx{G}{\comxxxx{p}{q}{\ell}{t}}{G'}
		}{
			\sttransexclxxx{G}{R}{G'}
		}
	\qquad
		\rulexxx{}{
			\sttransinclxx[\centernot]{G}{R}
		\qquad
			\sttransexclxxx{G}{R}{G'}
		}{
			\stTransexclxxx{G}{R}{G'}
		}
	\end{gathered}$

	\caption{Derived transition rules for global types}
	\label{fig:trans-glob-extra}
	\Description{}
\end{figure}

Furthermore, let $\smash{\sttransinclxxx{G}{R}{G'}}$ (resp.
$\smash{\sttransexclxxx{G}{R}{G'}}$) denote the existence of a transition from
$\stypex{G}$ to $\stypex{G'}$ in which each (resp. none) of the roles in
$\rnamex[]{R}$ participate. Let $\smash{\stTransexclxxx{G}{R}{G'}}$ denote that:
(1) none of the roles in $\rnamex[]{R}$ participate in none of the transitions
of $\stypex{G}$; (2) $\stypex{G}$ has a transition to $\stypex{G'}$. They are
the smallest relations induced by the rules in \Cref{fig:trans-glob-extra}.

Given a fixed set of roles, for each derived transition relation
$\smash{{\dashrightarrow} \in \bigcup
\setxx{\setx{{\xrightarrow{{\rnamex[]{R}}}},
{\xrightarrow{\overline{\rnamex[]{R}}}},
{\xRightarrow{{\rnamex[]{R}}}}}}{\rnamex[]{R} \subseteq \rnamex{\mathbb{R}}}}$,
we write ``${\stypex{G} \dashrightarrow}$'' instead of ``$\stypex{G}
\dashrightarrow \stypex{G'}$ for some $\stypex{G'}$'', we write ``${\stypex{G}
\centernot\dashrightarrow}$'' instead of ``$\stypex{G} \centernot\dashrightarrow
\stypex{G'}$ for each $\stypex{G'}$'', and we write ${\dashrightarrow}^*$ for
the reflexive transitive closure.

If $\smash{\sttransinclxx{G}{\setx{\rnamex[]{r}}}}$, then $\rnamex{r}$ is
\textit{enabled}. If
$\smash{\sttransinclxx[\centernot]{G}{\setx{\rnamex[]{r}}}}$, then $\rnamex{r}$
is \textit{disabled}. If there exist global actions $\stypex{\alpha_1}, \ldots,
\stypex{\alpha_n}$ such that $\smash{\sttransxxx{G}{\alpha_1}{} {\cdots}
\sttransxxx{}{\alpha_n}{} \!\!\! \sttransinclxx{}{\setx{\rnamex[]{r}}}}$, then
$\rnamex{r}$ is \textit{active} in $\stypex{G}$; otherwise, $\rnamex{r}$ is
\textit{inactive}. Let $\rnamex{r} \in \stypex{G}$ and $\rnamex{r} \notin \stypex{G}$ denote the activeness and inactiveness of $\rnamex{r}$ in $\stypex{G}$.

\subsection{Processes}

\subsubsection*{Syntax}

Regarding the syntax of processes:
\begin{itemize}
	\item Let $\dexprx{\mathbb{X}}$ denote the set of \textit{variables}, ranged
	over by $x$.

	\item Let $\dexprx{\mathbb{V}} = \setx{\dexprx{\datax{unit}},
	\dexprx{\datax{false}}, \dexprx{\datax{true}}, \dexprx{\datax{0}},
	\dexprx{\datax{1}}, \dexprx{\datax{2}}, \ldots}$ denote the set of \textit{values}, ranged over by $\dexprx{v}$.

	\item Let $\dexprx{\mathbb{E}} = \dexprx{\mathbb{X}} \cup \dexprx{\mathbb{V}}
	\cup \setx{\dexprx{\datax{2==3}}, \dexprx{\datax{x+1}}, \ldots}$ denote the set
	of \textit{expressions}, ranged over by $\dexprx{e}$.

	\item Let $\sexprx{\mathbb{P}}$ denote the set of \textit{processes}, ranged over by
	$\proc P$. It is induced by the following grammar:
	\begin{gather*}
		\sexprx{P} \GRAMMAR \sndxxxx{q}{\ell}{e}{P} \PIPE \rcvxxxxx{p}{\ell_i}{x_i \rev{\isa \dtypex{t_i}}}{P_i}{i \in I} \PIPE
		\letxxx{x}{e}{P} \PIPE \ifxxx{e}{P_1}{P_2} \PIPE \recxx{X}{P} \PIPE \sexprx{X} \PIPE \nil
	\end{gather*}
	Output process $\sndxxxx{q}{\ell}{e}{P}$ implements the send of a message
	labelled $\lnamex{\ell}$, with (the value of) expression $\dexprx{e}$ as the
	payload, to role $\rnamex{q}$, followed by $\sexprx{P}$. Input process
	$\rcvxxxxx{p}{\ell_i}{x_i \rev{\isa \dtypex{t_i}}}{P_i}{i \in I}$ implements the receive of the payload
	of a message labelled $\lnamex{\ell_j}$, from role $\rnamex{p}$, into variable
	$\dexprx{x_j}$ \rev{of type $\dtypex{t_j}$}, followed by $\sexprx{P_j}$, for some $j \in I$. Process
	$\letxxx{x}{e}{P}$ implements the binding of variable $\dexprx{x}$ to the value
	of expression $\dexprx{e}$ in $\sexprx{P}$. Process $\ifxxx{e}{P_1}{P_2}$
	implements a conditional choice. Processes $\recxx{X}{P}$ and $\sexprx{X}$
	implement recursion. As usual, we stipulate that each process is
	\textit{\rev{message-}guarded} (i.e., recursion variables occur only under sends/receives)
	and \textit{closed} (i.e., recursion variables are bound); these are simple
	syntactic checks.

	\item Let $\sexprx{\mathbb{C}}$ denote the set of \textit{families of
	processes}---``sessions''---ranged over by $\sexprx{C}$. It is induced by the
	following grammar:
	\begin{gather*}
		\sexprx{C} \GRAMMAR \atxx{r}{P} \PIPE \compxx{C_1}{C_2}
	\end{gather*}
	Session $\atxx{r}{P}$ implements role $\rnamex{r}$ as process $\sexprx{P}$.
	Session $\compxx{C_1}{C_2}$ implements the parallel composition. As usual, we
	stipulate that each session implements each role at most once (e.g.,
	$\compxx{\atxx{r}{P_1}}{\atxx{r}{P_2}}$ is ruled out); this is a simple
	syntactic check.
	We note that process creation and session creation are orthogonal to the
	contributions of this paper and thus we omit them.
\end{itemize}

Furthermore, let $\mathsf{obj}(\sexprx{P})$ denote the \textit{object} of
$\sexprx{P}$: it is the receiver if $\sexprx{P}$ is an output process, the
sender if $\sexprx{P}$ is an input process, and undefined otherwise. It is
induced by the following equations:
\begin{gather*}
	\mathsf{obj}(\sndxxxx{q}{\ell}{e}{P}) = \rnamex{q}
\qquad
	\mathsf{obj}(\rcvxxxxx{p}{\ell_i}{x_i}{P_i}{i \in I}) = \rnamex{p}
\end{gather*}

\subsubsection*{Operational semantics}

\begin{figure}[t]
	\centering$\begin{gathered}
		\rulexxx{}{
		}{
			\evalxx{v}{v}
		}
	\qquad
		\rulexxx{}{
			\evalxx{e_1}{v}
		\qquad
			\evalxx{e_2}{v}
		}{
			\dexprx{\datax{$e_1$==$e_2$}} \Downarrow \dexprx{\datax{true}}
		}
	\qquad
		\rulexxx{}{
			\evalxx{e_1}{v_1}
		\qquad
			\evalxx{e_2}{v_2}
		\qquad
			\dexprx{v_1} \neq \dexprx{v_2}
		}{
			\dexprx{\datax{$e_1$==$e_2$}} \Downarrow \dexprx{\datax{false}}
		}
	\qquad
		{\cdots}
	\end{gathered}$

	\caption{Evaluation rules for expressions (excerpt)}
	\label{fig:eval}
	\Description{}
\end{figure}

\begin{figure}[t]
	\centering$\begin{gathered}
		\rulexxx{$\rightarrow$P-Com}{
			\evalxx{e}{v}
		\qquad
			j \in I
		}{
			\setransxxx{\compxx{\atxx{p}{\sndxxxx{q}{\ell_j}{e}{P}}}{\atxx{q}{\rcvxxxxx{p}{\ell_i}{x_i\rev{\dtypex{\isa t_i}}}{P_i}{i \in I}}}}{\rev{\comxxxx{p}{q}{\ell_j}{t_j}}}{\compxx{\atxx{p}{P}}{\atxx{q}{P_j \substxx{x_j}{v}}}}
		}
	\\\\
		\rulexxx{$\rightarrow$P-Let}{
			\evalxx{e}{v}
		}{
			\setransxxx{\atxx{r}{\letxxx{x}{e}{P}}}{\rev{\uptau}}{\atxx{r}{P \substxx{x}{v}}}
		}
	\\\\
		\rulexxx{$\rightarrow$P-If1}{
			\evalxx{e}{\datax{true}}
		}{
			\setransxxx{\atxx{r}{\ifxxx{e}{P_1}{P_2}}}{\rev{\uptau}}{\atxx{r}{P_1}}
		}
	\qquad
		\rulexxx{$\rightarrow$P-If2}{
			\evalxx{e}{\datax{false}}
		}{
			\setransxxx{\atxx{r}{\ifxxx{e}{P_1}{P_2}}}{\rev{\uptau}}{\atxx{r}{P_2}}
		}
	\\\\
		\rulexxx{$\rightarrow$P-Rec}{
		}{
			\setransxxx{\atxx{r}{\recxx{X}{P}}}{\rev{\uptau}}{\atxx{r}{P \substxx{X}{\recxx{X}{P}}}}
		}
	\qquad
		\rev{\rulexxx{}{
			\setransxxx{C_1}{\rev{\alpha}}{C_1'}
		}{
			\setransxxx{\compxx{C_1}{C_2}}{\rev{\alpha}}{\compxx{C_1'}{C_2}}
		}}
	\\\\
		\rulexxx{}{
			\setransxxx{\compxx{C_2}{C_1}}{\rev{\alpha}}{\sexprx{C'}}
		}{
			\setransxxx{\compxx{C_1}{C_2}}{\rev{\alpha}}{\sexprx{C'}}
		}
	\qquad
		\rulexxx{}{
			\setransxxx{\compxx{(\compxx{C_1}{C_2})}{C_3}}{\rev{\alpha}}{\sexprx{C'}}
		}{
			\setransxxx{\compxx{C_1}{(\compxx{C_2}{C_3})}}{\rev{\alpha}}{\sexprx{C'}}
		}
	\qquad
		\rulexxx{}{
			\setransxxx{\compxx{C_1}{(\compxx{C_2}{C_3})}}{\rev{\alpha}}{\sexprx{C'}}
		}{
			\setransxxx{\compxx{(\compxx{C_1}{C_2})}{C_3}}{\rev{\alpha}}{\sexprx{C'}}
		}
	\end{gathered}$

	\caption{Transition rules for sessions}
	\label{fig:trans-proc}
	\Description{}
\end{figure}

Regarding the operational semantics of processes:
\begin{itemize}
	\item Let $\evalxx{e}{v}$ denote the evaluation of expression $\dexprx{e}$ to
	value $\dexprx{v}$. It is induced by the rules in \Cref{fig:eval} (excerpt);
	the rules are standard.

	\item Let $\setransxxx{C}{{\rev{\alpha}}}{C'}$ and $\setransxxx{C}{{\rev{\uptau}}}{C'}$ denote the transition from session $\sexprx{C}$
	to session $\sexprx{C'}$ \rev{through global action $\stypex{\alpha}$ or \textit{internal action} $\stypex{\uptau}$; we use these transition labels to formally relate the behaviour of sessions to that of global types when proving safety}. It is induced by the rules in \Cref{fig:trans-proc}:
	\begin{itemize}
		\item Rule \autorefrule{$\rightarrow$P-Com} states that an output process and
		a corresponding input process can make a transition by sending and receiving a
		message. We note that the communication is synchronous. In this rule,
		$\sexprx{P \substxx{\dexprx{x}}{\dexprx{v}}}$ denotes the substitution of
		$\dexprx{x}$ by $\dexprx{v}$ in $\sexprx{P}$.

		\item The remaining rules are standard. We note that parallel composition of
		sessions is explicitly commutative and associative (bottom three rules),
		\rev{similar to both the original MPST paper and the ``Less Is More''
		paper \cite{DBLP:conf/popl/HondaYC08,DBLP:journals/pacmpl/ScalasY19}
		(which rely on an auxiliary structural congruence relation that includes
		commutativity and associativity axioms to define the transition rules;
		we omitted such a relation for simplicity, as we do not need its full power).
		Equivalently, a session is a partial function from roles to processes.}
	\end{itemize}
\end{itemize}


\section{The Special Case: Typing with Global Types}\label{sec:global}

In this section, we present our results for the special case of our synthetic
approach to MPST.

\subsection{Type System}

Let $\denvx{\mathbf{\Gamma}}$ and $\senvx{\mathbf{\Delta}}$ denote the sets of
\textit{data type environments} and \textit{session type environments}, ranged
over by $\denvx{\varGamma}$ and $\senvx{\varDelta}$. They are induced by the
following grammar:
\begin{gather*}
	\denvx{\varGamma} \GRAMMAR \denvx{\varnothing} \PIPE \denvxxx{\varGamma}{x}{t}
\qquad
	\senvx{\varDelta} \GRAMMAR \senvx{\varnothing} \PIPE \senvxxx{\varDelta}{X}{G}
\end{gather*}
As usual, we consider type environments up to reordering of entries for
different variables (e.g., we can reorder
$\denvxxx{\denvxxx{\varGamma}{\datax{x}}{\datax{Unit}}}{\datax{y}}{\datax{Bool}}$ to $\denvxxx{\denvxxx{\varGamma}{\datax{y}}{\datax{Bool}}}{\datax{x}}{\datax{Unit}}$, but we cannot reorder $\denvxxx{\denvxxx{\varGamma}{\datax{x}}{\datax{Unit}}}{\datax{x}}{\datax{Bool}}$).

\begin{figure}[t]
	\centering$\begin{gathered}
		\rulexxx{}{
		}{
			\dtypedxxx{\varGamma, \expr{x} \typed \expt{t}}{x}{t}
		}
	\qquad
		\rulexxx{}{
		}{
			\dtypedxxx{\varGamma}{\datax{true}}{\datax{Bool}}
		}
	\qquad
		\rulexxx{}{
			\dtypedxxx{\varGamma}{e_1}{t_1}
		\qquad
			\dtypedxxx{\varGamma}{e_2}{t_2}
		}{
			\dtypedxxx{\varGamma}{e_1 \datax{==} e_2}{\datax{Bool}}
		}
	\qquad
		{\cdots}
	\end{gathered}$

	\caption{Typing rules for expressions (excerpt)}
	\label{fig:typing-expr}
	\Description{}
\end{figure}

Let $\dtypedxxx{\varGamma}{E}{S}$ denote well-typedness of expression $\expr{E}$
by payload type $\expt{S}$ in data type environment $\denvx{\varGamma}$. It is the smallest relation
induced by the rules in \Cref{fig:typing-expr} (excerpt); the rules are
standard.

\begin{figure}[t]
	\centering$\begin{gathered}
		\rulexxx{$\vdash$-Send}{
			\dtypedxxx{\varGamma}{e}{t}
		\qquad
			\stypedxxxx{\varGamma}{\varDelta}{\atxx{p}{P}}{G'}
		\qquad
			\sttransxxx{G}{\comxxxx{p}{q}{\ell}{t}}{G'}
		}{
			\stypedxxxx{\varGamma}{\varDelta}{\atxx{p}{\sndxxxx{q}{\ell}{e}{P}}}{G}
		}
	\\\\
		\rulexxx{$\vdash$-Recv}{
			\rev{
			\sttransxx{G}{\comxxxx{p}{q}{\ell'}{t'}}
			}
		\qquad
			\rev{
			\forall\,\sttransxxx{G}{\comxxxx{p}{q}{\ell}{\rev{t}}}{G'}\,.\,
			\big[
			\exists\,j \in I\,.\,
			\big[
			\lnamex{\ell_j} = \lnamex{\ell}\ \wedge\ \stypedxxxx{\denvxxx{\varGamma}{x_j}{\rev{t}}}{\varDelta}{\atxx{q}{P_j}}{G'}
			\big]
			\big]
			}
		}{
			\stypedxxxx{\varGamma}{\varDelta}{\atxx{q}{\rcvxxxxx{p}{\ell_i}{x_i}{P_i}{i \in I}}}{G}
		}
	\\\\
		\rulexxx{$\vdash$-Skip}{\begin{gathered}
			\text{(1) } \sttransinclxx[\centernot]{G}{\setx{\rnamex{r}}}
		\qquad
			\text{(2) }
			\rev{
			\forall\,\sttransexclxxx[][*]{G}{\setx{\rnamex{r}}}{G'}\,.\,
			\big[
			\exists\,\stypex{G''}\,.\,
			\big[
			\stTransexclxxx[][*]{G'}{\setx{\rnamex{r}}}{}
			\sttransinclxx{G''}{\setx{\rnamex{r}}}
			\big]
			\big]
			}
		\\[-.5\baselineskip]
			\text{(3) }
			\rev{
			\forall\,\stypex{G} = \stTransexclxxx[][*]{G'}{\setx{\rnamex{r}}}{} \sttransinclxx{G''}{\setx{\rnamex{r}}}\,.\,
			\big[
			\stypedxxxx{\varGamma}{\varDelta}{\atxx{r}{P}}{G''}
			\big]
			}
		\\[-.5\baselineskip]
			\text{(4) }
			\rev{
			\forall\,\sttransexclxxx[][*]{G}{\setx{\rnamex{r}}}{G'}\,.\,
			\big[
			\sttransinclxx{G'}{\setx{\rnamex{r}}}
			\ \vee\
			\sttransexclxxx[][*]{G'}{\setx{\rnamex{r}, {\color{black}\mathsf{obj}(\sexprx{P})}}}{} \!\!\! \sttransinclxx{}{\setx{\rnamex{r}, {\color{black}\mathsf{obj}(\sexprx{P})}}}
			\big]
			}
		\end{gathered}}{
			\stypedxxxx{\varGamma}{\varDelta}{\atxx{r}{P}}{G}
		}
	\\\\
		\rulexxx{$\vdash$-End}{
			\rev{
			\forall\,\sttransexclxxx[][*]{G}{\setx{\rnamex{r}}}{G'}\,.\,
			\big[
			\sttransinclxx[\centernot]{G'}{\setx{\rnamex{r}}}
			\big]
			}
		}{
			\stypedxxxx{\varGamma}{\varDelta}{\atxx{r}{\nil}}{G}
		}
	\qquad
		\rulexxx{$\vdash$-Let}{
			\dtypedxxx{\varGamma}{e}{t}
		\qquad
			\stypedxxxx{\denvxxx{\varGamma}{x}{t}}{\varDelta}{\atxx{r}{P}}{G}
		}{
			\stypedxxxx{\varGamma}{\varDelta}{\atxx{r}{\letxxx{x}{e}{P}}}{G}
		}
	\\\\
		\rulexxx{$\vdash$-If}{
			\dtypedxxx{\varGamma}{e}{\datax{Bool}}
		\qquad
			\stypedxxxx{\varGamma}{\varDelta}{\atxx{r}{P_1}}{G}
		\qquad
			\stypedxxxx{\varGamma}{\varDelta}{\atxx{r}{P_2}}{G}
		}{
			\stypedxxxx{\varGamma}{\varDelta}{\atxx{r}{\ifxxx{e}{P_1}{P_2}}}{G}
		}
	\\\\
		\rulexxx{$\vdash$-Rec}{
			\stypedxxxx{\varGamma}{\senvxxx{\varDelta}{X}{G}}{\atxx{r}{P}}{G}
		\qquad
			\rev{\text{$\proc{P}$ is message-guarded}}
		}{
			\stypedxxxx{\varGamma}{\varDelta}{\atxx{r}{\recxx{X}{P}}}{G}
		}
	\qquad
		\rulexxx{$\vdash$-Var}{
			\sttransexclxxx[][*]{G}{\setx{\rnamex{r}}}{G'}
		}{
			\stypedxxxx{\varGamma}{\senvxxx{\varDelta}{X}{G}}{\atxx{r}{X}}{G'}
		}
	\\\\
		\rulexxx{$\vdash$-Comp}{
			\mathsf{dom}(\sexprx{C_1}) \cap \mathsf{dom}(\sexprx{C_2}) = \emptyset
		\qquad
			\stypedxx{C_1}{G}
		\qquad
			\stypedxx{C_2}{G}
		}{
			\stypedxx{\compxx{C_1}{C_2}}{G}
		}
	\end{gathered}$

	\caption{Typing rules for processes}
	\label{fig:typing-proc}
	\Description{}
\end{figure}

Let $\stypedxxxx{\varGamma}{\varDelta}{C}{G}$ denote well-typedness of session
$\sexprx{C}$ by global type $\stypex{G}$ in type environments
$\denvx{\varGamma}$ and $\senvx{\varDelta}$. \rev{We write $\stypedxx{C}{G}$ instead of $\stypedxxxx{\varnothing}{\varnothing}{C}{G}$.} It is the smallest relation induced by the rules in
\Cref{fig:typing-proc}.
\rev{We first explain how the core, non-standard rules should be read and what they informally mean.
In \Cref{sect:glob:disc}, we provide a more in-depth discussion.}
\begin{itemize}
	\item Rule \autorefrule{$\vdash$-Send} states that, as an implementation of
	role $\rnamex{p}$ (sender), an output process is well-typed when: the payload
	is well-typed (first premise); the successor is well-typed (second premise);
	there exists a corresponding transition (third premise). More intuitively, this
	rule means that each send implemented needs to be specified, but not each send
	specified needs to be implemented.

	\item Rule \autorefrule{$\vdash$-Recv} states that, as an implementation of
	role $\rnamex{q}$ (receiver), an input process is well-typed when, for each
	transition \rev{(at least one)}, a corresponding branch exists (i.e., it has the specified message
	label) and is well-typed in an extended type environment (i.e., the variable
	has the specified payload type). More intuitively, this rule means that each
	receive specified needs to be implemented, but not each receive implemented
	needs to be specified.

	\item Rule \autorefrule{$\vdash$-Skip} states that, as an implementation of
	role $\rnamex{r}$, a process is well-typed when:
	\begin{enumerate}
		\item For the present $\stypex{G}$, a send or receive by $\rnamex{r}$
		\textit{is not} specified.

		\item For the future, a send or receive by $\rnamex{r}$ \textit{is} specified.

		That is, for each ``near future'' $\stypex{G'}$---reachable through
		zero-or-more transitions without $\rnamex{r}$, but $\rnamex{r}$ may have been
		enabled---there exists a ``distant future'' $\stypex{G''}$---reachable through
		another zero-or-more transitions without $\rnamex{r}$, and $\rnamex{r}$ must
		have been disabled---for which a send or receive by $\rnamex{r}$ is specified.
		At least one distant future exists (when $\stypex{G} = \stypex{G'}$).

		\item In each distant future $\stypex{G''}$, the process is well-typed. We
		note that the ``${\stypex{G} =}$'' part in this premise is technically redundant; we
		included it so that meta-variables $\stypex{G'}$ and $\stypex{G''}$ are bound
		in the same way as in premise (2).

		More intuitively, this premise means that an implementation of $\rnamex{r}$
		ignores all communications in which $\rnamex{r}$ does not participate.
		However, regardless of which other communications other processes engage in (ignored by $\rnamex{r}$), an
		implementation of $\rnamex{r}$ must behave in compliance with any possible
		future that may arise.

		\item In each near future, \rev{either $\rnamex{r}$ is enabled, or} $\rnamex{r}$ and
		$\mathsf{obj}(\sexprx{P})$ (i.e., the next communication partner of
		$\rnamex{r}$) cannot communicate with each other until either one of them has
		communicated with another process.

		More intuitively, this premise means that there needs to be some kind of
		causality: implementations of $\rnamex{r}$ and $\mathsf{obj}(\sexprx{P})$
		cannot start communicating with each other spontaneously: either it must
		already be possible, or it must happen in response to a communication of one of them with
		another process.
	\end{enumerate}

	We note that rule \autorefrule{$\vdash$-Skip} is not syntax-directed. This is
	different from existing type systems in the MPST literature, including in the
	classical approach and the ``Less Is More'' approach.

	\item Rule \autorefrule{$\vdash$-Rec} and rule \autorefrule{$\vdash$-Var} state
	that, as an implementation of role $\rnamex{r}$, a recursive process is
	well-typed when: the body is well-typed (premise of rule
	\autorefrule{$\vdash$-Rec}); the global types upon starting and finishing the
	body, $\stypex{G}$ and $\stypex{G'}$,  are reachable through transitions
	without $\rnamex{r}$ (premise of rule \autorefrule{$\vdash$-Var}). The latter
	is a generalisation of the usual equality condition on $\stypex{G}$ and
	$\stypex{G'}$ in typing rules for recursive processes in the MPST literature.
	Our relaxation enables typing more recursive processes. The following example
	demonstrates the usefulness.

	\begin{example}\label{exmp:lasso}
		The following global type and its LTS specify a protocol in which a
		$\lnamex{\datax{Foo}}$ message is communicated first from Alice to Bob and
		next, \textit{ad infinitum}, from Bob to Carol and Dave:
		\begin{gather*}
			\begin{gathered}\stypex{\rev{G}^\mathrm{Lasso}} = \stypex{\begin{aligned}[t]
			&
				\comxxxx{\rolex{a}}{\rolex{b}}{\datax{Foo}}{}
				\prewide
				\mux{\datax{X}}
				\prewide
			\\&\quad
				\comxxxx{\rolex{b}}{\rolex{c}}{\datax{Foo}}{}
				\prewide
			\\&\quad
				\comxxxx{\rolex{b}}{\rolex{d}}{\datax{Foo}}{}
				\prewide
			\\&\quad
				\datax{X}
			\end{aligned}}\end{gathered}
		\qquad
			\begin{gathered}\begin{tikzpicture}[x=.75cm, y=.75cm]
				\tikzstyle{state} = []
				\tikzstyle{trans} = [-stealth, rounded corners]
				\tikzstyle{label} = [inner sep=0pt, anchor=base, yshift=1.5mm, font=\footnotesize]
				\node [state] (G1) at (0,0) {$\stypex{\rev{G}_{\smash{1}}^\text{Lasso}}$};
				\node [state] (G2) at (3,0) {$\stypex{\rev{G}_{\smash{2}}^\text{Lasso}}$};
				\node [state] (G3) at (6,0) {$\stypex{\rev{G}_{\smash{3}}^\text{Lasso}}$};
				\draw [trans] ([xshift=-.5cm]G1.west) to (G1);
				\draw [trans] (G1) to node [label] {$\comxxxx{\rolex[]{a}}{\rolex[]{b}}{\datax[]{Foo}}{}$} (G2);
				\draw [trans] (G2) to ($(G2)+(0,+1)$) to node [label] {$\comxxxx{\rolex[]{b}}{\rolex[]{c}}{\datax[]{Foo}}{}$} ($(G3)+(0,+1)$) to (G3);
				\draw [trans] (G3) to ($(G3)+(0,-1)$) to node [label] {$\comxxxx{\rolex[]{b}}{\rolex[]{d}}{\datax[]{Foo}}{}$} ($(G2)+(0,-1)$) to (G2);
			\end{tikzpicture}\end{gathered}
		\end{gather*}
		The following derivation (excerpt for simplicity) states that, as an
		implementation of Dave, process
		$\sexprx{\recxx{\datax{X}}{\rcvxxx{\rolex{b}}{\datax{Foo}}{\datax{x}} \pre
		\datax{X}}}$ is well-typed by $\stypex{\rev{G}_{\smash{1}}^\text{Lasso}}$:
		\begin{gather*}
			\dfrac{
				\dfrac{
					\dfrac{
						\dfrac{
							\sttransexclxxx[][*]{\rev{G}_{\smash{1}}^\text{Lasso}}{\setx{\rolex[]{d}}}{\rev{G}_{\smash{2}}^\text{Lasso}}
						}{
							\stypedxxxx{\denvxxx{\varnothing}{\datax{x}}{\datax{Unit}}}{\senvxxx{\varnothing}{\datax{X}}{\rev{G}_{\smash{1}}^\text{Lasso}}}{\atxx{\rolex{d}}{\datax{X}}}{\rev{G}_{\smash{2}}^\text{Lasso}}
						}
						\,\text{\scriptsize\autorefrule{$\vdash$-Var}}
					\qquad
						{\cdots}
					}{
						\stypedxxxx{\varnothing}{\senvxxx{\varnothing}{\datax{X}}{\rev{G}_{\smash{1}}^\text{Lasso}}}{\atxx{\rolex{d}}{\rcvxxx{\rolex{b}}{\datax{Foo}}{\datax{x}} \pre \datax{X}}}{\rev{G}_{\smash{3}}^\text{Lasso}}
					}
					\,\text{\scriptsize\autorefrule{$\vdash$-Recv}}
				\qquad
					{\cdots}
				}{
					\stypedxxxx{\varnothing}{\senvxxx{\varnothing}{\datax{X}}{\rev{G}_{\smash{1}}^\text{Lasso}}}{\atxx{\rolex{d}}{\rcvxxx{\rolex{b}}{\datax{Foo}}{\datax{x}} \pre \datax{X}}}{\rev{G}_{\smash{1}}^\text{Lasso}}
				}
				\,\text{\scriptsize\autorefrule{$\vdash$-Skip}}
			}{
				\stypedxx{\atxx{\rolex{d}}{\recxx{\datax{X}}{\rcvxxx{\rolex{b}}{\datax{Foo}}{\datax{x}} \pre \datax{X}}}}{\rev{G}_{\smash{1}}^\text{Lasso}}
			}
			\,\text{\scriptsize\autorefrule{$\vdash$-Rec}}
		\end{gather*}
		This derivation crucially takes advantage of our relaxation: rule
		\autorefrule{$\vdash$-Var} does not require equality of the global types on
		the left-hand side and on the right-hand side of the turnstile, but the
		existence of a sequence of transitions between them is sufficient. \qed
	\end{example}

	In general, typing recursive processes is a non-trivial problem. We are
	currently working on further generalisations of rule
	\autorefrule{$\vdash$-Var}. For the purpose of this paper (notably: passing the
	``Less Is More'' benchmark), the current version of rule \autorefrule{$\vdash$-Var}
	already provides enough expressive power.
\end{itemize}
\rev{For the top-level session, as a well-formedness requirement, the type
system also checks that each role that occurs in the global type is implemented
as a process in the session.}

\rev{Just as in the classical approach to MPST, it is possible in our approach
to write global types that fundamentally cannot be implemented as well-typed
sessions; they are inherently ``unrealisable'' as distributed systems. In the
classical approach, such global types are ruled out by leaving the projection
onto at least one role undefined; thus, there are not enough local types to
check processes against. In contrast, in our approach, unrealisability manifests
through the standard notion of \textit{type inhabitation}.}
In particular, global types that
specify protocols that violate the \textit{Knowledge of Choice} (KC) principle
are uninhabited. Intuitively, KC demands that if the future of a protocol
depends on choices made in the past, then each role needs to be(come) aware of
those choices in a timely fashion. The following example demonstrates an
uninhabited global type.

\begin{example}
	The \textit{Confusion} protocol consists of roles \textit{Alice}, \textit{Bob},
	and \textit{Carol}. First, a $\lnamex{\datax{Foo}}$ message or a
	$\lnamex{\datax{Bar}}$ message is communicated from Alice to Bob. Next, a
	$\lnamex{\datax{Confusion}}$ message is communicated from Bob to Carol. Last, a
	message with the same label as the one that was communicated from Alice to Bob
	is communicated from Carol to Alice. While Alice and Bob are aware of the
	choice between $\lnamex{\datax{Foo}}$ and $\lnamex{\datax{Bar}}$, Carol is not:
	regardless of the choice, she always receives a $\lnamex{\datax{Confusion}}$
	message.

	The following global type specifies the Confusion protocol:
	\begin{gather*}
		\stypex{G^\text{Conf}} = \begin{tree}
			\branchx{
				\comxx{\rolex{a}}{\rolex{b}} \holex{h1}
			}
			\branchxx[h1]{
				\mtypexx{\datax{Foo}}{}
				\prewide
				\comxxxx{\rolex{b}}{\rolex{c}}{\datax{Confusion}}{}
				\prewide
				\comxxxx{\rolex{c}}{\rolex{a}}{\datax{Foo}}{}
				\prewide
				\one
			}{
				\mtypexx{\datax{Bar}}{}
				\prewide
				\comxxxx{\rolex{b}}{\rolex{c}}{\datax{Confusion}}{}
				\prewide
				\comxxxx{\rolex{c}}{\rolex{a}}{\datax{Bar}}{}
				\prewide
				\one
			}
		\end{tree}
	\end{gather*}
	This global type is uninhabited. The problem is that any well-typed
	implementation of Carol would have to start with receiving a
	$\lnamex{\datax{Confusion}}$ message:
	\begin{gather*}
		\sexprx{P_{\smash{\rolex[][]{c}}}^\text{Conf}} = \sexprx{\rcvxxx{\rolex{b}}{\datax{Confusion}}{\datax{\_}} \prewide P'}
	\end{gather*}
	However, $\sexprx{P'}$ must now be well-typed by both
	$\sexprx{\comxxxx{\rolex{c}}{\rolex{a}}{\datax{Foo}}{} \pre \one}$ and
	$\sexprx{\comxxxx{\rolex{c}}{\rolex{a}}{\datax{Bar}}{} \pre \one}$, so it has
	to be both $\sexprx{\sndxxx{\rolex{a}}{\datax{Foo}}{} \pre \nil}$ and
	$\sexprx{\sndxxx{\rolex{a}}{\datax{Bar}}{} \pre \nil}$, which is a
	contradiction. \qed
\end{example}

\noindent
\rev{We note that unrealisability implies unprojectability and uninhabitation,
but not the other way around: projectability and inhabitation are conservative
in the sense that they reject more global types than just the unrealisable ones.
For now, the exact relation between projectability and inhabitation is unknown,
but we conjecture that the former is strictly subsumed by the latter.}

\startrev
\subsection{In-Depth Discussion of the Design of the Typing Rules}
\label{sect:glob:disc}

\subsubsection{Rule \textup{\protect\autorefrule{$\vdash$-Skip}}}

The most complicated rule of the type system is rule
\autorefrule{$\vdash$-Skip}. One apparent complication is that it looks ahead
multiple transitions of the global type instead of only a single one. The reason
why we chose the multi-transition design is that it makes dealing with cycles
significantly easier. The key insight is that, ultimately, we need to reason
about the \textit{reachable successors} of a global type (``near futures'' and
``distant futures''), subject to additional conditions along the way. This can
be directly expressed by looking ahead multiple transitions, but only
indirectly---using substantial additional bookkeeping as part of the typing
judgment---by looking ahead a single transition at a time. Essentially, the
complexity of dealing with cycles is pushed into the computation of the
transitive closure, for which existing algorithms can be straightforwardly
adapted (as is done in our prototype language and tooling in VS Code).

When quantifying over reachable successors by looking ahead multiple
transitions, a subtle point that needs to be addressed is that the domain is
non-empty: at least one reachable successor needs to exist. This is the purpose
of premise 2. It ensures that the universal quantification in premise 3 has a
non-empty domain \textit{in every possible successor that is reachable after
unrelated communications}. This is essential for soundness (i.e., if an empty
domain were allowed, then premise 3 would be vacuously true, which would
erroneously mean that $\proc{P}$ could be, or do, anything).

More generally, regarding the efficacy of the four premises of
\autorefrule{$\vdash$-Skip}, the theorems later on in this paper establish that
they are \textit{sufficient} (in the technical sense) to prove type soundness
(as also confirmed by our Agda formalisation). We also show that the premises
are sufficiently liberal to pass the ``Less Is More'' benchmark (i.e., for each
example in that benchmark, an inhabited global type exists). Whether or not the
premises are also \textit{necessary} remains for now an
open question.

\subsubsection{Rule \textup{\protect\autorefrule{$\vdash$-Var}}}

According to rule \autorefrule{$\vdash$-Var}, process variable $\proc{X}$ is
typable by global type $\stypex{G'}$ when $\stypex{G'}$ is reachable from the
global type $\stypex{G}$ stored for $\proc{X}$ in the session type environment.
As $\proc{X}$ can stand for any process, one might expect that a generalisation
for any process is sound as well:
\begin{gather*}
	\rulexxx{$\vdash$-Forward}{
		\stypedxxxx{\varGamma}{\varDelta}{\atxx{r}{P}}{G}
		\qquad
		\sttransexclxxx[][*]{G}{\setx{\rnamex{r}}}{G'}
	}{
		\stypedxxxx{\varGamma}{\varDelta}{\atxx{r}{P}}{G'}
	}
\end{gather*}
Indeed, this rule is admissible: it is a direct consequence of
\Cref{lemma:unrelated-step}, which we present in the next section. Informally,
that lemma states that the well-typedness of a process that implements role
$\rnamex{r}$ is preserved by any global type transition that does not invole
$\rnamex{r}$. Rule \autorefrule{$\vdash$-Forward} is then established by
a straightforward inductive argument on a sequence of transitions
that do not involve $\rnamex{r}$.

\subsubsection{Rule \textup{\protect\autorefrule{$\vdash$-End}}}

In the operational semantics of global types, we do not distinguish between
successful and abnormal termination: any global type that has no outgoing
transitions is considered successfully terminated. If all recursion is guarded
(as stipulated), then the only global type without outgoing transitions is
$\one$, which specifies succesful termination, so we do not need additional
expressive power to represent abnormal termination. In general, however, it can
be useful.

Adding an explicit notion of successful termination---independent of the
presence\slash absence of outgoing transitions---is a non-trivial problem to
solve, though. In particular, there does not seem to be an obvious way to define
a notion of ``global final state'' that can be used in rule
\autorefrule{$\vdash$-End}. For instance, reconsider global type
$\stypex{G^\mathrm{Lasso}}$ from \Cref{exmp:lasso}:
\begin{gather*}
	\stypex{G^\mathrm{Lasso}} = \stypex{G_1^\mathrm{Lasso}} = \stypex{
		\comxxxx{\rolex{a}}{\rolex{b}}{\datax{Foo}}{}
		\prewide
		\stypex{G_2^\mathrm{Lasso}}}
\qquad
	\stypex{G_2^\mathrm{Lasso}} = \stypex{
		\mux{\datax{X}}
		\prewide
		\comxxxx{\rolex{b}}{\rolex{c}}{\datax{Foo}}{}
		\prewide
		\comxxxx{\rolex{b}}{\rolex{d}}{\datax{Foo}}{}
		\prewide
		\datax{X}}
\end{gather*}
A well-typed implementation of $\rolex{a}$ is
$\sexprx{\sndxxx{\rolex{b}}{\datax{Foo}}{} \pre \nil}$. But, after applying rule
\autorefrule{$\vdash$-Send}, we would need to type-check $\sexprx{\nil}$ against
global type $\stypex{G_2^\mathrm{Lasso}}$, which cannot be a ``global final
state''. In other words, participants may finish their communications in a
protocol before the protocol as a whole terminates (if ever). A possible
solution could be to define a separate set of ``local final states'' for each
role, but doing so might have deep consequences that require careful study in
future work.

\finishrev

\subsection{Main Theoretical Result: Type Soundness}

The main theoretical result of the special case of our synthetic approach to
MPST is \textit{type soundness}: well-typedness implies safety and liveness.
Formally, we prove type soundness in terms of \textit{progress} and
\textit{preservation}. Progress means that well-typed sessions \rev{eventually
either terminate or perform another communication. In particular, it is
impossible for well-typed sessions to diverge into an infinite sequence of
internal transitions, as all recursion variables in well-typed processes must be
message-guarded: at least one send or receive must happen before each recursive
call. Thus, well-typed sessions are live (i.e., progress is exactly strong
enough to formally define liveness).} Preservation means that well-typedness is
preserved by transitions of sessions, \rev{and that these transitions are
allowed by the global types. In particular, if a well-typed session makes a
transition through a communication, then the global type can make a transition
with exactly the same communication. Thus, well-typed sessions are safe (i.e.,
preservation is exactly strong enough to formally define safety)}. We show the
proofs of these theorems in \Cref{sec:behaviour}.

\startrev
\begin{agdaenv}{theorem}{Progress}{https://github.com/SyntheticBehaviouralTypes/srmpst-formalisation/blob/f4b15c0074bc496a31a44c913b92863a46a32ee7/Safety.agda\#L632-L663}\label[theorem]{thm:progress-glob}%
	If $\stypedxx{C}{G}$, then: \textup{(1)} not $\setransxxx{C}{\uptau}{}
	\!\!\! \setransxxx{}{\uptau}{} {\cdots}$; \textup{(2)}
	$\setransxxx{C}{\uptau}{} {\cdots}
	\setransxxx{}{\uptau}{\compxxx{\nil}{{\cdots}}{\nil}}$, or
	$\setransxxx{C}{\uptau}{} {\cdots} \setransxxx{}{\uptau}{} \!\!\!
	\setransxxx{}{\alpha}{C'}$, for some $\sexprx{C'}$.
\end{agdaenv}

\begin{agdaenv}{theorem}{Preservation}{https://github.com/SyntheticBehaviouralTypes/srmpst-formalisation/blob/f4b15c0074bc496a31a44c913b92863a46a32ee7/Safety.agda\#L182-L210}\label[theorem]{thm:preservation-glob}
	Suppose $\stypedxx{C}{G}$:
	\begin{itemize}
		\item If $\setransxxx{C}{\alpha}{C'}$, then $\stypedxx{C'}{G'}$ and
		$\sttransxxx{G}{\alpha}{G'}$, for some $\stypex{G'}$.

		\item If $\setransxxx{C}{\uptau}{C'}$, then $\stypedxx{C'}{G}$.
	\end{itemize}
\end{agdaenv}

We note that our notions of progress and preservation do not prevent
\textit{starvation}: while a non-ter\-mi\-nat\-ing session as-a-whole is
guaranteed to alway eventually perform another communication, without
\textit{fairness}, individual processes might get stuck waiting for a message
that is never sent.

\finishrev

\subsection{Main Practical Result: Passing the ``Less Is More'' Benchmark}\label{sect:glob:benchm}

The main practical result is that the synthetic approach of this paper passes
the ``Less Is More'' benchmark of Scalas and Yoshida
\citep{DBLP:journals/pacmpl/ScalasY19}. This is a set of four challenging example
protocols that demonstrate limitations of the classical approach to MPST
(\Cref{fig:mpst:popl08}); it served as a motivation for the ``Less Is More''
approach (\Cref{fig:mpst:popl19}). The type system of this section is the first
one to support the example protocols in a fully compositional manner. This means
that processes are all individually type-checked, without the need for
whole-system reconstruction and analysis (e.g., the model checking step in the ``Less Is More'' approach).

\begin{figure}[t]
	\begin{minipage}{\linewidth}
		\fbox{\begin{minipage}{\linewidth-7pt}
			\noindent \textbf{Roles:} Server ($\rolex{s}$), Client ($\rolex{c}$),
			Authorisation Service ($\rolex{a}$) \smallbreak

			\noindent \textbf{Protocol:} Server tells Client it can continue the session
			by logging in, or it cancels the session. In the former case, Client tells
			Authorisation Service its password, after which Authorisation Service tells
			Server whether the login succeeded. In the latter case, Client tells
			Authorisation Service to quit. \smallbreak

			\noindent\strut\rlap{\textbf{Global type:}}\hfill\smash{$\stypex{
				\begin{tree}
					\branchx{
						\comxx{\rolex{s}}{\rolex{c}} \holex{h1}
					}
					\branchxx[h1]{
						\mtypexx{\datax{Login}}{}
						\prewide
						\comxxxx{\rolex{c}}{\rolex{a}}{\datax{Passwd}}{\datax{Str}}
						\prewide \rolex{a} \unbuf \rolex{s} \isa \lnamex{\datax{Auth}}(\expt{\datax{Bool}})
						\prewide
						\one
					}{
						\mtypexx{\datax{Cancel}}{}
						\prewide
						\comxxxx{\rolex{c}}{\rolex{a}}{\datax{Quit}}{}
						\prewide
						\one
					}
				\end{tree}}
			$}\hfill\strut\smallbreak

			\noindent \textbf{Well-typed session:}
			\smallbreak\strut\hfill$\sexprx{\begin{aligned}
			&
				\atxx{\rolex{s}}{\sndxxx{\rolex{c}}{\datax{Cancel}}{} \prewide \nil}
			\\{\mid{}}&
				\atxx{\rolex{c}}{
					\rcvxx{\rolex{s}}{
						\mexprxx{\datax{Login}}{\datax{\_}}
						\prewide
						\sndxxx{\rolex{a}}{\datax{Passwd}}{\datax{"asdf"}}
						\prewide
						\nil
					\commawide
						\mexprxx{\datax{Cancel}}{\datax{\_}}
						\prewide
						\sndxxx{\rolex{a}}{\datax{Quit}}{}
						\prewide
						\nil
					}
				}
			\\{\mid{}}&
				\atxx{\rolex{a}}{
					\rcvxx{\rolex{c}}{
						\mexprxx{\datax{Passwd}}{\datax{x}}
						\prewide
						\sndxxx{\rolex{s}}{\datax{Auth}}{\datax{x=="asdf"}}
						\prewide
						\nil
					\commawide
						\mexprxx{\datax{Quit}}{\datax{\_}}
						\prewide
						\nil
					}
				}
			\end{aligned}}$\hfill\strut
		\end{minipage}}

		\subcaption{OAuth2 Fragment}
		\label{fig:benchmark:1}
	\end{minipage}

	\medbreak

	\begin{minipage}{\linewidth}
		\fbox{\begin{minipage}{\linewidth-7pt}
			\noindent \textbf{Roles:} Alice ($\rolex{a}$), Store ($\rolex{s}$),
			Bob ($\rolex{b}$) \smallbreak

			\noindent \textbf{Protocol:} Alice asks Store for a quote of an item. Store
			tells Alice the price. Alice asks Bob to split the price or to cancel the
			session. In the former case, Bob tells Alice whether or not he is willing to
			split. If he is, then Alice tells Store that the purchase goes through, but
			if not, then she asks Bob again to split the price or to cancel the session.
			In the latter case, Alice tells Store that no purchase will be made.
			\smallbreak

			\noindent \textbf{Global type:}
			\smallbreak\strut\hfill$\stypex{
				\begin{tree}
					\branchx{
						\comxxxx{\rolex{a}}{\rolex{s}}{\datax{Query}}{\datax{Str}}
						\prewide
						\comxxxx{\rolex{a}}{\rolex{s}}{\datax{Price}}{\datax{Int}}
						\prewide
						\mux{\datax{X}}
						\prewide
						\comxx{\rolex{a}}{\rolex{b}} \holex{h1}
					}
					\branchxx[h1]{
						\mtypexx{\datax{Split}}{\datax{Int}}
						\prewide
						\comxx{\rolex{b}}{\rolex{a}} \holex{h2}
					}{
						\mtypexx{\datax{Cancel}}{}
						\prewide
						\comxxxx{\rolex{a}}{\rolex{s}}{\datax{No}}{}
						\prewide
						\one
					}
					\branchxx[h2]{
						\mtypexx{\datax{Yes}}{}
						\prewide
						\comxxxx{\rolex{a}}{\rolex{b}}{\datax{Buy}}{}
						\prewide
						\one
					}{
						\mtypexx{\datax{No}}{}
						\prewide \datax{X}
					}
				\end{tree}
			}$\hfill\strut\smallbreak

			\noindent \textbf{Well-typed session:}
			\smallbreak\strut\hfill$\sexprx{\begin{aligned}
			&
				\atxx{\rolex{a}}{
					\sndxxx{\rolex{s}}{\datax{Item}}{\datax{"tapl"}}
					\prewide
					\rcvxxx{\rolex{s}}{\datax{Price}}{\datax{x}}
					\prewide
					\sndxxx{\rolex{b}}{\datax{Cancel}}{}
					\prewide
					\sndxxx{\rolex{s}}{\datax{No}}{}
					\prewide
					\nil
				}
			\\{\mid{}}&
				\atxx{\rolex{s}}{
					\rcvxxx{\rolex{a}}{\datax{Item}}{\datax{y}}
					\prewide
					\sndxxx{\rolex{a}}{\datax{Price}}{\datax{20}}
					\prewide
					\rcvxx{\rolex{s}}{
						\mexprxx{\datax{Buy}}{\datax{\_}}
						\prewide
						\nil
					\commawide
						\mexprxx{\datax{No}}{\datax{\_}}
						\prewide
						\nil
					}
				}
			\\{\mid{}}&
				\atxx{\rolex{b}}{
					\rcvxx{\rolex{a}}{
						\mexprxx{\datax{Split}}{\datax{z}}
						\prewide
						\sndxxx{\rolex{a}}{\datax{Yes}}{}
						\prewide
						\nil
					\commawide
						\mexprxx{\datax{Cancel}}{\datax{\_}}
						\prewide
						\nil
					}
				}
			\end{aligned}}$\hfill\strut
		\end{minipage}}

		\subcaption{Recursive Two-Buyers}
		\label{fig:benchmark:2}
	\end{minipage}

	\caption{``Less Is More'' benchmark \citep[Fig. 4]{DBLP:journals/pacmpl/ScalasY19} \rev{-- spread over two pages}}
	\label{fig:benchmark}
	\Description{}
\end{figure}

\addtocounter{figure}{-1}

\begin{figure}[p]
	{\phantomsubcaption}
	{\phantomsubcaption}
	\begin{minipage}{\linewidth}
		\fbox{\begin{minipage}{\linewidth-7pt}
			\noindent \textbf{Roles:} Mapper ($\rolex{m}$), Worker 1 ($\rolex{w1}$),
			Worker 2 ($\rolex{w2}$), Reducer ($\rolex{r}$) \smallbreak

			\noindent \textbf{Protocol:} Mapper tells Worker 1 and Worker 2 to each
			process a datum. Worker 1 and Worker 2 tell Reducer the results of their
			processing. Reducer tells Master to enter another iteration of
			mapping/reducing or to stop. In the latter case, Mapper tells Worker 1 and
			Worker 2 to stop, too. \smallbreak

			\noindent \textbf{Global type:}
			\smallbreak\strut\hfill$\stypex{
				\begin{tree}
					\branchx{
						\mux{\datax{X}}
						\prewide
						\comxxxx{\rolex{m}}{[\rolex{w1}, \rolex{w2}]}{\datax{Datum}}{\datax{Int}}
						\prewide
						\comxxxx{[\rolex{w1}, \rolex{w2}]}{\rolex{r}}{\datax{Result}}{\datax{Int}}
						\prewide
						\comxx{\rolex{r}}{\rolex{m}} \holex{h1}
					}
					\branchxx[h1]{
						\mtypexx{\datax{Continue}}{\datax{Int}}
						\prewide
						\datax{X}
					}{
						\mtypexx{\datax{Stop}}{}
						\prewide
						\comxxxx{\rolex{m}}{[\rolex{w1}, \rolex{w2}]}{\datax{Stop}}{}
						\prewide
						\one
					}
				\end{tree}
			}$\hfill\strut\smallbreak

			We write ``$\comxxxxx{p}{[q_1, q_2]}{\ell}{t}{G}$'' and ``$\comxxxxx{[p_1,
			p_2]}{q}{\ell}{t}{G}$'' instead of
			``$\comxxxxx{p}{q_1}{\ell}{t}{\comxxxxx{p}{q_2}{\ell}{t}{G}}$'' and
			``$\comxxxxx{p_1}{q}{\ell}{t}{\comxxxxx{p_2}{q}{\ell}{t}{G}}$''. \smallbreak

			\noindent \textbf{Well-typed session:}
			\smallbreak\strut\hfill$\sexprx{\begin{aligned}
			&
				\atxx{\rolex{m}}{
					\recx{\datax{X}}
					\prewide
					\sndxxx{[\rolex{w1}, \rolex{w2}]}{\datax{Datum}}{\datax{123}}
					\prewide
					\rcvxx{\rolex{r}}{
						\mexprxx{\datax{Continue}}{\datax{z}}
						\prewide
						\datax{X}
					\commawide
						\mexprxx{\datax{Stop}}{\datax{\_}}
						\prewide
						\sndxxx{[\rolex{w1}, \rolex{w2}]}{\datax{Stop}}{}
						\prewide
						\nil
					}
				}
			\\{\mid{}}&
				\atxx{\rolex{w1}}{P_{\smash{\rolex[][]{w1}}}}
				\mid
				\atxx{\rolex{w2}}{P_{\smash{\rolex[][]{w2}}}}
				\mid
				\atxx{\rolex{r}}{
					\rcvxxx{\rolex{w1}}{\datax{Result}}{\datax{y1}}
					\prewide
					\rcvxxx{\rolex{w2}}{\datax{Result}}{\datax{y1}}
					\prewide
					\sndxxx{\rolex{m}}{\datax{Stop}}{}
					\prewide
					\nil
				}
			\end{aligned}}$\hfill\strut\smallbreak
			where:
			\smallbreak\strut\hfill$\sexprx{\begin{gathered}
				P_{\smash{\rolex[][]{w}i}} = \begin{tree}
					\branchx{
						\rcvx{\rolex{m}} \holex{h1}
					}
					\branchxx[h1]{
						\mexprxx{\datax{Datum}}{\datax{x}}
						\prewide
						\sndxxx{\rolex{r}}{\datax{Result}}{\datax{x}}
						\prewide
						\recx{\datax{X}}
						\prewide
						\rcvxx{\rolex{m}}{
							\mexprxx{\datax{Datum}}{\datax{x}}
							\prewide
							\sndxxx{\rolex{r}}{\datax{Result}}{\datax{x}}
							\prewide
							\datax{X}
						\commawide
							\mexprxx{\datax{Stop}}{\datax{\_}}
							\prewide
							\nil
						}
					}{
						\mexprxx{\datax{Stop}}{\datax{\_}}
						\prewide
						\nil
					}
				\end{tree}
			\end{gathered}}$\hfill\strut\smallbreak
			We write ``$\sndxxxx{[q_1, q_2]}{\ell}{e}{P}$'' instead of ``$\sndxxxx{q_1}{\ell}{e}{\sndxxxx{q_2}{\ell}{e}{P}}$''.
		\end{minipage}}

		\subcaption{Recursive Map/Reduce ($n\,{=}\,2$)}
		\label{fig:benchmark:3}
	\end{minipage}

	\medbreak

	\begin{minipage}{\linewidth}
		\fbox{\begin{minipage}{\linewidth-7pt}
			\noindent \textbf{Roles:} Starter ($\rolex{s}$), Workers A1, B1, C1
			($\rolex{wa1}$, $\rolex{wb1}$, $\rolex{wc1}$), Workers A2, B2, C2
			($\rolex{wa2}$, $\rolex{wb2}$, $\rolex{wc2}$) \smallbreak

			\noindent \textbf{Protocol:} Starter tells Worker A1 and Worker A2 to each
			process a datum. In parallel:
			\begin{itemize}
				\item Worker A1 tells Worker B1 to process the datum or to stop. In the
				former case, Worker B1 tells Worker C1 to process the datum, after which
				Worker C1 tells Worker A1 the result, after which Worker A1 again tells
				Worker B1 to process the datum or to stop. In the latter case, Worker B1
				tells Worker C1 to stop, too.

				\item Workers A2, B2, C2 follow the same sub-protocol as Workers A1, B1, C1,
				independently.
			\end{itemize}
			\smallbreak

			\noindent \textbf{Global type:}
			\smallbreak\strut\hfill$\stypex{
				\comxxxx{\rolex{s}}{\rolex{wa1}}{\datax{Datum}}{\datax{Int}}
				\prewide
				\comxxxx{\rolex{s}}{\rolex{wa2}}{\datax{Datum}}{\datax{Int}}
				\prewide
				(\parxx{G_1}{G_2})
			}$\hfill\strut\smallbreak
			where:
			\smallbreak\strut\hfill$\stypex{\begin{gathered}
				G_i = \begin{tree}
					\branchx{
						\mux{X}
						\prewide
						\comxx{\rolex{wa}\text{$i$}}{\rolex{wb}\text{$i$}} \holex{h1}
					}
					\branchxx[h1]{
						\mtypexx{\datax{Datum}}{\datax{Int}}
						\prewide
						\comxxxx{\rolex{wb}\text{$i$}}{\rolex{wc}\text{$i$}}{\datax{Datum}}{\datax{Int}}
						\prewide
						\comxxxx{\rolex{wc}\text{$i$}}{\rolex{wa}\text{$i$}}{\datax{Result}}{\datax{Int}}
						\prewide
						\datax{X}
					}{
						\mtypexx{\datax{Stop}}{}
						\prewide
						\comxxxx{\rolex{wb}\text{$i$}}{\rolex{wc}\text{$i$}}{\datax{Stop}}{}
						\prewide
						\one
					}
				\end{tree}
			\end{gathered}}$\hfill\strut\smallbreak

			\noindent \textbf{Well-typed session:}
			\smallbreak\strut\hfill$\sexprx{\begin{aligned}
			&
				\atxx{\rolex{s}}{
					\sndxxx{\rolex{wa1}}{\datax{Datum}}{\datax{123}}
					\prewide
					\sndxxx{\rolex{wa2}}{\datax{Datum}}{\datax{456}}
					\prewide
					\nil
				}
				\mid
				C_1
				\mid
				C_2
				\qquad
				C_i =
				\atxx{\rolex{wa}\text{$i$}}{P_{\smash{\rolex[][]{wa}i}}}
				\mid
				\atxx{\rolex{wb}\text{$i$}}{P_{\smash{\rolex[][]{wb}i}}}
				\mid
				\atxx{\rolex{wc}\text{$i$}}{P_{\smash{\rolex[][]{wc}i}}}
			\end{aligned}}$\hfill\strut\smallbreak
			where:
			\smallbreak\strut\hfill$\sexprx{\begin{aligned}
				P_{\smash{\rolex[][]{wa}i}} &=
					\rcvxxx{\rolex{s}}{\datax{Datum}}{\datax{x}}
					\prewide
					\sndxxx{\rolex{wb}\text{$i$}}{\datax{Stop}}{}
					\prewide
					\nil
			\\
				P_{\smash{\rolex[][]{wb}i}} &= \begin{tree}
					\branchx{
						\rcvx{\rolex{wa}\text{$i$}} \holex{h1}
					}
					\branchxx[h1]{
						\mexprxx{\datax{Datum}}{\datax{x}}
						\prewide
						\sndxxx{\rolex{wc}\text{$i$}}{\datax{Datum}}{\datax{x}}
						\prewide
						\recx{\datax{X}}
						\prewide
						\rcvxx{\rolex{wa}\text{$i$}}{
							\mexprxx{\datax{Datum}}{\datax{x}}
							\prewide
							\sndxxx{\rolex{wc}\text{$i$}}{\datax{Datum}}{\datax{x}}
							\prewide
							\datax{X}
						\commawide
							\mexprxx{\datax{Stop}}{\datax{\_}}
							\prewide
							P_{\smash{\rolex[][]{wb}i}}'
						}
					}{
						\mexprxx{\datax{Stop}}{\datax{\_}}
						\prewide
						P_{\smash{\rolex[][]{wb}i}}'
					\qquad
						P_{\smash{\rolex[][]{wb}i}}' =
							\sndxxx{\rolex{wc}\text{$i$}}{\datax{Stop}}{}
							\prewide
							\nil
					}
				\end{tree}
			\\
				P_{\smash{\rolex[][]{wc}i}} &= \begin{tree}
					\branchx{
						\rcvx{\rolex{wb}\text{$i$}} \holex{h1}
					}
					\branchxx[h1]{
						\mexprxx{\datax{Datum}}{\datax{x}}
						\prewide
						\sndxxx{\rolex{wc}\text{$i$}}{\datax{Result}}{\datax{x}}
						\prewide
						\recx{\datax{X}}
						\prewide
						\rcvxx{\rolex{wb}\text{$i$}}{
							\mexprxx{\datax{Datum}}{\datax{x}}
							\prewide
							\sndxxx{\rolex{wc}\text{$i$}}{\datax{Result}}{\datax{x}}
							\prewide
							\datax{X}
						\commawide
							\mexprxx{\datax{Stop}}{\datax{\_}}
							\prewide
							\nil
						}
					}{
						\mexprxx{\datax{Stop}}{\datax{\_}}
						\prewide
						\nil
					}
				\end{tree}
			\end{aligned}}$\hfill\strut
		\end{minipage}}

		\subcaption{Independent Multiparty Workers ($n\,{=}\,2$)}
		\label{fig:benchmark:4}
	\end{minipage}

	\caption{``Less Is More'' benchmark \citep[Fig. 4]{DBLP:journals/pacmpl/ScalasY19} \rev{-- spread over two pages}}
	\label{fig:benchmark-more}
	\Description{}
\end{figure}

\Cref{fig:benchmark} (spread over two pages) defines, for each example
protocol in the ``Less Is More'' benchmark, a global type and a well-typed
session. There are two kinds of example protocols:
\begin{itemize}
	\item \textit{OAuth2 Fragment} (\Cref{fig:benchmark:1}), \textit{Recursive
	Map/Reduce} (\Cref{fig:benchmark:3}), and \textit{Independent Multiparty
	Workers} (\Cref{fig:benchmark:4}) are protocols that can be specified by a
	projectable global type, but the resulting family of local types is
	inconsistent.

	\item \textit{Recursive Two-Buyers} (\Cref{fig:benchmark:2}) is a protocol that
	can be specified by a global type, but it is not projectable (neither using
	plain projection, nor using full projection). Thus, this is the first time that the
	safety and liveness of implementations of Recursive Two-Buyers can be proved
	using a global type.
\end{itemize}


\section{The General Case: Typing with LTSs}\label{sec:behaviour}


As demonstrated in \Cref{sect:overv:vscode}, the synthetic approach can be generalised---beyond global types---%
to define the typing of sessions even
without discussing the syntax of the types themselves. After all, an
important observation from the typing rules of
\Cref{fig:typing-proc} is that \emph{no rule relies on the
  syntactic structure of global types}. This is the essence of
our synthetic approach to MPST. Two questions arise naturally from this
observation:
\begin{enumerate}
\item Can we consider that the rules in \Cref{fig:typing-proc}
  refer to a generic, semantic notion of behaviour that does
  not depend on a particular syntactic structure?
  \item What are the properties that are required so
    these semantic objects still allow our type system to guarantee safety and liveness?
\end{enumerate}

\noindent
Regarding the first question,
as shown in the previous sections, we see an LTS as a classifier for a
session. This LTS must model the
communications among all processes that participate in the session.
Regarding the second question,
\Cref{sec:behaviours} defines a \emph{well-behaved multiparty LTS} as
an LTS that exhibits the shape and properties required
for well-typedness to imply safety and liveness
in our type system.

Naturally, global types from MPST presentations in the literature can
be seen as syntactic objects that support all the requirements of well-behaved
multiparty LTSs. \Cref{sec:prop-behaviours} describes how global types
in \Cref{sec:global}
(following \citep{Yoshida:2020}) intrinsically constitute
well-behaved MLTSs in synthetic MPST.

\subsection{Well-Behaved Multiparty LTSs (WB-MLTS)}\label{sec:behaviours}

Well-behaved multiparty LTSs consist of transitions of the form
$\gstep B {\alpha} {B'}$ that satisfy the set of properties
below. First, we introduce the definition of MLTSs.

\begin{definition}[Multiparty Labelled Transition System (MLTS)]
  An MLTS is an LTS
  $(\gt{\mathcal{B}}, \stypex{A}, \xrightarrow{\phantom{\alpha}})$,
  with a set of states $\gt{B},
  \ldots \in \gt{\mathcal{B}}$, and global action labels
  $\stypex{\alpha} \in \stypex{A}$ of the form $\comxxxx{\Rp}{\Rq}{\lbl}{t}$, with
  $\Rp \neq \Rq$.
\end{definition}

The typing judgement and typing rules of \Cref{fig:typing-proc} are then
parameterised by MLTSs:
$\poft \Gamma \Delta \vargu P B \Rr$.
However, simply type-checking against an arbitrary MLTS does not guarantee progress
and preservation. To provide these stronger guarantees, we need to restrict to
MLTSs that satisfy a set of \emph{well-behavedness conditions}, that  specify the
criteria that transitions of MLTSs must satisfy. This relies on the notion of
receiver disjointness.

\begin{definition}[Receiver Disjointness]
  Two global actions $\stypex{\alpha_1} = \comxxxx{\Rp}{\Rq}{\lbl_1}{t_1}$ and
  $\stypex{\alpha_2} = \comxxxx{\Rr}{\Rs}{\lbl_2}{t_2}$ are receiver-disjoint,
  $\stypex{\alpha_1} \indep \stypex{\alpha_2}$, iff
  $\Rq \notin \setx{\Rr, \Rs}$ and $\Rs \notin \setx{\Rp, \Rq}$.
\end{definition}

\noindent Intuitively, if two global actions are receiver-disjoint, then they
should be able to be reordered. The idea is that, in a concurrent system,
receivers are by definition independent from each other, so the order in which a
sender sends messages to them does not matter. We are now ready to define
well-behaved multiparty LTSs.

\begin{definition}[Well-Behaved Multiparty LTS (WB-MLTS)]\label{def:mb}
  A WB-MLTS is an MLTS
  $(\gt{\mathcal{B}}, \dlbl{A}, \xrightarrow{\phantom{\alpha}})$ that satisfies
  \textbf{all} of the following conditions for any state $\stypex{B}$:

  \begin{enumerate}
  \item \textbf{Sender determinacy:} For all $\gstep B {\alpha_1}{B_1}$ and
  $\gstep B {\alpha_2}{B_2}$, then either
  $\stypex{\alpha_1} \indep \stypex{\alpha_2}$, or there exists
  $\Rp$, $\Rq$, $\dlbl{\lbl_1}$, $\dlbl{\lbl_2}$, $\dtypex{t_1}$, and $\dtypex{t_2}$ such that
  $\stypex{\alpha_1} = \comxxxx{\Rp}{\Rq}{\lbl_1}{t_1}$ and
  $\stypex{\alpha_2} = \comxxxx{\Rp}{\Rq}{\lbl_2}{t_2}$.

  \item \textbf{Determinism:} For all $\gstep B {\alpha}{B_1}$ and
  $\gstep B {\alpha}{B_2}$, then $\gt{B_1} = \gt{B_2}$.

  \item \textbf{Conditional commutativity:} For all $\gstep B
  {\comxxxx{\Rr}{\Rs}{\lbl_1}{t_1}}{B_1} \gstep {} {\comxxxx{\Rp}{\Rq}{\lbl_2}{t_2}}{B'}$, if there exist $\lnamex{\ell}$ and $\dtypex{t}$ such that $\gstep B {\comxxxx{\Rp}{\Rq}{\lbl}{t}}{}$ and
  $\setx{\Rp, \Rq} \cap \setx{\Rr, \Rs} = \emptyset$, then there exists a
  $\gt{B_2}$ such that $\gstep B {\comxxxx{\Rp}{\Rq}{\lbl_2}{t_2}}{B_2} \gstep {} {\comxxxx{\Rr}{\Rs}{\lbl_1}{t_1}} {B'}$.

  \item \textbf{Diamond (confluence for reorderable global actions):} For all
  $\gstep {B}{\alpha_1}{B_1}$, and $\gstep {B}{\alpha_2}{B_2}$,
  if $\stypex{\alpha_1} \indep \stypex{\alpha_2}$,
  then there exists a $\gt{B'}$ such that $\gstep {B_1}{\alpha_2}{B'}$
  and $\gstep {B_2}{\alpha_1}{B'}$.
  \end{enumerate}
\end{definition}

Intuitively, sender determinacy states that, if two global actions are possible
in a state, then these actions cannot have different senders but the same
receiver. That is, the sender is fixed.
Conditional commutativity states that an alternative in a choice cannot become available
for two roles $\Rp$ and $\Rq$ after unrelated communications. In other
words, for a global action $\comxxxx{\Rp}{\Rq}{\lbl_2}{t_2}$ to become available,
either $\Rp$ or $\Rq$ must have received a message enabling this choice. \rev{This
condition ensures that well-formed MLTSs
do not specify ``bad'' protocols in which actions at one process can enable
actions at another process without any interaction between those two processes.
After all, in the absence of covert communication between them, it is impossible to implement
such processes. Thus, such protocols are ruled out by conditional commutativity.}

From WB-MLTS's well-behavedness criteria, we prove a series of lemmas that are then
used to establish the standard \textbf{progress} and \textbf{preservation}
properties. The most important of these lemmas are:
(1) if we have two processes $\proc{P_{\sf p}}$ and $\proc{P_{\sf q}}$ well-typed with regards to $\gt{B}$ as
roles $\Rp$ and $\Rq$, and $\proc{P_{\sf p}}$ is ready to send $\lnamex{\lbl_j}$ to
$\Rq$, and $\proc{P_{\sf q}}$ is ready to receive from $\Rp$, then the state $\gt{B}$
must accept global action $\comxxxx{\Rp}{\Rq}{\lbl_j}{t_j}$; (2) the continuations
of the output/input processes are still well typed; and (3)
a well-typed process is still well-typed after an unrelated global action.

\begin{agdaenv}{lemma}{}{https://github.com/SyntheticBehaviouralTypes/srmpst-formalisation/blob/f4b15c0074bc496a31a44c913b92863a46a32ee7/Safety.agda\#L98-L110}\label[lemma]{lemma:wt-comm}
  If $\poftemptyenvs \vargu {\sndxxxx{q}{\ell_j}{e}{P}} B \Rp$,
  and $\poftemptyenvs \vargu {\rcvxxxxx{p}{\ell_i}{x_i}{P_i}{i \in I}} B \Rq$,
  then there exists a $\gt{B'}$ such that $\gstep B {\comxxxx{\Rp}{\Rq}{\lbl_j}{t_j}}{B'}$.
\end{agdaenv}

\begin{agdaenv}{lemma}{Inversions of \protect\autorefrule{$\vdash$-Send} and \protect\autorefrule{$\vdash$-Recv}}{https://github.com/SyntheticBehaviouralTypes/srmpst-formalisation/blob/f4b15c0074bc496a31a44c913b92863a46a32ee7/Safety.agda\#L136-L150}\label[lemma]{lemma:send-recv-inv}
  Suppose a state $\gt{B}$ such that $\gstep B {\comxxxx{\Rp}{\Rq}{\lbl_j}{t_j}}{B'}$:
  \begin{enumerate}
    \item If $\poftemptyenvs \vargu {\sndxxxx{q}{\ell_j}{e}{P}} B \Rp$,
      then  $\poftemptyenvs \vargu {P} {B'} \Rp$
    \item If $\poftemptyenvs \vargu {\rcvxxxxx{p}{\ell_i}{x_i}{P_i}{i \in I}} B \Rq$,
      then $\poft {\denvxxx{\varnothing}{x_j}{t_j}} \varnothing \vargu {P_j} {B'} \Rq$.
  \end{enumerate}
\end{agdaenv}

\begin{agdaenv}{lemma}{}{https://github.com/SyntheticBehaviouralTypes/srmpst-formalisation/blob/f4b15c0074bc496a31a44c913b92863a46a32ee7/Definitions.agda\#L194-L224}\label[lemma]{lemma:unrelated-step}
  Suppose a state $\gt{B}$, a role $\Rr$, and a global action $\dlbl{\alpha}$ such
  that $\Rr$ does not occur in $\dlbl{\alpha}$. If $\gstep B {\alpha} {B'}$, and
$\poft \varGamma \varDelta \vargu P B \Rr$, then $\poft \varGamma \varDelta \vargu P {B'} \Rr$
\end{agdaenv}

\noindent We now state the main theorems.

\startrev
\begin{agdaenv}{theorem}{Progress}{https://github.com/SyntheticBehaviouralTypes/srmpst-formalisation/blob/f4b15c0074bc496a31a44c913b92863a46a32ee7/Safety.agda\#L632-L663}\label[theorem]{thm:progress}%
  Suppose a state $\stypex{B}$ of a WB-MLTS. If $\stypedxx{C}{B}$, then:
  \textup{(1)} not $\setransxxx{C}{\uptau}{} \!\!\! \setransxxx{}{\uptau}{}
  {\cdots}$; \textup{(2)} $\setransxxx{C}{\uptau}{} {\cdots}
  \setransxxx{}{\uptau}{\compxxx{\nil}{{\cdots}}{\nil}}$, or
  $\setransxxx{C}{\uptau}{} {\cdots} \setransxxx{}{\uptau}{} \!\!\!
  \setransxxx{}{\alpha}{C'}$, for some $\sexprx{C'}$.
\end{agdaenv}

\begin{agdaenv}{theorem}{Preservation}{https://github.com/SyntheticBehaviouralTypes/srmpst-formalisation/blob/f4b15c0074bc496a31a44c913b92863a46a32ee7/Safety.agda\#L182-L210}\label[theorem]{thm:preservation}
	Suppose a state $\stypex{B}$ of a WB-MLTS and $\stypedxxxx{}{}{C}{B}$:
	\begin{itemize}
		\item If $\setransxxx{C}{\alpha}{C'}$, then $\stypedxx{C'}{B'}$ and
	  $\sttransxxx{B}{\alpha}{B'}$, for some $\stypex{B'}$ (i.e., $\stypex{B'}$ is
	  also a state of the same WB-MLTS).

		\item If $\setransxxx{C}{\uptau}{C'}$, then $\stypedxx{C'}{B}$.
	\end{itemize}
\end{agdaenv}
\finishrev

\rev{
Finally, we note that as long as the MLTSs have finitely many states, the type
system is decidable.
First, all the typing rules in \Cref{fig:typing-proc} are structural except
\autorefrule{$\vdash$-Skip} and, as long as the MLTS has finitely many states,
all of their premises are decidable, and the domain of any universal
quantification is finite.
In rule \autorefrule{$\vdash$-Skip}, the size of the process in the premises
does not grow; \autorefrule{$\vdash$-Skip}'s premises (1), (2), and (4) are
also decidable for any finite-state MLTS; and the universal quantification
of (3) is also finite.
Note, also, that \autorefrule{$\vdash$-Skip} cannot be applied twice in a row:
premise (1) becomes false after one use of this rule, after which, one
structural rule must be used.
Therefore, type-checking must terminate, and our VS Code extension is
implemented following this approach.
Details about algorithmic and performance aspects are covered in
\ifdefined\WITHAPPENDIX\Cref{sect:behav:decid}\else{the appendix \cite{techreport}}\fi.
}

\subsection{Global Types as Multiparty Behaviours}\label{sec:prop-behaviours}

\rev{%
We prove that the
global types of~\Cref{sec:global-types}  satisfy all of the conditions of
\Cref{def:mb}, as a consequence of the operational semantics of global types.
\Cref{cor:full-gt} below is then a consequence of
\Cref{thm:preservation,thm:progress,thm:prop-behaviours}.}

\begin{agdaenv}{theorem}{}{https://github.com/SyntheticBehaviouralTypes/srmpst-formalisation/blob/f4b15c0074bc496a31a44c913b92863a46a32ee7/Definitions/GlobalTypesWPar.agda\#L1911}\label[theorem]{thm:prop-behaviours}
  \strut \rev{Any global type $\gt {G}$ satisfies the well-behavedness conditions of \Cref{def:mb}.}



\end{agdaenv}
%

\begin{agdaenv}{corollary}{Progress and Preservation of Well-typed Sessions with Global Types}{https://github.com/SyntheticBehaviouralTypes/srmpst-formalisation/blob/f4b15c0074bc496a31a44c913b92863a46a32ee7/FullGT.agda}\label[corollary]{cor:full-gt}
  A well-typed session with a grammatical global type satisfies
  progress and preservation.
\end{agdaenv}


\section{Formalisation of Synthetic MPST in Agda} 
\label{sec:formalisation}

A big advantage of the synthetic approach to MPST is that it leads to a simpler
formalisation in proof assistants. We justify this claim by presenting a
formalisation of the type system in Sections~\ref{sec:global}
and~\ref{sec:behaviour}, and a comparison with respect to similar formalisations
of MPST in the literature. While our formalisation was done in Agda, it
should be straightforward to port it to other proof assistants.

\subsection{An Agda Encoding of the Synthetic Approach}

The core part of our formalisation is the encoding of well-behaved MLTSs
(\Cref{def:mb}). In Agda, we encode them in terms of records parameterised by
the number of participants in the protocol: (1) record \verb|BTheory| encodes
MLTSs, and (2) record \verb|BT-Prop| encodes the well-behavedness properties.
Their encoding in Agda is straightforward, in that it relies on a direct
encoding of an LTS as a relation between two states and an action. The
remaining definitions are also encoded as expected, and they can be encoded
in a similar fashion in other proof assistants.

The encoding of the type system is done within a module that is parameterised by
well-behaved MLTSs, i.e., an Agda module that takes as a parameter a record of
type \verb|B : BTheory|, and a record that proves that it is well-behaved
\verb|BP : BT-Prop B|. The type system itself is defined as a relation between
process names, process terms, and well-behaved MLTSs.

The full proofs of progress and preservation are done for arbitrary
well-behaved MLTSs in about 650 LOC of Agda code.
In general, the key lemmas are also a direct encoding of the ones presented in
\Cref{sec:behaviour}. The most difficult proof in our system is showing that any
global type has a well-behaved MLTS (roughly 2000 LOC), and even in this case, the
majority of the proof is about dealing with binders, renaming, etc., which is
tedious but not intellectually complicated.

In our experience, with the
synthetic approach to MPST, there is no need to massage the definitions to make the
proofs simpler/more natural, unlike when mechanising the classical approach to MPST.
We further elaborate this point in the next subsection.

\subsection{Comparison with the Mechanisation of Classical MPST}\label{sec:formalisation:comparison}

The biggest difficulty in mechanising classical MPST is dealing with the
\emph{projection} operator and \emph{recursion}. Specifically, the hardest part
is showing that if a (possibly) recursive global type is projectable, then the
corresponding family \textit{of local types} is indeed safe and live. Once this is
proved, though, it tends to be straightforward to show that type-checking
against a safe and live family of local types entails safety and liveness of
the well-typed family of processes.

In contrast, the synthetic approach avoids dealing with projection altogether,
but we need to deal with a more complex typing relation, where the most complex
rule is \autorefrule{$\vdash$-Skip}. The question that we address in this
section is: why is it the case that dealing with \autorefrule{$\vdash$-Skip}
leads to much simpler proofs than showing that projectability leads to safety
and liveness of local types? We will review the most representative
mechanisations to illustrate this. There are two main approaches to discuss.

\subsubsection*{\bf Zooid and related approaches.}
Zooid~\cite{DBLP:conf/pldi/Castro-Perez0GY21} relies on a notion of coinductive
projection and unrolling of global/local types to guarantee the correspondence
between a global type, and the projection of all of the roles in the global
type. This is similar to other work that subsumes the proofs in Zooid, e.g., by
Tirore et al.~\cite{DBLP:conf/ecoop/TiroreBC25}. There are several challenges in
using a coinductive notion of projection.
Firstly, the use of coinductive relations often results in cumbersome proofs
within proof assistants.
Second, deciding a coinductive
projection relation is not straightforward. The most common approach in the
literature is to use a more restrictive syntactic projection function, and then
show that this syntactic projection is contained within the coinductive
projection. Note, however, that the more complex syntactic projection is, the
harder it is to mechanise.

For example, \citet{DBLP:conf/pldi/Castro-Perez0GY21}
only formalise syntactic projection that uses \emph{plain merge}. This
restriction rules out all of the examples in this paper, as well a the majority
of the examples in our Agda formalisation. Similarly, \citet{DBLP:conf/itp/TiroreBC23}
mechanise in Rocq a \emph{sound and complete} projection, that handles $\mu$-binders
correctly, so that their syntactic projection exactly corresponds to a notion
of coinductive projection. However, it uses \emph{plain merge} as well, as does the authors' follow-up work~\cite{DBLP:conf/ecoop/TiroreBC25}. This further illustrates the difficulty of dealing with syntactic projection, while it also indicates the
complexities of supporting the more expressive notion of \emph{full merge}.
To our knowledge, full merge has never been mechanised yet.

\rev{%
\Citet{DBLP:conf/cav/LiSWZ23} showed that a \emph{complete} projection relation
-- where every implementable global type is projectable -- can be obtained via
automata-theoretic methods, with their notion of implementability later
formalised in Rocq~\cite{DBLP:conf/itp/LiW25}. Their framework separates
synthesis from implementability checking, the latter decided by a set of
\emph{Coherence Conditions}. In essence, their synthesis and implementability
together correspond to (a more expressive form of) the classical projection.
Thus, their approach does not avoid the challenges associated with defining and
reasoning about a projection relation. An interesting open problem is clarifying
the precise relationship between Li et~al.’s Coherence Conditions and our
notions of Well-Behavedness and type checking: this could lead to a unified and
more expressive projection-less approach.
}

\subsubsection*{\bf Multiparty GV}
MPGV~\cite{DBLP:journals/pacmpl/JacobsBK22a} allows for multiple sessions, and
the mechanisation builds on top of a complex mechanisation of
\textit{connectivity graphs}~\cite{DBLP:journals/pacmpl/JacobsBK22} to deal with
session interleaving, and guarantee that a series of invariants are preserved.
Global types in MPGV are restricted to plain merge,
and still rely on a notion of coinductive projection, which causes the same
difficulties as the Zooid approach.
\rev{MPGV also offers a coinductive, global-type-free (i.e.\ bottom-up)
formulation of consistency, and the authors prove that projectability implies
consistency. It provides a compositional and mechanised framework supporting
multiple interleaved sessions. In contrast, our work develops a top-down form
of compositionality, where processes are type-checked directly against richer
global specifications within a single-session setting.}

\bigbreak
The synthetic approach completely abstracts away the syntax used to encode
recursive protocols, and avoids completely the need to deal with folding\slash
unfolding, and projection.  Instead, MLTSs can have cycles, but the presence or
absence of such cycles does not complicate the formalisation.

One might expect that the ability of
\autorefrule{$\vdash$-Skip} to postpone type checking in the synthetic approach
would increase proof complexity. The main reason this increase does not arise
lies in the conditions under which a state is considered ``skippable.'' First,
all conditions of \autorefrule{$\vdash$-Skip} must hold in any ``near-future''
state reachable without involving the role currently being type-checked. This
guarantees that \autorefrule{$\vdash$-Skip} can be reapplied in each such state,
thereby simplifying the proof of \Cref{lemma:unrelated-step}. Another potential source
of complexity is the need to reason about permutations of actions; however, this
does not arise in our formalisation. The rule \autorefrule{$\vdash$-Skip} cannot
be applied to two processes that are, respectively, ready to send and to receive
from each other. In this sense, its conditions impose a form of determinism on
the application of typing rules, effectively eliminating many potentially
cumbersome proof cases. This simplification is illustrated in the proof of
\Cref{lemma:wt-comm}.

Thus, with the synthetic approach, \textit{the absence} of the need to reason
about the projectability of (possibly) recursive global types \textit{does} make
the formalisation significantly simpler, while \textit{the presence} of the need
to reason about rule \autorefrule{$\vdash$-Skip} \textit{does not} make it
significantly more complex.

\section{Prototype Language and Tooling of Synthetic MPST in VS Code} \label{sec:proto}

As demonstrated in \Cref{sect:overv:vscode}, we developed a prototype language
and tooling of the synthetic approach to MPST as an extension of \textit{VS
Code}, including a dedicated \textit{LSP server}.

To develop this prototype, we use the \textit{Rascal} meta-programming
language~\cite{DBLP:conf/scam/KlintSV09}. Among other features, Rascal has core
support to write context-free grammars (for defining concrete syntax), algebraic
data types (for defining abstract syntax), and advanced pattern matching on
grammar rules and ADT constructors. Together with standard programming
abstractions, these features aim to simplify the implementation of parsers, type
checkers, interpreters, and code generators.

Leveraging Rascal, the implementation of the
type-checking algorithm is done in about 200 LOC, and it relies on a graph
representation of protocols. The key insight is that, as long as the
protocol can be represented as a finite-state MLTS, then the conditions for
our rules are decidable.


\section{Additional Related Work}\label{sect:relw}

\newcommand\labelenum[1]{%
  \textcolor{Gray}{\sffamily\bfseries\upshape\mathversion{bold}#1}}
\newcommand{\nosep}{%
  \partopsep=0pt%
  \topsep=0pt%
  \itemsep=0pt%
  \parsep=0pt%
}

In addition to the related work discussed in
\Cref{sect:intr:sota,sec:formalisation:comparison}, the following contributions
in the literature are relevant to this paper, too. In particular,
starting from the introduction of MPST~\cite{DBLP:conf/popl/HondaYC08}
there is a substantial lineage of papers that seek to improve the expressiveness of
the MPST method. Below, we focus on two main aspects: first, using
the synthetic approach to behavioural typing~\cite{DBLP:conf/ecoop/JongmansF23} to simplify
MPST theory by removing projection and merge. And second, enabling the
use of more powerful classifiers (types) for sessions (i.e., WB-MLTSs)
to be able to type more protocols.

Using the operational semantics of types is the key ingredient of the
synthetic approach. This was first studied in the context of
\emph{multiparty compatibility} (MC) \cite{DBLP:conf/icalp/DenielouY13} and
extensions
\cite{DBLP:conf/concur/BocchiLY15,DBLP:conf/popl/LangeTY15,DBLP:conf/cav/LangeY19}.
The idea is to interpret local types as \emph{communicating finite
  state machines} (CFSM) \cite{DBLP:journals/jacm/BrandZ83}.
Multiparty compatibility, then, is a predicate on the joint state
space of the CFSMs to ensure safety and liveness.
As such, MC is a \textit{bottom-up technique} (from local view to a global view), whereas the synthetic approach in this paper is a \textit{top-down technique}.


A different, but related, technique is the notion of well-behaved
local types as studied by~\citet{DBLP:conf/ecoop/JongmansF23}, of
which a rudimentary version (without processes and type checking) was
studied by Jongmans and Yoshida \cite{DBLP:conf/esop/JongmansY20}. The
key difference between their work and ours is that they rely on local
types against which processes are type-checked. In contrast, in this
paper, we type processes directly against global types. As a result,
for the first time in the MPST literature, we obtain a notion of type
inhabitation that is independent of auxiliary concepts such as
projectability and/or well-formedness. In particular, global types
that specify \textit{unrealisable protocols} are uninhabited.

Another version of well-behaved global types was studied by Gheri et al.
\cite{DBLP:conf/ecoop/GheriLSTY22}, in the context of choreography
automata \cite{DBLP:conf/coordination/BarbaneraLT20}, but it is
limited to projection (no type checking).


Our approach avoids issues with merge by avoiding projection
altogether. However, there are several non-traditional techniques for
projection in the MPST literature. Lopez et al.
\cite{DBLP:conf/oopsla/LopezMMNSVY15} capture projection in a
decidable type equivalence. Castellani et al.
\cite{DBLP:journals/corr/abs-2203-12876} and Hamers et al.
\cite{DBLP:conf/tacas/HamersJ20} do not use projection at all, but
type-check families of processes against global types
(non-compositional).


\rev{\citet{DBLP:journals/pacmpl/ScalasY19},
\citet{DBLP:conf/lics/GlabbeekHH21}, and \citet{10.1145/3661814.3662085}} presented examples of safe and
live families of processes that are unsupported by the MPST method due
to limited expressiveness of global types. Scalas and Yoshida address
the issue by developing an MPST theory without global types (only
local types), while van Glabbeek et al. address the issue by
developing improved merging. In contrast, in this
paper, we address the issue by
proposing an expressive type system to verify protocol implementations purely against global
specifications of behaviour (i.e., global types and WB-MLTSs). We believe that having a singular global
specification has intrinsic value as a programming artefact that
comprehensively defines protocols from a system-wide perspective.
\rev{Finally, Peters and Yoshida study
  the expressivity of a session calculus typable by a collection of
  local types with mixed choice, that we do not address directly in this paper.
  To support this, the main challenge is to relax the sender determinacy condition
  without breaking soundness.}

\section{Conclusion and Future Work}\label{sect:concl}

\subsubsection*{Summary}
We have presented the synthetic approach to MPST. The main theoretical
result is that well-typedness implies safety and liveness. The main practical
result is that our type system is expressive enough to pass
the ``Less Is More'' benchmark compositionally (i.e., we support at least
all
the challenging examples of \citet{DBLP:journals/pacmpl/ScalasY19}). This has been an open problem for several years. Our complete
formalisation in Agda, together with examples, demonstrates that the synthetic approach leads
to simpler formalisations in proof assistants. Furthermore, a key practical
advantage of the synthetic approach is its ability to extend the expressiveness of
MPST by purely relying on global protocol specifications: well-behaved multiparty LTSs in general, and global types as a special case. That is, we
showed that a simple form of classical MPST satisfies the necessary
well-behavedness conditions to ensure safety and liveness within the synthetic
approach.

\startrev
\subsubsection*{Discussion}

For simplicity, our approach uses a synchronous communication semantics. The main
complication with asynchronous communication semantics is that multiparty LTSs may no
longer be finite, which may affect decidability. We believe that a careful
application of \textit{run-time global types} in the Zooid approach
\cite{DBLP:conf/pldi/Castro-Perez0GY21} might be adapted to a synthetic setting
to address this issue.

In principle, a top-down approach like ours inhabits all process systems
provable with a bottom-up approach like ``Less Is More''. The compared
expressivity is difficult to establish at a theoretical level, though. For
example, ``Less Is More'' typing contexts and their reduction semantics can be
used as multiparty LTSs. The issue is to determine if well-behavedness together
with type inhabitation is equivalent to consistency property $\varphi$ in the
``Less Is More'' approach. Since well-behavedness is weaker than the typing
context properties in that approach, our intuition is that both approaches are
equally expressive.

Sharing a global view among all processes may not always be acceptable. For instance, in a ring
protocol, it may be desirable to hide the size of the ring from each of the
processes. To address this, we are currently studying ways to avoid exploring
unrelated transitions, by relying on strong confluence. This may allow a form of
lightweight projection -- only used for type checking -- where we remove
irrelevant transitions from the LTS to hide information. The lightweight projection may also enable
potential optimisations to the type checking algorithm.

\finishrev

\subsubsection*{Future work}

These results open the door to several promising extensions. One direction is to
identify the largest class of syntactic protocol descriptions that satisfy our
well-behavedness criteria, potentially removing the need to prove
well-behavedness when implementing protocols. Another is to generalise these
criteria further, increasing the expressiveness of our type system \rev{(e.g studying
global types with mixed choice in the style of \cite{DBLP:conf/ecoop/JongmansF23} or \cite{10.1145/3661814.3662085}.)} Finally, we
aim to extend our multiparty LTSs with verification conditions that allow properties
typically established via model checking to be verified directly through type
checking.


\bibliographystyle{ACM-Reference-Format}
\bibliography{references}

@inproceedings{10.1145/3661814.3662085,
author = {Peters, Kirstin and Yoshida, Nobuko},
title = {Separation and Encodability in Mixed Choice Multiparty Sessions},
year = {2024},
isbn = {9798400706608},
publisher = {Association for Computing Machinery},
address = {New York, NY, USA},
url = {https://doi.org/10.1145/3661814.3662085},
doi = {10.1145/3661814.3662085},
booktitle = {Proceedings of the 39th Annual ACM/IEEE Symposium on Logic in Computer Science},
articleno = {62},
numpages = {15},
location = {Tallinn, Estonia},
series = {LICS '24}
}

@inproceedings{DBLP:conf/podc/Milner82,
  author       = {Robin Milner},
  title        = {Four Combinators for Concurrency},
  booktitle    = {{PODC}},
  pages        = {104--110},
  publisher    = {{ACM}},
  year         = {1982}
}

@article{DBLP:journals/tocl/Glabbeek24,
  author       = {Rob van Glabbeek},
  title        = {Comparing the Expressiveness of the {\(\pi\)}-calculus and {CCS}},
  journal      = {{ACM} Trans. Comput. Log.},
  volume       = {25},
  number       = {1},
  pages        = {1:1--1:58},
  year         = {2024}
}

@InProceedings{Yoshida:2020,
author="Yoshida, Nobuko
and Gheri, Lorenzo",
editor="Hung, Dang Van
and D{\textasciiacute}Souza, Meenakshi",
title="A Very Gentle Introduction to Multiparty Session Types",
booktitle="Distributed Computing and Internet Technology",
year="2020",
publisher="Springer International Publishing",
address="Cham",
pages="73--93",
isbn="978-3-030-36987-3"
}

@misc{techreport,
	author = "David {Castro-Perez} and Francisco Ferreira and Sung-Shik Jongmans",
	title = {A Synthetic Reconstruction of Multiparty Session Types (with Appendix)},
	year = {2025},
  note = {To be published on Arxiv}
}

@inproceedings{DBLP:conf/esop/HondaVK98,
  author       = {Kohei Honda and
                  Vasco Thudichum Vasconcelos and
                  Makoto Kubo},
  title        = {Language Primitives and Type Discipline for Structured Communication-Based
                  Programming},
  booktitle    = {{ESOP}},
  series       = {Lecture Notes in Computer Science},
  volume       = {1381},
  pages        = {122--138},
  publisher    = {Springer},
  year         = {1998}
}

@inproceedings{DBLP:conf/concur/Honda93,
  author       = {Kohei Honda},
  title        = {Types for Dyadic Interaction},
  booktitle    = {{CONCUR}},
  series       = {Lecture Notes in Computer Science},
  volume       = {715},
  pages        = {509--523},
  publisher    = {Springer},
  year         = {1993}
}

@article{DBLP:journals/pacmpl/UdomsrirungruangY25,
  author       = {Thien Udomsrirungruang and
                  Nobuko Yoshida},
  title        = {Top-Down or Bottom-Up? Complexity Analyses of Synchronous Multiparty
                  Session Types},
  journal      = {Proc. {ACM} Program. Lang.},
  volume       = {9},
  number       = {{POPL}},
  pages        = {1040--1071},
  year         = {2025}
}

@inproceedings{DBLP:conf/ecoop/JongmansF23,
  author       = {Sung{-}Shik Jongmans and
                  Francisco Ferreira},
  title        = {Synthetic Behavioural Typing: Sound, Regular Multiparty Sessions via
                  Implicit Local Types},
  booktitle    = {{ECOOP}},
  series       = {LIPIcs},
  volume       = {263},
  pages        = {42:1--42:30},
  publisher    = {Schloss Dagstuhl - Leibniz-Zentrum f{\"{u}}r Informatik},
  year         = {2023}
}

@book{Hintikka1973-HINLLA,
	author = {Jaakko Hintikka},
	title = {Logic, Language-Games and Information: Kantian Themes in the Philosophy of Logic},
	publisher = {Oxford, England: Oxford, Clarendon Press},
	year = {1973}
}

@article{VanBenthem1974-VANHOA-3,
	title = {Hintikka on Analyticity},
	year = {1974},
	journal = {Journal of Philosophical Logic},
	doi = {10.1007/bf00257484},
	author = {J. F. A. K. Van Benthem},
	pages = {419--431},
	publisher = {Springer},
	volume = {3},
	number = {4}
}

@inproceedings{DBLP:conf/coordination/BarbaneraLT20,
  author    = {Franco Barbanera and
               Ivan Lanese and
               Emilio Tuosto},
  title     = {Choreography Automata},
  booktitle = {{COORDINATION}},
  series    = {LNCS},
  volume    = {12134},
  pages     = {86--106},
  publisher = {Springer},
  year      = {2020}
}

@inproceedings{DBLP:conf/ecoop/GheriLSTY22,
  author    = {Lorenzo Gheri and
               Ivan Lanese and
               Neil Sayers and
               Emilio Tuosto and
               Nobuko Yoshida},
  title     = {Design-By-Contract for Flexible Multiparty Session Protocols},
  booktitle = {{ECOOP}},
  series    = {LIPIcs},
  volume    = {222},
  pages     = {8:1--8:28},
  publisher = {Schloss Dagstuhl - Leibniz-Zentrum f{\"{u}}r Informatik},
  year      = {2022}
}

@inproceedings{DBLP:conf/lics/GlabbeekHH21,
  author    = {Rob van Glabbeek and
               Peter H{\"{o}}fner and
               Ross Horne},
  title     = {Assuming Just Enough Fairness to make Session Types Complete for Lock-freedom},
  booktitle = {{LICS}},
  pages     = {1--13},
  publisher = {{IEEE}},
  year      = {2021}
}

@inproceedings{DBLP:conf/tacas/HamersJ20,
  author    = {Ruben Hamers and
               Sung{-}Shik Jongmans},
  title     = {Discourje: Runtime Verification of Communication Protocols in Clojure},
  booktitle = {{TACAS} {(1)}},
  series    = {LNCS},
  volume    = {12078},
  pages     = {266--284},
  publisher = {Springer},
  year      = {2020}
}

@inproceedings{DBLP:conf/popl/HondaYC08,
author = {Honda, Kohei and Yoshida, Nobuko and Carbone, Marco},
title = {Multiparty asynchronous session types},
year = {2008},
isbn = {9781595936899},
publisher = {Association for Computing Machinery},
address = {New York, NY, USA},
url = {https://doi.org/10.1145/1328438.1328472},
doi = {10.1145/1328438.1328472},
booktitle = {Proceedings of the 35th Annual ACM SIGPLAN-SIGACT Symposium on Principles of Programming Languages},
pages = {273–284},
numpages = {12},
location = {San Francisco, California, USA},
series = {POPL '08}
}

@article{DBLP:journals/ftpl/AnconaBB0CDGGGH16,
  -author    = {Davide Ancona and
               Viviana Bono and
               Mario Bravetti and
               Joana Campos and
               Giuseppe Castagna and
               Pierre{-}Malo Deni{\'{e}}lou and
               Simon J. Gay and
               Nils Gesbert and
               Elena Giachino and
               Raymond Hu and
               Einar Broch Johnsen and
               Francisco Martins and
               Viviana Mascardi and
               Fabrizio Montesi and
               Rumyana Neykova and
               Nicholas Ng and
               Luca Padovani and
               Vasco T. Vasconcelos and
               Nobuko Yoshida},
  author    = {Davide {Ancona et al.}},
  title     = {Behavioral Types in Programming Languages},
  journal   = {Foundations and Trends in Programming Languages},
  volume    = {3},
  number    = {2-3},
  pages     = {95--230},
  year      = {2016}
}

@inproceedings{DBLP:conf/oopsla/LopezMMNSVY15,
  author    = {Hugo A. L{\'{o}}pez and
               Eduardo R. B. Marques and
               Francisco Martins and
               Nicholas Ng and
               C{\'{e}}sar Santos and
               Vasco Thudichum Vasconcelos and
               Nobuko Yoshida},
  title     = {Protocol-based verification of message-passing parallel programs},
  booktitle = {{OOPSLA}},
  pages     = {280--298},
  publisher = {{ACM}},
  year      = {2015}
}

@article{DBLP:journals/csur/HuttelLVCCDMPRT16,
  author    = {Hans H{\"{u}}ttel and
               Ivan Lanese and
               Vasco T. Vasconcelos and
               Lu{\'{\i}}s Caires and
               Marco Carbone and
               Pierre{-}Malo Deni{\'{e}}lou and
               Dimitris Mostrous and
               Luca Padovani and
               Ant{\'{o}}nio Ravara and
               Emilio Tuosto and
               Hugo Torres Vieira and
               Gianluigi Zavattaro},
  title     = {Foundations of Session Types and Behavioural Contracts},
  journal   = {{ACM} Comput. Surv.},
  volume    = {49},
  number    = {1},
  pages     = {3:1--3:36},
  year      = {2016}
}

@article{DBLP:journals/pacmpl/ScalasY19,
  author    = {Alceste Scalas and
               Nobuko Yoshida},
  title     = {Less is more: multiparty session types revisited},
  journal   = {Proc. {ACM} Program. Lang.},
  volume    = {3},
  number    = {{POPL}},
  pages     = {30:1--30:29},
  year      = {2019}
}

@article{DBLP:journals/jacm/BrandZ83,
  author    = {Daniel Brand and
               Pitro Zafiropulo},
  title     = {On Communicating Finite-State Machines},
  journal   = {J. {ACM}},
  volume    = {30},
  number    = {2},
  pages     = {323--342},
  year      = {1983}
}

@inproceedings{DBLP:conf/icalp/DenielouY13,
  author    = {Pierre{-}Malo Deni{\'{e}}lou and
               Nobuko Yoshida},
  title     = {Multiparty Compatibility in Communicating Automata: Characterisation
               and Synthesis of Global Session Types},
  booktitle = {{ICALP} {(2)}},
  series    = {LNCS},
  volume    = {7966},
  pages     = {174--186},
  publisher = {Springer},
  year      = {2013}
}

@inproceedings{DBLP:conf/concur/BocchiLY15,
  author    = {Laura Bocchi and
               Julien Lange and
               Nobuko Yoshida},
  title     = {Meeting Deadlines Together},
  booktitle = {{CONCUR}},
  series    = {LIPIcs},
  volume    = {42},
  pages     = {283--296},
  publisher = {Schloss Dagstuhl - Leibniz-Zentrum fuer Informatik},
  year      = {2015}
}

@inproceedings{DBLP:conf/cav/LangeY19,
  author    = {Julien Lange and
               Nobuko Yoshida},
  title     = {Verifying Asynchronous Interactions via Communicating Session Automata},
  booktitle = {{CAV} {(1)}},
  series    = {LNCS},
  volume    = {11561},
  pages     = {97--117},
  publisher = {Springer},
  year      = {2019}
}

@article{DBLP:journals/jlp/GhilezanJPSY19,
  author    = {Silvia Ghilezan and
               Svetlana Jaksic and
               Jovanka Pantovic and
               Alceste Scalas and
               Nobuko Yoshida},
  title     = {Precise subtyping for synchronous multiparty sessions},
  journal   = {J. Log. Algebraic Methods Program.},
  volume    = {104},
  pages     = {127--173},
  year      = {2019}
}

@inproceedings{DBLP:conf/popl/LangeTY15,
  author    = {Julien Lange and
               Emilio Tuosto and
               Nobuko Yoshida},
  title     = {From Communicating Machines to Graphical Choreographies},
  booktitle = {{POPL}},
  pages     = {221--232},
  publisher = {{ACM}},
  year      = {2015}
}

@article{DBLP:journals/pacmpl/JacobsBK22a,
  author    = {Jules Jacobs and
               Stephanie Balzer and
               Robbert Krebbers},
  title     = {Multiparty {GV:} functional multiparty session types with certified
               deadlock freedom},
  journal   = {Proc. {ACM} Program. Lang.},
  volume    = {6},
  number    = {{ICFP}},
  pages     = {466--495},
  year      = {2022}
}

@article{DBLP:journals/corr/abs-2203-12876,
  author    = {Ilaria Castellani and
               Mariangiola Dezani{-}Ciancaglini and
               Paola Giannini},
  title     = {Asynchronous Sessions with Input Races},
  journal   = {CoRR},
  volume    = {abs/2203.12876},
  year      = {2022}
}

@inproceedings{DBLP:conf/esop/JongmansY20,
  author    = {Sung{-}Shik Jongmans and
               N. Yoshida},
  title     = {Exploring Type-Level Bisimilarity towards More Expressive Multiparty
               Session Types},
  booktitle = {{ESOP}},
  series    = {LNCS},
  volume    = {12075},
  pages     = {251--279},
  publisher = {Springer},
  year      = {2020}
}

@inproceedings{DBLP:conf/pldi/Castro-Perez0GY21,
  author       = {David Castro{-}Perez and
                  Francisco Ferreira and
                  Lorenzo Gheri and
                  Nobuko Yoshida},
  editor       = {Stephen N. Freund and
                  Eran Yahav},
  title        = {Zooid: a {DSL} for certified multiparty computation: from mechanised
                  metatheory to certified multiparty processes},
  booktitle    = {{PLDI} '21: 42nd {ACM} {SIGPLAN} International Conference on Programming
                  Language Design and Implementation},
  address      = {Virtual Event, Canada, June 20-25, 2021},
  pages        = {237--251},
  publisher    = {{ACM}},
  year         = {2021},
  url          = {https://doi.org/10.1145/3453483.3454041},
  doi          = {10.1145/3453483.3454041},
  bibsource    = {dblp computer science bibliography, https://dblp.org}
}

@inproceedings{DBLP:conf/ecoop/TiroreBC25,
  author       = {Dawit Legesse Tirore and
                  Jesper Bengtson and
                  Marco Carbone},
  editor       = {Jonathan Aldrich and
                  Alexandra Silva},
  title        = {Multiparty Asynchronous Session Types: {A} Mechanised Proof of Subject
                  Reduction},
  booktitle    = {39th European Conference on Object-Oriented Programming, {ECOOP} 2025},
  address      = {June 30 to July 2, 2025, Bergen, Norway},
  series       = {LIPIcs},
  volume       = {333},
  pages        = {31:1--31:30},
  publisher    = {Schloss Dagstuhl - Leibniz-Zentrum f{\"{u}}r Informatik},
  year         = {2025},
  doi          = {10.4230/LIPICS.ECOOP.2025.31},
  bibsource    = {dblp computer science bibliography, https://dblp.org}
}

@inproceedings{DBLP:conf/itp/TiroreBC23,
  author       = {Dawit Legesse Tirore and
                  Jesper Bengtson and
                  Marco Carbone},
  editor       = {Adam Naumowicz and
                  Ren{\'{e}} Thiemann},
  title        = {A Sound and Complete Projection for Global Types},
  booktitle    = {14th International Conference on Interactive Theorem Proving, {ITP}
                  2023, July 31 to August 4, 2023, Bia{\l}ystok, Poland},
  series       = {LIPIcs},
  volume       = {268},
  pages        = {28:1--28:19},
  publisher    = {Schloss Dagstuhl - Leibniz-Zentrum f{\"{u}}r Informatik},
  year         = {2023},
  url          = {https://doi.org/10.4230/LIPIcs.ITP.2023.28},
  doi          = {10.4230/LIPICS.ITP.2023.28},
  bibsource    = {dblp computer science bibliography, https://dblp.org}
}

@inproceedings{DBLP:conf/cav/LiSWZ23,
  author       = {Elaine Li and
                  Felix Stutz and
                  Thomas Wies and
                  Damien Zufferey},
  editor       = {Constantin Enea and
                  Akash Lal},
  title        = {Complete Multiparty Session Type Projection with Automata},
  booktitle    = {Computer Aided Verification - 35th International Conference, {CAV}
                  2023, July 17-22, 2023, Proceedings, Part {III}},
  series       = {Lecture Notes in Computer Science},
  address      = {Paris, France},
  volume       = {13966},
  pages        = {350--373},
  publisher    = {Springer},
  year         = {2023},
  doi          = {10.1007/978-3-031-37709-9\_17},
  bibsource    = {dblp computer science bibliography, https://dblp.org}
}

@inproceedings{DBLP:conf/itp/LiW25,
  author       = {Elaine Li and
                  Thomas Wies},
  editor       = {Yannick Forster and
                  Chantal Keller},
  title        = {Certified Implementability of Global Multiparty Protocols},
  booktitle    = {16th International Conference on Interactive Theorem Proving, {ITP}
                  2025, September 28 to October 1, 2025, Reykjavik, Iceland},
  series       = {LIPIcs},
  volume       = {352},
  pages        = {15:1--15:20},
  publisher    = {Schloss Dagstuhl - Leibniz-Zentrum f{\"{u}}r Informatik},
  year         = {2025},
  url          = {https://doi.org/10.4230/LIPIcs.ITP.2025.15},
  doi          = {10.4230/LIPICS.ITP.2025.15},
  bibsource    = {dblp computer science bibliography, https://dblp.org}
}

@article{DBLP:journals/pacmpl/JacobsBK22,
  author       = {Jules Jacobs and
                  Stephanie Balzer and
                  Robbert Krebbers},
  title        = {Connectivity graphs: a method for proving deadlock freedom based on
                  separation logic},
  journal      = {Proc. {ACM} Program. Lang.},
  volume       = {6},
  number       = {{POPL}},
  pages        = {1--33},
  year         = {2022},
  doi          = {10.1145/3498662},
  bibsource    = {dblp computer science bibliography, https://dblp.org}
}

@inproceedings{DBLP:conf/scam/KlintSV09,
  author       = {Paul Klint and
                  Tijs van der Storm and
                  Jurgen J. Vinju},
  title        = {{RASCAL:} {A} Domain Specific Language for Source Code Analysis and
                  Manipulation},
  booktitle    = {Ninth {IEEE} International Working Conference on Source Code Analysis
                  and Manipulation, {SCAM} 2009, Edmonton, Alberta, Canada, September
                  20-21, 2009},
  pages        = {168--177},
  publisher    = {{IEEE} Computer Society},
  year         = {2009},
  url          = {https://doi.org/10.1109/SCAM.2009.28},
  doi          = {10.1109/SCAM.2009.28},
  bibsource    = {dblp computer science bibliography, https://dblp.org}
}

\ifdefined\WITHAPPENDIX
 \appendix

\startrev
\section{On Decidability and Efficiency}
\label{sect:behav:decid}

\subsection{Algorithm}
\label{sect:behav:decid:algo}

As long as MLTSs have finitely many states, the type system is decidable. To
substantiate this claim, we describe a type checking algorithm that succesfully
applies typing rules to all the processes of the session-under-analysis or
fails. Either way, it terminates. The type checker in our VS Code extension uses
this algorithm.

First, we note that the type system in \Cref{fig:typing-proc} is \textit{almost}
syntax-directed: all typing rules are, except rule \autorefrule{$\vdash$-Skip}.
However, premise 4 of rule \autorefrule{$\vdash$-Skip} demands that $\sexprx{P}$
is an output or input process (because $\mathsf{obj}(\sexprx{P})$ must be
defined), while premise 1 demands that the role implemented by $\sexprx{P}$
cannot send or receive (which is the opposite of rules
\autorefrule{$\vdash$-Send} and \autorefrule{$\vdash$-Recv}). Thus, it is
impossible for rule \autorefrule{$\vdash$-Skip} to be applicable at the same
time as the other rules (i.e., the rules are mutually exclusive). The
type checking algorithm leverages this insight by proceeding in two phases:
\begin{itemize}
  \item \textbf{Phase 1:} Apply all rules except \autorefrule{$\vdash$-Skip} in
  a bottom-up, syntax-directed fashion until a process is reached for which none
  of these rules are applicable.

  \item \textbf{Phase 2:} When no other rules apply, then check if
  \autorefrule{$\vdash$-Skip} is applicable. If so, then type checking continues
  with phase 1. Otherwise, type checking fails.
\end{itemize}
To see that each phase 1 and each phase 2 individually terminates, we first note
that all the premises of all the rules---including
\autorefrule{$\vdash$-Skip}---that require us to ``query'' an MLTS for the
presence/absence of particular transitions are decidable. This is a direct
consequence of the assumption that MLTSs have finitely many states (hence,
finitely many transitions to exhaustively consider). Algorithmically, the type
checker may compute the state space of an MLTS on-the-fly, by need, as it tries
to apply typing rules. There are two situations: a rule requires the computation
of a single transition, or multiple transitions in sequence. Rules
\autorefrule{$\vdash$-Send} and \autorefrule{$\vdash$-Recv} are an example of
the former. In this situation, it suffices to check whether the current state
admits the transition (which is the leitmotif of synthetic typing). In the case
of rule \autorefrule{$\vdash$-Skip}, sequences of transitions are computed
incrementally until a state is reached that allows the current process to
perform an action. With rule \autorefrule{$\vdash$-End}, the stop condition when
computing sequences of transitions is when a state in which the current process
is allowed to perform an action, is no longer reachable.

What remains to be shown, then, is that the type checking algorithm does not
diverge in an infinite sequence of phases 1 and 2. First, we note that phase 1
is \textit{structural} in the sense that with each rule application, the process
becomes smaller. Thus, it is impossible to have an infinite sequence of phase 1.
The key insight, now, is that it is never possible to apply rule
\autorefrule{$\vdash$-Skip} twice in a row. This is because after applying rule
\autorefrule{$\vdash$-Skip} once, there must be a remaining action, so one of
the structural rules must apply (i.e., as part of premise 3, due to the way
$\stypex{G''}$ is constructed, it is impossible to have another application of
rule \autorefrule{$\vdash$-Skip}, as premise 1 is false for $\stypex{G''}$).
Thus, after each phase 2, there must be a phase 1: it is also impossible to have
an infinite sequence of phase 2. So, all in all, we apply the structural rules
eagerly, until we type check the whole process, with the occasional
\autorefrule{$\vdash$-Skip} when no structural rules apply, after which another
structural rule must apply.%
\footnote{%
  \rev{Morally, \autorefrule{$\vdash$-Skip} could be removed and added to the rules
  for sending and receiving. This would reify the notion of applying all the
  structural rules first and then ``skipping''. We chose to separate
  ``skipping'' from rules \autorefrule{$\vdash$-Send} and
  \autorefrule{$\vdash$-Recv} since this leads to a simpler presentation.}%
}

\subsection{Performance}

Regarding performance, first, we note that using the algorithm of
\Cref{sect:behav:decid:algo}, type checking against an arbitrary MLTS is
polynomial in the number of states of that LTS.

In the special case, when the MLTS arises from a global type \textbf{without
parallel composition}, the size of the MLTS itself is linear in the size of that
global type. Thus, the complexity of type checking against a global type is
polynomial. In particular, such global types do not pose a particular
computational challenge when it comes to rule \autorefrule{$\vdash$-Skip}. To
further illustrate this, consider the following global type:
\begin{gather*}
	\stypex{\begin{tree}
			\branchx{
				\comxx{\rolex{a}}{\rolex{b}} \holex{h1}
			}
			\branchxxx[h1]{
				\mtypexx{\datax{Foo$1$}}{\datax{Nat}}
				\prewide
				\comxxxx{\rolex{b}}{\rolex{c}}{\datax{Bar}}{\datax{Bool}}
				\prewide
				\comxxxx{\rolex{c}}{\rolex{d}}{\datax{Baz}}{\datax{Bool}}
				\prewide
				\one
			}{
        \enspace\smash{{\vdots}}
      }{
				\mtypexx{\datax{Foo$N$}}{\datax{Nat}}
				\prewide
				\comxxxx{\rolex{b}}{\rolex{c}}{\datax{Bar}}{\datax{Bool}}
				\prewide
				\comxxxx{\rolex{c}}{\rolex{d}}{\datax{Baz}}{\datax{Bool}}
				\prewide
				\one
			}
		\end{tree}}
\end{gather*}
In this example, each of the $N$ branches has the same continuation, so the MLTS
is a DAG. Only this one continuation needs to be explored by rule
\autorefrule{$\vdash$-Skip} to type-check implementations of roles $\rnamex{c}$
and $\rnamex{d}$ (instead of $N$ continuations). If the continuations were
different, in contrast, then these processes would need to be checked against
each of those continuations. This is the same, though, with the classical
approach and the ``Less Is More'' approach. However, when the MLTS arises from a
global type \textbf{with parallel composition}, the size of the MLTS can be
exponential (due to the interleavings of the branches), so type checking becomes
more computationally costly. Most of the global types in the literature lack
parallel composition, though.

In general, compared to model checking in the ``Less Is More'' approach,
querying an MLTS in the synthetic approach comes with more control of any
potential exponential growth, in the sense that it is syntactically
confined (i.e., in our case, only parallel composition may trigger it, while in
the ``Less Is More'' approach, it may happen during model checking at any
point). So from the start, we are in a better position already. Moreover, given
our requirement that the types have the diamond property and that parallel
composition requires disjoint participants in the branches, we also expect that
we could use this information to cull the state space for performance (which
would bring a form of de facto projection to this work).

\subsection{Outlook: Beyond Finite MLTSs}

We see several opportunities to support MLTSs with infinitely many states. In
the most general case, we could admit behavioural specifications that generate
such MLTSs and then generate additional proof obligations (e.g., the fact that
certain states are reachable when applying \autorefrule{$\vdash$-Skip}).

Furthermore, going from a well-behaved formalism, these proof obligations could
be automated in the type checker. For instance, consider the use of pushdown
automata (or a form of context-free global types) as protocol specifications
with infinite state space, but for which reachability is still decidable. In
that case, the type checker can perform the reachability tests for us---possibly
using external tools that are optimised for this kind of analysis---so the
proof obligations could be automatically discharged. Depending on the nature of
the extension, the type system may remain decidable or become undecidable (but
still sound).

We note that a similar approach can be adopted when the formalism is unknown to
generate only well-behaved MLTSs (unlike with global types, for which
\Cref{thm:prop-behaviours} establishes that well-behavedness is guaranteed). In
that case, it needs to be proved separately for each MLTS that it is
well-behaved. The usage of external tools, such as model checkers, could be
useful in this case, too. However, depending on additional restrictions that we
might place on MLTSs, one may be able to internalise and automate the
well-behavedness checking and avoid depending on an external tool.

\finishrev

\fi

\end{document}
\endinput